\documentclass[useAMS,usenatbib]{mn2e}
\usepackage{txfonts}
\usepackage{color,epsfig,amssymb,longtable}
\usepackage{array,colortbl,lscape}
\usepackage{graphicx}
\usepackage{subfigure}
\usepackage{float}
\usepackage{epstopdf}
\usepackage{lscape} 
\usepackage{rotating}
\usepackage{pdflscape}

\newcommand{\halpha}{H$_{\alpha}$}
\newcommand{\hbeta}{H$_{\beta}$}
\newcommand{\hgamma}{H$_{\gamma}$}
\newcommand{\hdelta}{H$_{\delta}$}
\newcommand{\Msun}{M$_{\odot}$}
\bibpunct[,]{(}{)}{;}{a}{,}{,}

\begin{document}

\title[{\sc PopStar} III: Photometric properties of young star clusters 
and mixed populations] {{\sc PopStar} Evolutionary Synthesis Models III: 
Photometric properties of young star clusters and mixed populations}

\author[Garc{\'{\i}}a-Vargas, Moll{\'{a}} and Mart\'{\i}n-Manj\'{o}n]
{M.L. Garc{\'{\i}}a-Vargas$^{1}$ \thanks{e-mail:marisa.garcia@fractal-es.com},
M.~Moll{\'{a}}$^{2}$, and M.L. Mart\'{\i}n-Manj\'{o}n$^{3}$  \\
$^{1}$ FRACTAL SLNE, C/ Tulip\'{a}n 2, p13, 1A , 28231 Las Rozas de Madrid,
(Spain)\\
$^{2}$ Departamento de Investigaci\'{o}n B\'{a}sica, CIEMAT,
Avda. Complutense 40, 28040, Madrid, (Spain) \\
$^{3}$ Departamento de F\'{\i}sica Te\'{o}rica, Universidad Aut\'onoma de
Madrid. Cantoblanco. E-28049 Madrid. Spain. }

\date{Accepted Received ; in original form }
\pagerange{\pageref{firstpage}--\pageref{lastpage}} \pubyear{2010}

\maketitle

\begin{abstract}
This is the third paper of a series reporting the results from the {\sc 
PopStar} evolutionary synthesis models.  The main goal of this work is 
to present and discuss the synthetic photometric properties of Single 
Stellar Populations (SSPs) resulting from our {\sc PopStar} code. 
Colours in the Johnson and SDSS systems, \halpha\ and \hbeta\ 
luminosities and equivalent widths, and ionising region size, have been 
computed for a wide range of metallicity (Z $=$ 0.0001 - 0.05) and age 
(0.1\,Myr to 20\,Gyr).  We calculate the evolution of the cluster and 
the region geometry in a consistent manner. We demonstrate the 
importance of the contribution of emission lines to broader-band 
photometry when characterising stellar populations, through the 
presentation of both contaminated and non-contaminated colours (in both 
the Johnson and SDSS systems).  The tabulated colours include stellar 
and nebular components, in addition to line emission.  The main 
application of these models is the determination of physical properties 
of a given young ionising cluster, when only photometric observations 
are available; for an isolated star forming region, the young star 
cluster models can be used, free from the contamination of any 
underlying background stellar population. In most cases, however, the 
ionising population is usually embedded in a large and complex system, 
and the observed photometric properties result from the combination of a 
young star-forming burst and the underlying older population of the 
host. Therefore, the second objective of this paper is to provide a grid 
of models useful in the interpretation of mixed regions where the 
separation of young and old populations is not sufficiently reliable. We 
describe the set of PopStar Spectral Energy Distributions (SEDs), and 
the derived colours for mixed populations where an underlying host 
population is combined in different mass ratios with a recent ionising 
burst. These colours, together with other common photometric parameters, 
such as the \halpha\ radius of the ionised region, and Balmer line 
equivalent widths and luminosities, allow one to infer the physical 
properties of star-forming regions even in the absence of spectroscopic 
information.
\end{abstract}


\section{Introduction}

For many years, broad-band photometry was used as the primary tool in 
deriving ages for both single bursts and mixed stellar populations. 
Broadly speaking, blue colours were taken to be indicative of young 
populations, while red colours were associated with older populations.  
Only dust was thought to complicate this picture, through the general 
reddening of the population. From an observational perspective, few 
works have considered the potential contaminating impact of gas emission 
upon broadband colours. For example, \cite{arls08} analysed carefully, 
colour-colour diagrams, taking into account this correction, and thereby 
proposing a mechanism for separating the contribution from the 
underlying stellar population, starburst population, and line emission, 
for better quantitative comparison with evolutionary models of stellar 
populations. From a theoretical perspective, colours were calculated 
typically for the stellar component only or, in a few cases, by taking 
into account the nebular continuum. Only recently, several works 
\citep[][ and references therein]{mmetal08,reines10} have started to 
quantify the role of line emission from ionised gas surrounding young 
star clusters in the contamination of broad-band colours. These latter 
authors examined two young massive clusters in NGC~4449, demonstrating 
that the contribution of both nebular continuum and line emission is 
essential to reproduce the observed broad-band fluxes.  The comparison 
of model spectral energy distributions with and without nebular continuum 
and/or emission lines shows that the inferred stellar mass can change by 
up to a factor of $\sim$2.5, depending upon the filter in question. 
\citet{atek11} also find that nebular lines can contaminate the total 
broadband flux by $\sim$0.3~mag (median, but up to $\sim$1~mag, in 
certain cases).  This effect is important since low-redshift galaxies 
with active star formation may mimic the colour-selection criteria used 
in some high-redshift dropout surveys. This potential impact on the mass 
and age of galaxies is also important for cosmological studies which use 
these values as clues to disentangling the temporal history of galaxy 
assembly.  Despite this importance, to our knowledge there is no 
available systematic grid of models which includes the effect of 
emission lines on broadband colours spanning a wide range in age and 
metallicity.  Our work has been designed to fill this gap.

Over the past two decades, a wealth of powerful ground-based facilities, 
particularly in the visible, has driven the use of spectroscopic 
techniques to infer the underlying physical properties of ionising star 
clusters.  High signal-to-noise spectra were required, in order to 
derive the nebular electron density and electronic temperature, and from 
these gas parameters, the abundance content. In the absence of a 
consistent grid of models against which to compare, photometric 
information has not been used as the main tool for deriving star-forming 
properties. In terms of observations, there were few consistent samples 
of H{\sc ii} regions (or small to medium-size star-forming objects) that 
would allow us to test models and from which we could extract 
statistical conclusions. Moreover, these existing observational samples 
were usually biased towards bright and low-to-intermediate metallicity 
regions. This subject has been exhaustively discussed by \citet[][ 
hereinafter Paper II]{mmetal10}, in which we presented the emission line 
spectra for our models as a function of the physical properties of the 
ionising clusters and compared the resulting diagnostic diagrams with a 
complete spectroscopic sample of H{\sc ii} regions for which abundances 
were derived consistently using an appropriate empirical calibration.

The recent generation of deep surveys, taken with mid-to-large aperture 
telescopes, have now released complete photometric catalogues of star 
forming regions and galaxies at different redshifts. This has motivated 
us to deliver a set of models which can aid in the derivation of the 
physical properties of stellar populations, without the necessity of 
spectroscopic data. With this in mind, we have computed the colours and 
other common photometric parameters, such \halpha\ radius, \halpha\ and 
\hbeta\ equivalent widths and \halpha\ and \hbeta\ luminosities. We are 
convinced that these models will be a powerful tool in the 
interpretation of star-forming region photometric data, providing the 
means to infer their mass, age, and metallicity when embedded within 
more complex and evolved systems.

We have computed colours previously with the {\sc PopStar} SEDs in both 
Johnson and SDSS systems \cite[][ hereafter, Paper I]{mgvb09}.  We 
present now the colour evolution of star forming regions formed by a 
SSP, including the contribution of the strongest emission lines from the 
ionised nebula, which can be used to derive the physical properties of 
the ionising clusters. These SSP models can be used directly to compare 
with specific, usually small and detailed, photometric samples, where 
colours have been corrected for the contamination from an underlying 
population. However, this is not the general case, since the subtraction 
of the underlying population delivers results that are not always 
reliable; furthermore, this approach is not the most commonly used one 
when a large amount of data, like the output of extremely large surveys, 
are analysed.  Such surveys are used to obtain statistical conclusions 
about the star forming regions, and therefore the use of appropriate 
theoretical models is essential. For this reason, in addition to the SSP 
models, we have also computed a complete set of photometric properties 
for mixed populations, to simulate local star-forming regions embedded 
in more complex star systems. These models are particularly useful when 
the subtraction of an underlying population is not sufficiently 
reliable.

This paper describes the photometrical model and demonstrates how 
photometry can be an alternative and powerful tool in the delivery of 
inferred ages and metallicities for star forming regions, as well as how 
it can give an estimation of the mass-ratio of the star-forming region 
to that of the underlying population. The models presented here are 
applicable to nearby/local star forming regions only, since the 
contribution of the emission lines varies with redshift. It is obvious 
that the most important emission lines contaminate different filters, 
with different transmittance, as a function of redshift.  We outline the 
importance of an extension of this work to higher redshift, to study the 
impact of emission lines on broadband photometry of samples at different 
redshift. Dust re-emission and near-IR photometry will be also included 
to extend the use of the models to large samples covering a wide range 
of redshifts.

Section 2 summarises the main properties of the grid of evolutionary 
synthesis models used to compute the magnitudes of pure SSPs, the method 
to calculate the emission line intensities of the photo-ionised nebula 
and, consequently, the associated contaminated colours. Section 3 
presents and analyses the results of our models for young ionising SSPs. 
These models can be used for deriving the physical properties of the 
ionising populations in regions whose observed colours have been 
previously de-contaminated from the underlying populations. Section 4 
describes the set of models for mixed populations, applicable to 
characterising star-forming regions where these bursts are placed on an 
underlying population and where we cannot separate both components 
observationally. Finally, Section 5 summarises our results.

\section{SSP Colour Calculations}
\subsection{{\sc PopStar} Model Summary}

A detailed description of {\sc PopStar} can be found in Paper I. {\sc 
PopStar} provides a set of evolutionary synthesis models for SSPs, 
covering a wide range in age and metallicity.  The basic grid is 
composed of SSPs for six different initial mass functions (IMFs).  For 
this work, we have used only a Salpeter IMF \citep{sal55} with lower and 
upper mass limits of 0.15 and 100~M$_{\odot}$, respectively. We have not 
included binaries nor mass segregation.

The isochrones employed are updated versions of those from \citet{bgs98} 
for six metallicities: Z = 0.0001, 0.0004, 0.004, 0.008, 0.02, and 0.05. 
The use of very low metallicity models (Z=0.0001) was not included in 
comparable work prior to that of {\sc PopStar}. The age ranges from 
$\log{\tau}=5.00$ to 10.30 with a variable time resolution reaching 
$\Delta(\log{\tau})=0.01$ in the youngest populations. Again, details of 
the isochrones are described in Paper~I.

Stellar atmosphere models are taken from \citet{lcb97}, due to its 
expansive coverage in effective temperature, gravity, and metallicities, 
for stars with Teff $\leq 25000$K.  For O, B, and WR stars, the NLTE 
blanketed models of \citet{snc02} (for metallicities Z $=0.001$, 0.004, 
0.008, 0.02, and 0.04) are used. There are 110 models for O-B stars, 
calculated by \citet{phl01}, with 25000\,K $<$ Teff $\leq 51500$\,K and 
$2.95 \leq \log{g} \leq 4.00$, and 120 models for WR stars (60 WN and 60 
WC), from \citet{hm98}, with 30000\,K $\leq T^{*} \leq 120000$\,K and 
$1.3\,R_{\odot}\leq R^{*}\leq 20.3\,R_{\odot}$ for WN, and with $ 
40000\,K \leq T^{*} \leq 140000$\,K and $ 0.8\,R_{\odot}\leq R^{*}\leq 
9.3 \,R_{\odot}$ for WC. T$^{*}$ and R$^{*}$ are the temperature and the 
radius at a Roseland optical depth of 10. The assignment of the 
appropriate WR model is consistently made by using the relationships 
between opacity, mass loss, and velocity wind, as described in Paper~I. 
For post-AGB and planetary nebulae (PN) with T$_{\rm eff}$ between 
50000\,K and 220000\,K, the NLTE models from \citet{rau03} are taken. For 
higher temperatures, {\sc PopStar} uses black-bodies. The use of these 
latter models modifies the resulting intermediate age SEDs.

For each cluster, the total mechanical energy from stellar winds and 
supernova has been calculated. We have used this mechanical energy to 
calculate the H{\sc ii} region's inner radius (see \S2.2).  We have also 
computed the nebular continuum emission from hydrogen and helium (He and 
He$^{+}$) free-free, free-bound, and 2-photon continuum emission.

The SEDs (stellar+nebular) corresponding to the six metallicities and 
ages up to 20\,Myr have been introduced to the photo-ionisation code 
CLOUDY \citep{fer98} to obtain the emission line spectra (see Paper II 
for details). We take seven possible cluster masses: 1.2 $\times\ 
10^{4}$, 2.0 $\times\ 10^{4}$, 4.0 $\times\ 10^{4}$, 6.0 $\times\ 
10^{4}$, 1.0 $\times\ 10^{5}$, 1.5 $\times\ 10^{5}$ and 2.0 $\times\ 
10^{5}$ \Msun\, selected to cover the \halpha\ luminosity range observed 
for medium-to-large H{\sc ii} regions. The gas surrounding the cluster 
is assumed to have the same chemical composition as the stars of the 
cluster. For each metallicity, the element abundances heavier than 
helium have been scaled by a constant factor, with respect to the 
hydrogen content, according to the solar abundances from \citet{gre98} 
and depleted when necessary (see Table 2 from Paper II). The models 
assume a bubble geometry, with an ionised bounded nebula whose size is 
given by the cluster evolution. Therefore, the internal radius of the 
shell is the distance at which the ionised gas is deposited in that 
nebula by the cluster's mechanical energy.  Be aware that the observed 
radius of an H{\sc ii} region is not necessarily the inner radius of the 
ionised region (see \S2.2). The effect of dust here has not been 
included.

\begin{figure}
\centering
\subfigure{\includegraphics[width=0.485\textwidth,angle=0]{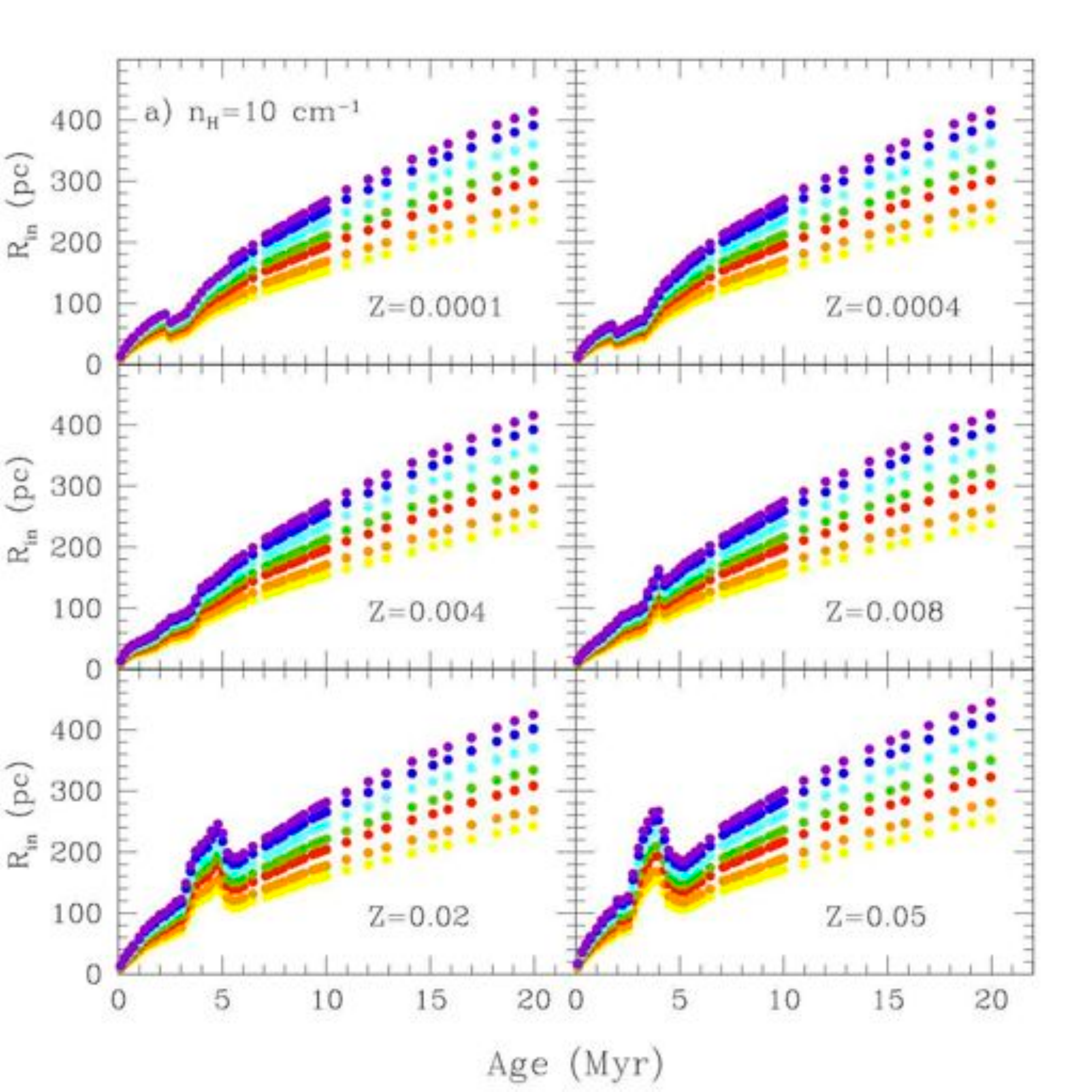}}
\subfigure{\includegraphics[width=0.485\textwidth,angle=0]{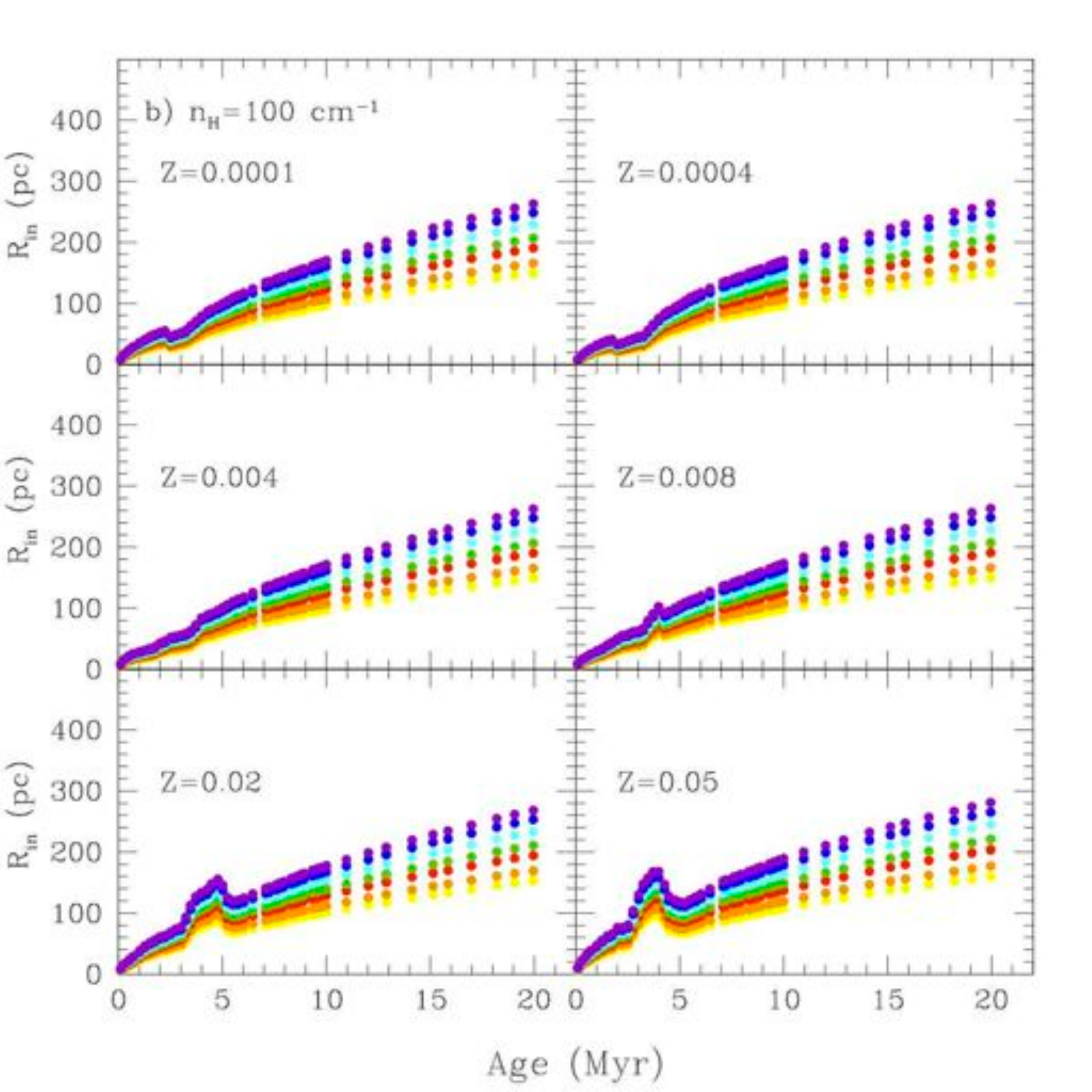}}
\caption{Region inner radius evolution: $\rm R_{in}$ (pc) {\sl vs}
cluster age (in\,Myr) for six different metallicities as labelled and
different cluster mass, plotted in different colours. Cluster
masses of 0.12, 0.20, 0.40, 0.60, 1.00, 1.50 and 2.00 $\times$ 10$^5$
\Msun\ have been plotted in yellow, orange, red, green, cyan, blue and
purple, respectively. The IMF is that of Salpeter with $\rm m_{low}=0.15
M_{\odot}$ and $\rm m_{up}= 100 M_{\odot}$. Panel (a) shows models
using a value of the gas electron density of n$_{e}$ = 10 cm$^{-3}$, while in panel (b) 
they are computed with n$_{e}$ = 100 cm$^{-3}$.}
\label{rin_age}
\end{figure}

The hydrogen density has been considered constant throughout the nebula 
and equal to the electron density for complete ionisation. We have 
generated models assuming two different values of electron density, 
n$_{H}$: 10~cm$^{-3}$ and 100~cm$^{-3}$, in order to quantify the impact 
of density on the emitted spectrum.  A density of 10 cm$^{-3}$ is more 
appropriate for small-medium isolated H{\sc ii} regions while 100 
cm$^{-3}$ is more appropriate for H{\sc ii} galaxies, large 
circumnuclear H{\sc ii} regions, often found around the nuclei of 
starbursts, or AGN. Although the constant density hypothesis is not 
realistic when detailed nebular studies are done, it can be considered 
representative when the integrated spectrum of the nebula is analysed. 
Results from the photo-ionisation models and their application to 
spectroscopic observations of H{\sc ii} regions are widely discussed in 
Paper~II.

The number of ionising photons, Q(H), and the \halpha\ and \hbeta\ 
luminosities have been calculated in Paper I. The equivalent width of 
\halpha\ and \hbeta\ are here calculated from the single population plus 
the nebular continuum (in the case of SSPs) or from the sum of all 
stellar populations plus nebular continuum, in the case of mixed 
populations (\S4). From the computed SEDs, we previously calculated 
their corresponding total (stellar + nebular) uncontaminated colours in 
Paper I; now, using the CLOUDYs output, we compute the colours including 
the contamination from emission lines, as described in \S2.3.

\subsection{Basic Properties of H{\sc ii} Regions: Sizes and Luminosities}

Throughout this work, we assume a scenario in which sufficient gas 
exists to be ionised (i.e., matter-bounded models have not been 
considered) and in which the birth of a star cluster, which we consider 
to be placed in the centre of a spherical region, is produced.  In fact, 
we will not be able to detect the region in the visible until the 
neutral gas has started to be ionised, allowing detection of the 
emission lines (in particular the Balmer lines). For that, we need to 
have not only ionising photons but also certain nebula conditions, 
including a given gas density and an optical depth for the emitting region. 
These conditions occur around 0.5 - 1.0\,Myr after cluster formation. As 
cluster evolves, the mechanical energy of the massive stars winds begins 
to sweep the gas away, compressing it and producing a shell. The 
wind-driven shell begins then to evolve with an initial phase of free 
expansion, followed by an adiabatic expansion phase, and then the 
swept-up material collapses to a thin, cold shell as a result of 
radiative cooling \citep[e.g.]{gib94,gtt}.

\begin{figure}
\centering
\resizebox{\hsize}{!}{\includegraphics[width=0.45\textwidth,angle=-90]{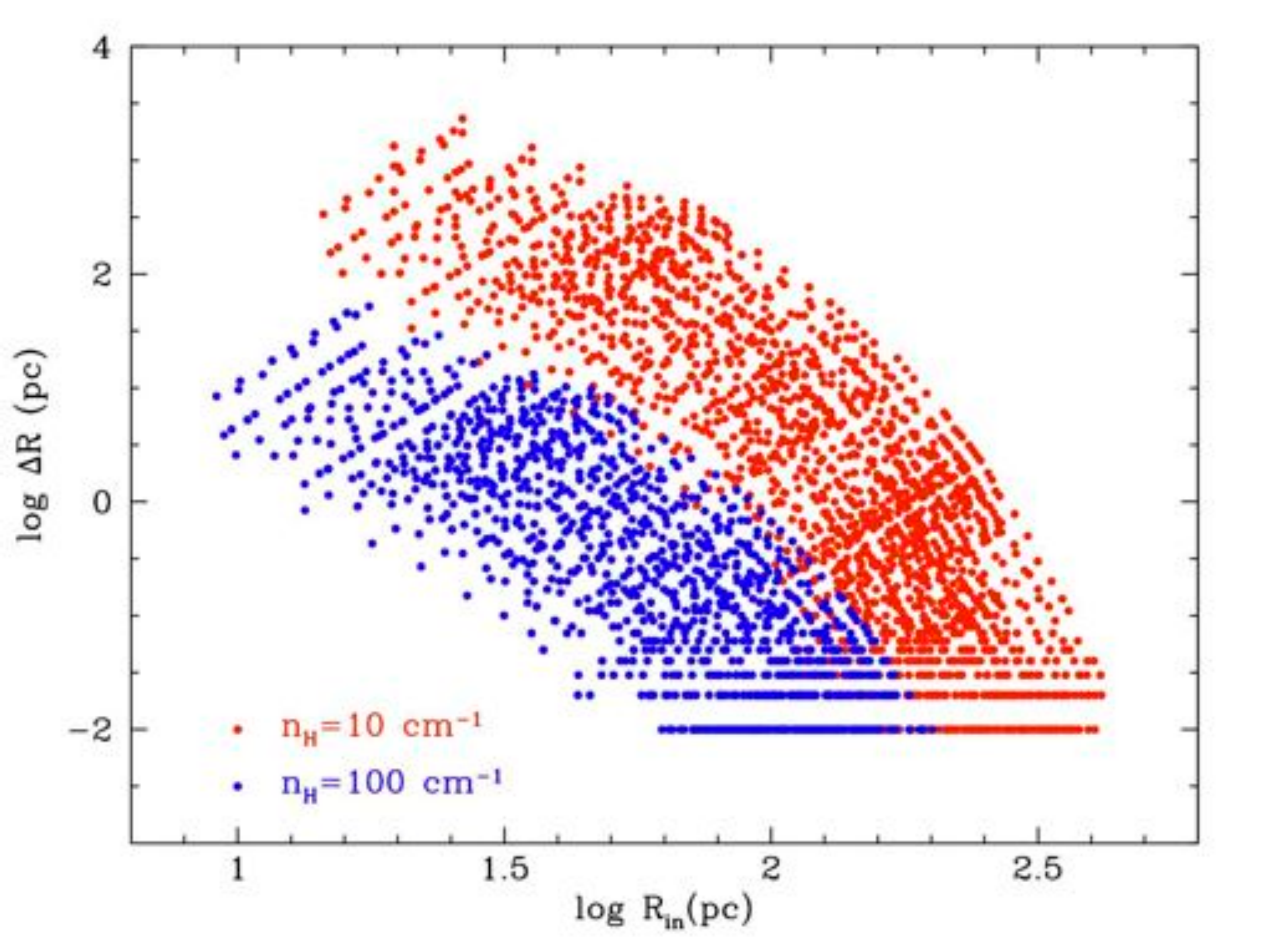}}
\caption{Shell thickness {\sl vs} inner radius $\rm R_{in}$ (pc) for
all metallicities, cluster masses and ages up to 20\,Myr. Models with
n$_{e}$ = 10 cm$^{-3}$ are plotted in red while models with n$_{e}$ =
100 cm$^{-3}$ are plotted in blue.}
\label{deltar-rin}
\end{figure}
\begin{table}
\caption{Computed radii for the modelled stellar clusters of $\rm Z = 0.008$ at selected ages for n$_{e}$ = 10 cm$^{-3}$. Complete table 1 for n$_{e}$ = 10 and 100 cm$^{-3}$ can be found online}
\begin{tabular}{ccrrcr}
\multicolumn{6}{c}{n$_{e}$ = 10 cm$^{-3}$}\\
\hline
Z & log$\tau$ &  $\rm M_{cl}$ & $\rm R_{in}$ &  $\Delta$ R &  $\rm R_{out}$ \\
   &   (yr)    &  (10$^{4}$\Msun ) &  (pc)  &  (pc) & (pc) \\
\hline   
 0.008 &  6.30&      1.2  &       45.440  &     1.72e+01 &        62.649 \\    
 0.008 &  6.30&      2.0  &       50.320  &     2.34e+01 &        73.717 \\    
 0.008 &  6.30&      4.0  &       57.810  &     3.55e+01 &        93.261 \\    
 0.008 &  6.30&      6.0  &       62.690  &     4.53e+01 &       108.005 \\    
 0.008 &  6.30&     10.0  &       69.440  &     6.16e+01 &       131.017 \\    
 0.008 &  6.30&     15.0  &       75.300  &     7.85e+01 &       153.832 \\    
 0.008 &  6.30&     20.0  &       79.760  &     9.33e+01 &       173.100 \\    
 0.008 &  6.40&      1.2  &       51.980  &     1.10e+01 &        62.994 \\    
 0.008 &  6.40&      2.0  &       57.570  &     1.50e+01 &        72.541 \\    
 0.008 &  6.40&      4.0  &       66.130  &     2.27e+01 &        88.873 \\    
 0.008 &  6.40&      6.0  &       71.720  &     2.90e+01 &       100.717 \\    
 0.008 &  6.40&     10.0  &       79.440  &     3.94e+01 &       118.846 \\    
 0.008 &  6.40&     15.0  &       86.150  &     5.02e+01 &       136.399 \\    
 0.008 &  6.40&     20.0  &       91.250  &     5.99e+01 &       151.115 \\    
\hline 
\label{radius}
\end{tabular}
\end{table}

\begin{figure*}
\centering
\subfigure{\includegraphics[width=0.485\textwidth,angle=0]{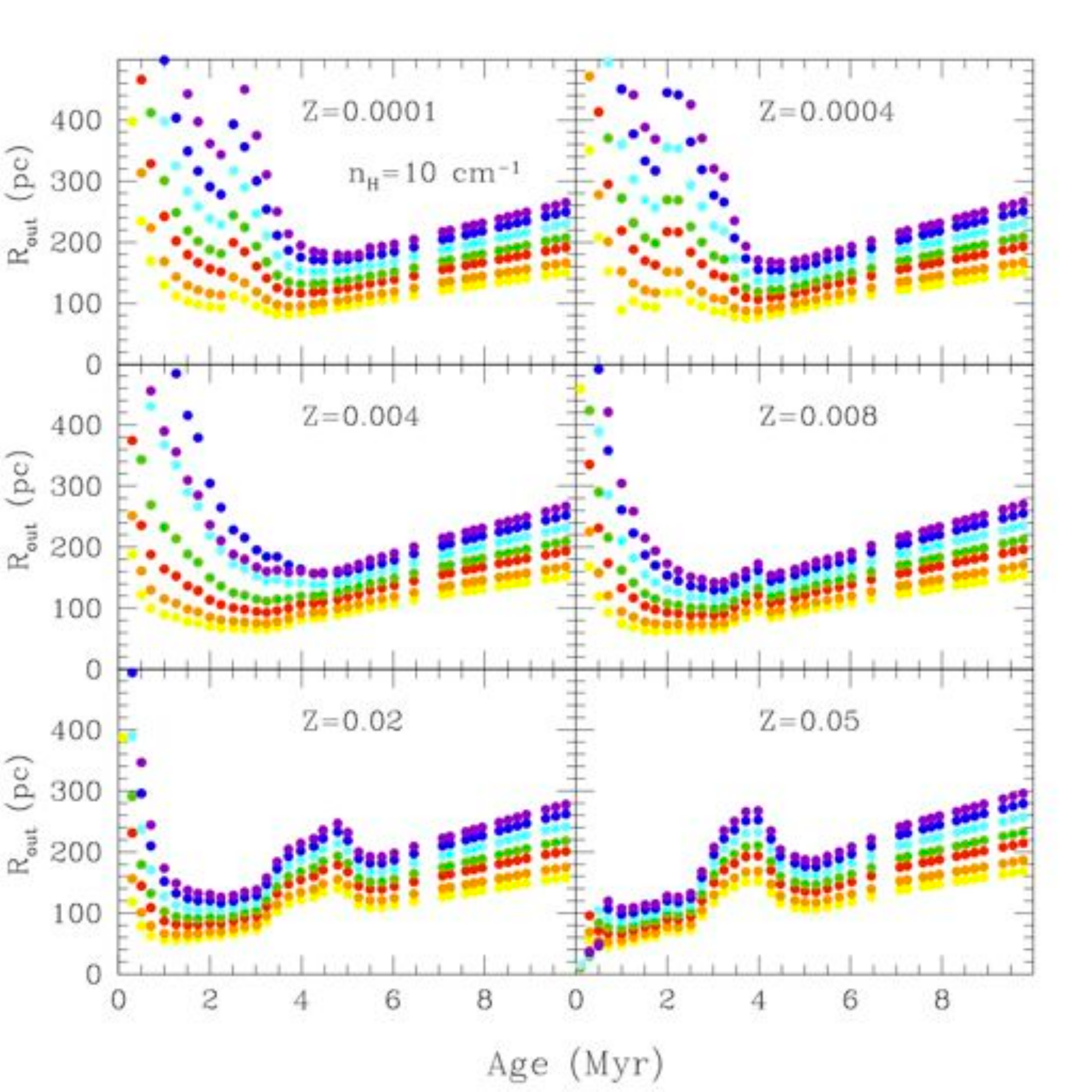}}
\subfigure{\includegraphics[width=0.485\textwidth,angle=0]{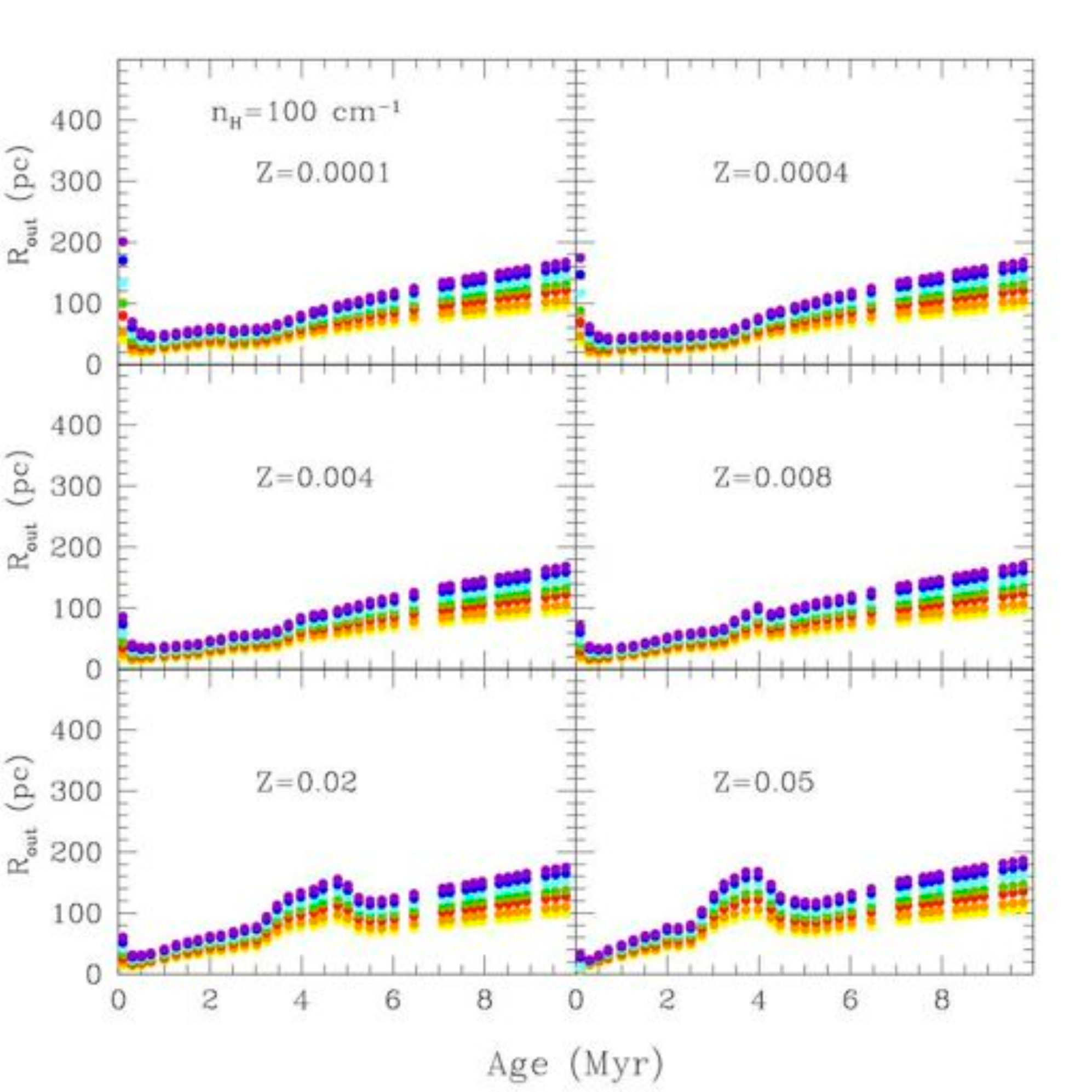}}
\caption{Region external radius evolution: $\rm R_{out}$ (pc) {\sl vs}
cluster age (Myr) for six different metallicities, all ages up to
20\,Myr and different cluster's mass, using the same colour coding that in
Fig.~\ref{rin_age}}
\label{rout_age}
\end{figure*}

At this stage the gas traps the ionisation front and the radiative phase 
begins. In this phase, the ionising photons are absorbed and the region 
cools via emission in the Balmer lines. In this process, the radius of 
the shock, limiting the inner border of the shell, R$_{in}$, evolves as 
\citep{cmw75}:

\begin{center}
R$_{in} = 1.6(\epsilon/n)^{1/5} t^{3/5}$ (pc)
\end{center} 

\noindent
where $\epsilon$ is the total injected mechanical energy per unit time 
in units of 10$^{36}$ ergs s$^{-1}$, $n$ is the interstellar medium 
density in units of cm$^{-3}$, and $t$ is the age of the shell in units 
of 10$^{4}$ yr. We have used the radius computed by the above equation 
as given for each time $t$ and 1~\Msun\ in Table 4 from Paper ~I, scaled 
to the stellar cluster mass of each model. We have extrapolated this 
bubble geometry to a shell structure formed by the combined effects of 
the mechanical energy deposition from the massive stars winds and SN 
explosions belonging to the ionising cluster. The ionised gas is assumed 
to be located in a thin spherical shell at that distance R$_{in}$ from 
the ionising source (called R$_{s}$ in Paper I). The shell inner radius 
results can be seen in Table~\ref{radius}.

The stacking of the material is caused by the shock wave providing a 
compression of the surrounding gas and therefore an increase of the 
recombinations ($\propto$ n$^{2}$). Due to the balance between the 
ionising flux and the number of recombinations in the whole region, the 
presence of a shell more massive and larger will affect the overall size 
of the H{\sc ii} region and the ionisation front, which will be trapped 
inside the shell where all the photons will be employed. This happens 
when the following condition is fulfilled:

\begin{equation}
 Q(H) = 4 \pi R_{in}^{2} \Delta R \beta_{2} n_{H}^{2}
\end{equation} 

\noindent
where n$_{H}$ is the density of the swept-up material and $\beta_{2}$ is 
the hydrogen recombination coefficient for levels higher than the 
fundamental one. Once the ionised front is trapped, the previously 
ionised material will recombine (corresponding to the initial swept-up 
mass (now in the shell and pushed by the shocked wind) which equates to 
the mass originally contained inside R$_{in}$).

\begin{figure}
\resizebox{\hsize}{!}{\includegraphics[width=0.40\textwidth,angle=0]{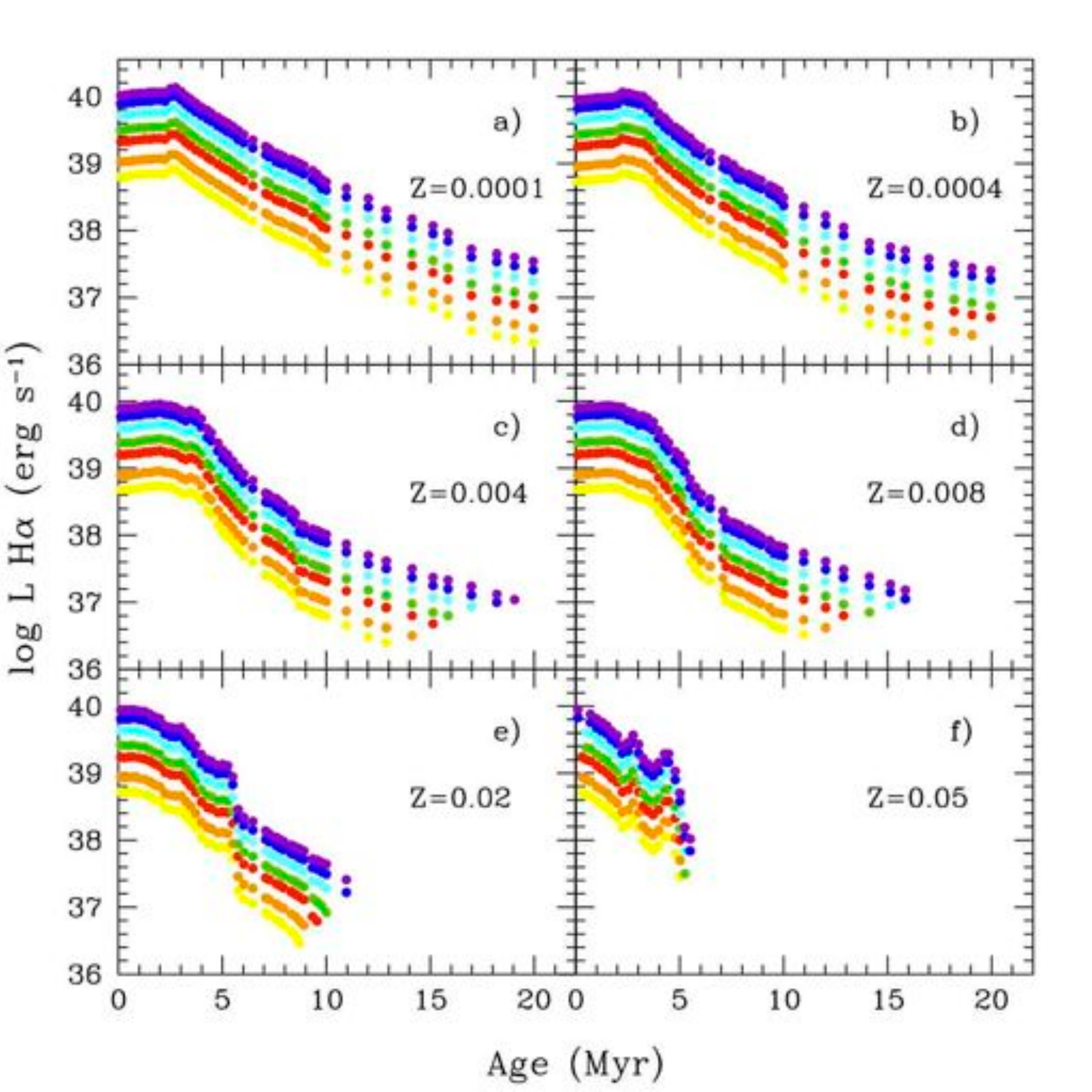}}
\caption{\halpha\ evolution: log L \halpha\ (erg/s) {\sl vs} cluster
age (Myr) for six different metallicities and different cluster
masses, plotted using the same colour coding that in
Fig.~\ref{rin_age}.}
\label{lha_age}
\end{figure}

\begin{figure}
\resizebox{\hsize}{!}{\includegraphics[width=0.48\textwidth,angle=-90]{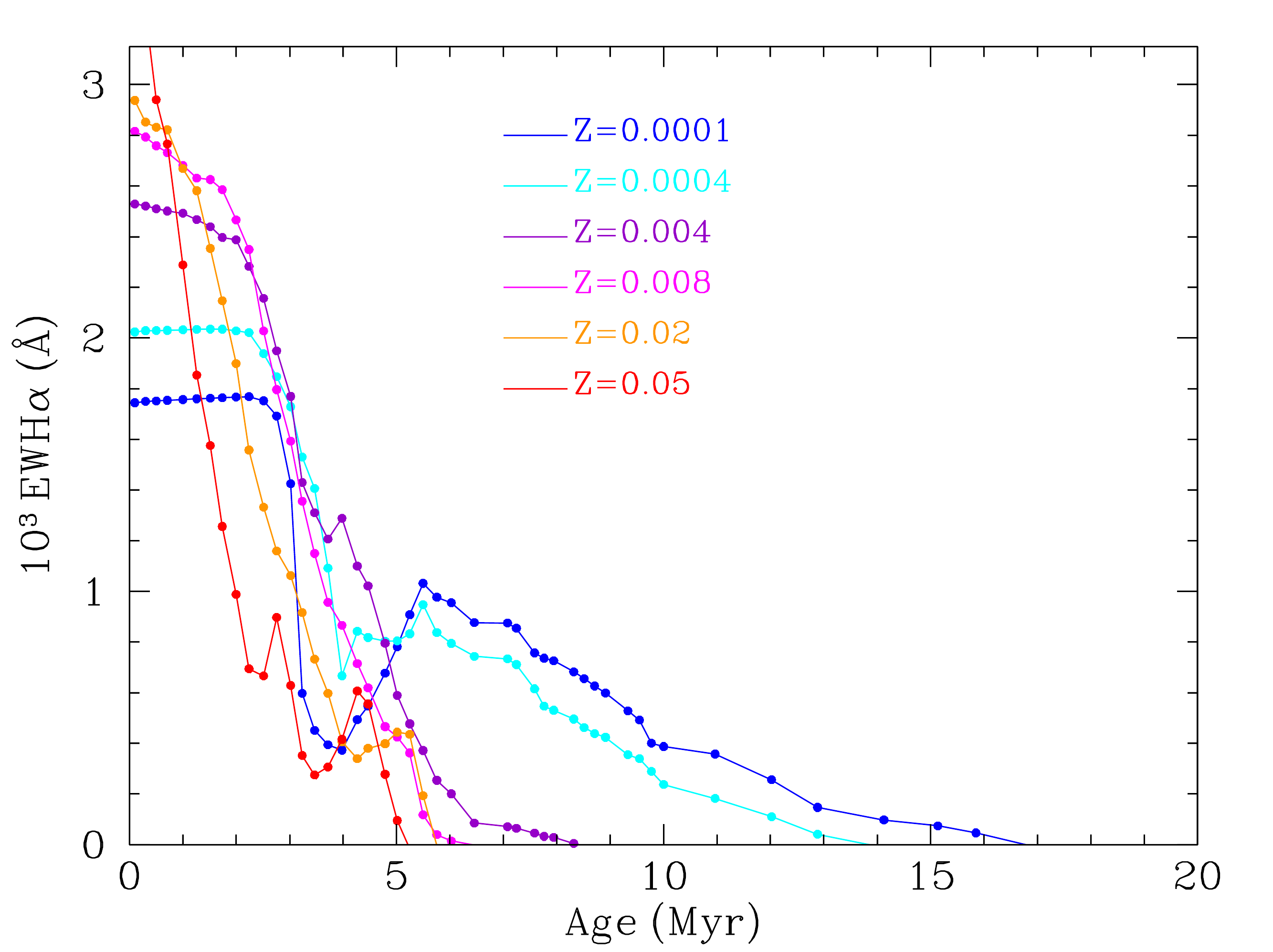}}
\caption{EW \halpha\ evolution: EW \halpha\ (\AA) {\sl vs} cluster age
(Myr) for six different metallicities as labelled.}
\label{ewha_age}
\end{figure}

We have therefore calculated the shell thickness, $\Delta$R, and the
outer radius, R$_{out}$ (the sum of the inner radius plus
the shell thickness). Table~\ref{radius} partially shown here summarises the results for the
modelled stellar clusters of $\rm Z = 0.008$ at selected ages and
n$_{e} =$ 10 cm$^{-3}$. The complete Table~\ref{radius},
 available in electronic format, for all ages and metallicities, is computed
for two values of the ionising gas electron density, since this
parameter influences both the inner radius and the shell thickness:
n$_{e} =$ 10\,cm$^{-3}$ and n$_{e} =$ 100\,cm$^{-3}$.
In each one, columns are: the
metallicity Z, the logarithm of the age, log $\tau$ (in yr), the
cluster mass, $\rm M_{cl}$ (in units of 10$^{4}$\, \Msun), the H{\sc ii}
region inner radius, $\rm R_{in}$ (in pc), the shell thickness,
$\Delta R$ (in pc) and the total radius of the region, R$_{out}$ (in
pc), this latter radius being the most appropriate when comparing with
real photometric radii observed in most H{\sc ii} regions (usually
measured from \halpha\ images). Both tables are available in
electronic format.

\begin{figure*}
\centering
\subfigure{\includegraphics[width=0.485\textwidth]{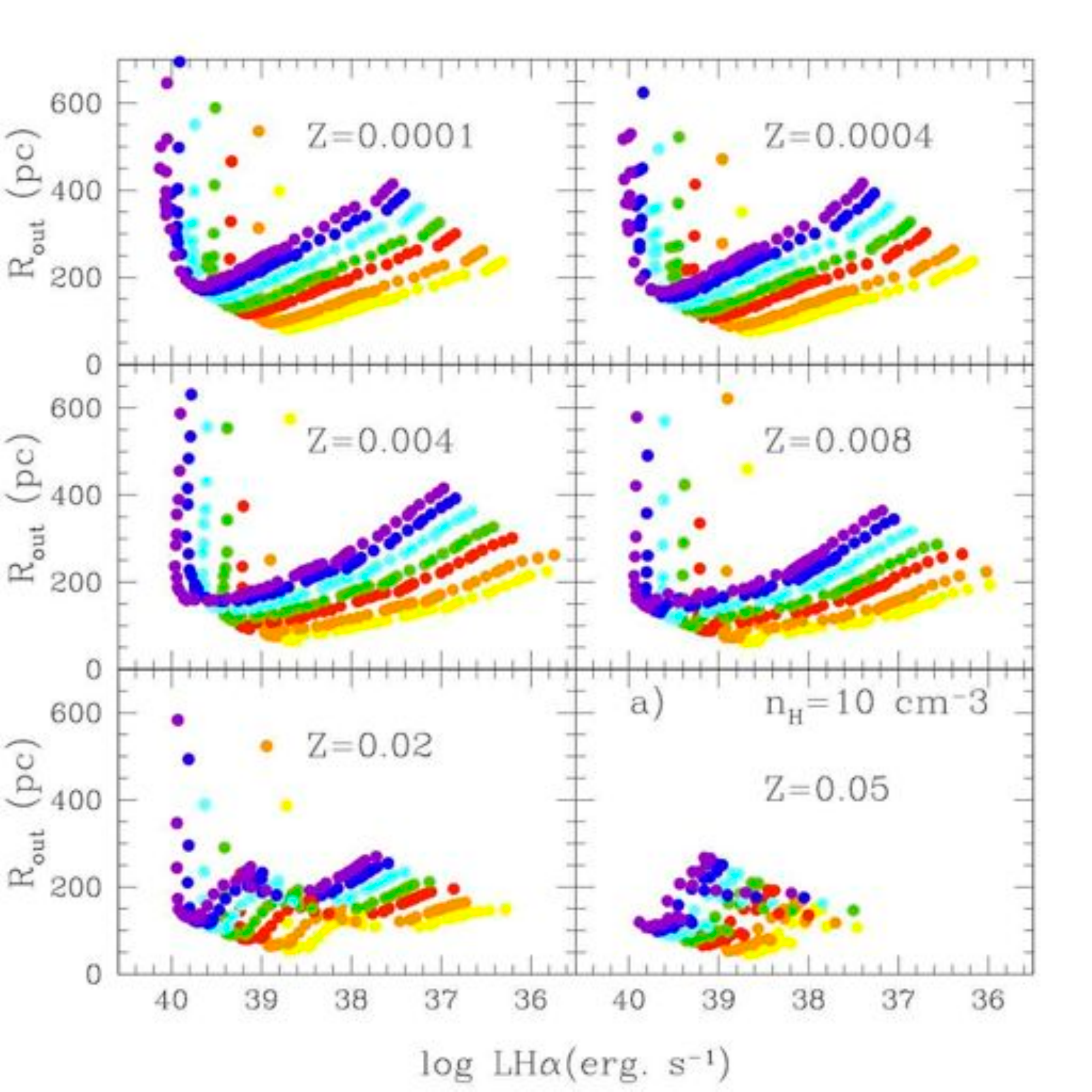}}
\subfigure{\includegraphics[width=0.485\textwidth]{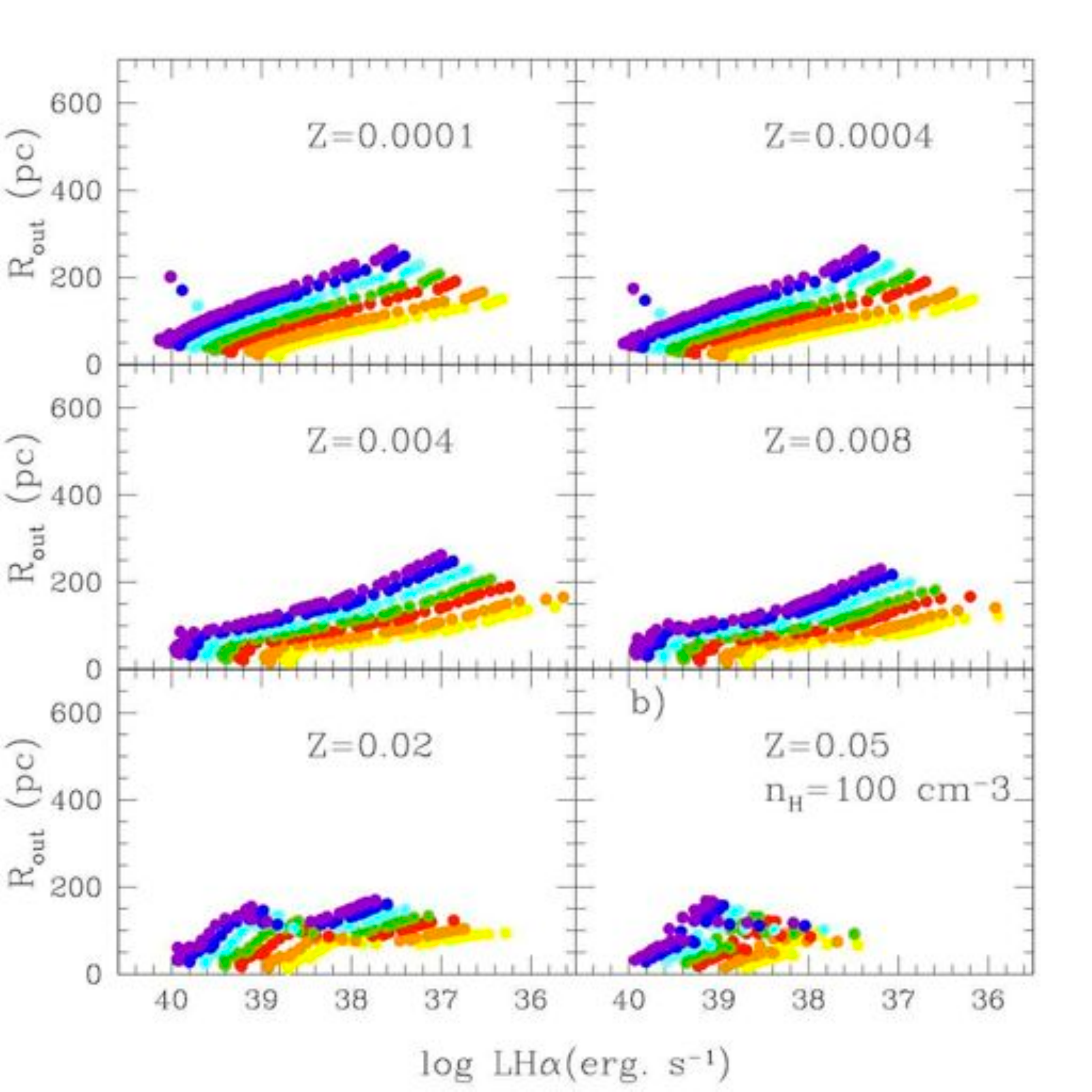}}
\caption{\halpha\ evolution: R (pc) {\sl vs} log (L \halpha) (erg.s$^{-1}$)
for six different metallicities and different cluster masses.
 Cluster masses have been plotted using the same colour coding that in
Fig.~\ref{rin_age}. Panel (a) on the right shows models computed for
a gas electron density of n$_{e}$ = 10\,cm$^{-3}$,  while panel (b) on the left uses a value of 
n$_{e}$ = 100\,cm$^{-3}$}
\label{rout_lhalpha}
\end{figure*}

The evolution of $\rm R_{in}$ is plotted in Fig.~\ref{rin_age}. At the 
beginning of the cluster's evolution, the inner radius is small and the 
shell thickness is very large, as shown in Fig.~\ref{deltar-rin}, so 
that photons cannot escape. At a certain age (around 0.5\,Myr) we start 
to detect the region by the emission line luminosity classifying it like 
a classical H{\sc ii} region. As the cluster evolves and its mechanical 
energy increases, the inner radius becomes larger while the shell 
becomes thinner. We will still see the region with a different emission 
line spectrum resulting from the cluster evolution (that changes both 
the ionising spectrum and the region geometry). We identify the observed 
object as an H{\sc ii} region (by definition) if we can detect hydrogen 
in emission, and this happens even after the emission line spectrum of 
forbidden lines is over and up to 20\,Myr (in average because there is a 
metallicity dependence). 

The external radius of the region, which we 
would identify as the observed one in most photometric observations, is 
plotted against age in Fig.~\ref{rout_age}. In this figure we can see 
the differences compared to Fig.~\ref{rin_age}. This radius starts with 
a value of around hundreds of pc, smaller for n$_{e} =$ 100\,cm$^{-3}$, 
and then decreases until SNe explosions begin to appear, which increase 
again the size of the region. Once the SNe explosions start to appear in the region, they can change 
the appearance and/or the geometry, however, we then should be more 
cautious when interpreting the observations based on our models, since 
the emission line spectrum will be the result of the shock and the 
photo-ionisation mechanisms. Our models do not include a shock 
contribution that can affect some emission lines, including [OI]6300 \AA.

Besides the evolution of the radius of the region measured on the 
\halpha\ images, we show in Fig.~\ref{lha_age} and Fig.~\ref{ewha_age} 
the evolution of the intensity and equivalent width of \halpha. In 
Fig.~\ref{lha_age} we see that \halpha\ luminosity maintains a high 
level for ages longer than 5\,Myr and it maintains a detectable 
intensity until 20\,Myr for the lowest abundances. This fact will have 
an impact on the colours, as we will see in the next section. The 
equivalent width shows a similar decreasing dependence with age, but 
also shows high values after the first 5.5~Myr for Z=0.0001 and 
Z=0.0004, while it falls to zero for the other abundances.

Fig~\ref{rout_lhalpha} shows the outer radius of the region as a 
function of the logarithm of \halpha\ luminosity. Models for different 
cluster masses are represented with different colours, as before. We see 
that H{\sc ii} regions may be quite large in size and luminosities for 
the lowest abundances compared with the metal-rich regions. Therefore if 
these regions are not observed, we need to consider potential 
observational selection effects. On the contrary, the metal-rich H{\sc 
ii} regions are much smaller, implying difficulty in observing these regions.

\begin{figure}
\resizebox{\hsize}{!}{\includegraphics[width=0.45\textwidth,angle=0]{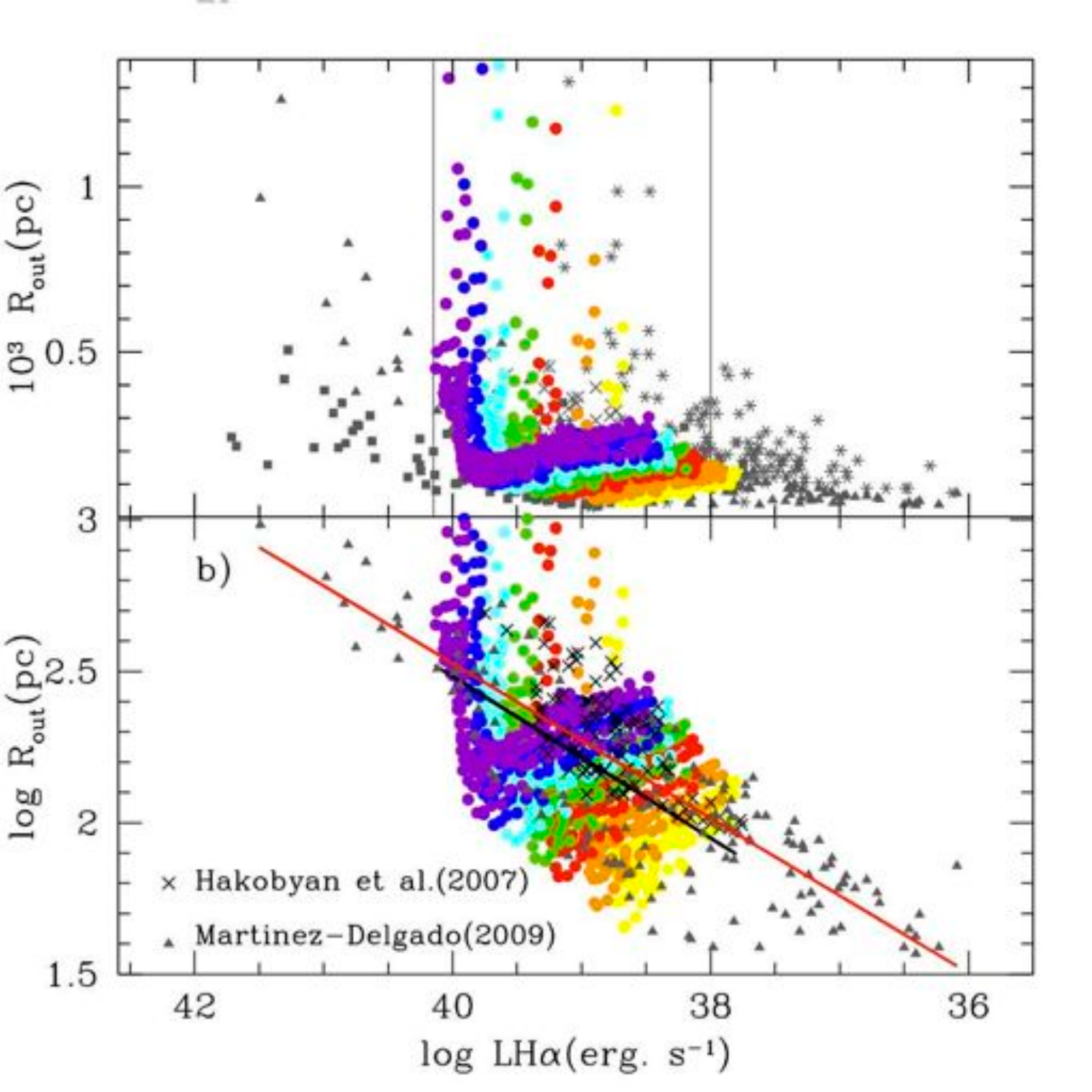}}
\caption{Relation of the outer radius $R_{out}$ {\sl vs} the \halpha\
luminosity compared with observations. 
The coloured full dots are our models for $\rm n_{H}=10\, cm^{-1}$, selecting those
with $\Delta R \ge 0.5$ \,pc. 
In panel a) grey symbols represent the observational data
from \citet{mayya94,fer04,hak07,ismael} as asterisks, filled squares,
crosses, and filled triangles, respectively. In panel b), with both axis in logarithmic scale, the (grey)
triangles are data from \citet{ismael} and the (black) crosses are from \citet{hak07}. 
The red and black lines are the corresponding least-squares fit to the data set and models, respectively.}
\label{rout-lha-obs}
\end{figure}

It seems that an inverse correlation between size and luminosity arises 
from these plots for luminosities lower than $10^{40}$~erg~s$^{-1}$ if 
the evolution of one cluster mass is followed, with larger radii for 
longer evolutionary times, when the H$\alpha$ luminosity decreases. 
Observations instead show a positive correlation between size and 
luminosity. In fact our results are restricted due to the cluster masses 
selected in our computations. In Fig.~\ref{rout-lha-obs}a) we show as 
coloured points our results (with the same code than in 
Fig.~\ref{rout_lhalpha}) only for models with $\Delta R > 0.5$ \,pc. 
This selection of models constrains the resulting luminosities to the 
range $10^{38}$--$10^{40}$ erg~s$^{-1}$. We compare these models with 
data from \citet{mayya94, fer04,hak07} and \citet{ismael}, shown as grey 
symbols. We see that some data fall out of the region defined by our 
models. In fact, observational points (which proceed from H{\sc ii} 
regions of different galaxies) show an abrupt decrease at a given 
luminosity, followed by a smooth increase that indicates the transition 
time in which massive star winds disappear and SNe start to increase the 
size of the bubble. However, some observational points seem to affect 
this change at an \halpha\ luminosity higher than our models have.  
Thus, most observations from \citet{fer04} show a behaviour that would 
be reproduced by models with a cluster mass larger than 2 $\times\ 
10^{5}$\,\Msun, our maximum cluster mass, or with a different IMF. Differences in the mass limits or slope of the IMF will change the mass distribution in the cluster generation, with direct consequences 
on the number of ionizing photons and L\halpha. This effect was discussed in Paper I (Moll\'{a} et~al. 2009). 
On the other end, many data 
from \citet{mayya94} need a cluster mass smaller than 1.2 $\times\ 
10^{4}$\,\Msun, the lowest limit of the models computed here. These 
latter observations show a similarly strong increase around L\halpha 
$\sim 10^{38}$\,erg.s$^{-1}$, slightly smaller than that shown by the 
lowest cluster mass model. Therefore in panel b), we compare our models 
only with the other two sets of data from \citet{hak07} and 
\citet{ismael} fitted by ionised clusters whose masses are within our 
model range. Doing so, we see that the observed correlation between size 
and luminosity is reproduced with our models, the dispersion for is 
driven by the range in cluster masses.  In that panel the red line is 
the least-squares straight fit to the data while the black line is the 
corresponding one for models. However, we warn that this correlation is 
not totally due to an evolutionary effect - as it has been argued in 
previous works - but a combination of the evolutionary state (age) and 
the mass of the cluster. A model with a given cluster mass shows a 
decrease in the size, while decreasing the \halpha\ luminosity, before 
the SNe explosion time. Then, a new increase in the size of the region 
is seen (see Fig~6), while the H$\alpha$ luminosity continues 
decreasing.  Only when all cluster masses are included in the same plot 
(Fig. 7b) does the correlation between radius and luminosity arise 
clearly (e.g. Fig.7b), showing, in the plane log $\rm R_{out}-log 
LH\alpha$, a slope similar to the observed one.

\subsection{Colour Calculation}
\begin{table}
\caption{Transmission value T of Johnson and SDSS filters. The complete table can be found in the online version.}
\begin{tabular}{cc}
\hline
$\lambda$ & T \\
($\AA$) &    \\
\hline
   3050. &     0.000 \\
   3100. &     0.020 \\
   3150. &     0.077 \\
   3200. &     0.135 \\
   3250. &     0.204 \\
   3300. &     0.282 \\
   3350. &     0.385 \\
   3400. &     0.493 \\
\hline
\label{trans}
\end{tabular}
\end{table}

\begin{figure}
\subfigure{\includegraphics[width=0.35\textwidth,angle=-90]{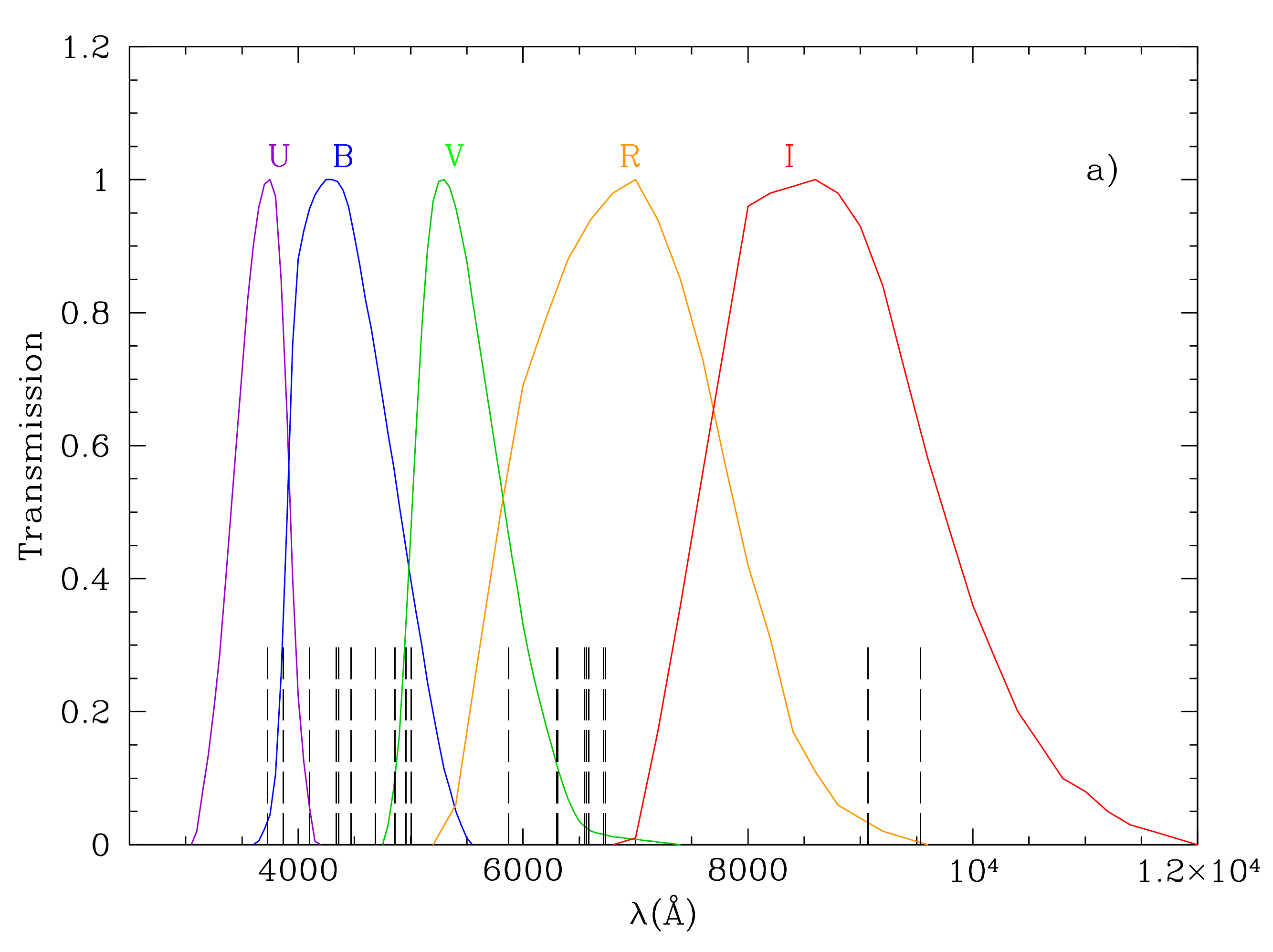}}
\subfigure{\includegraphics[width=0.35\textwidth,angle=-90]{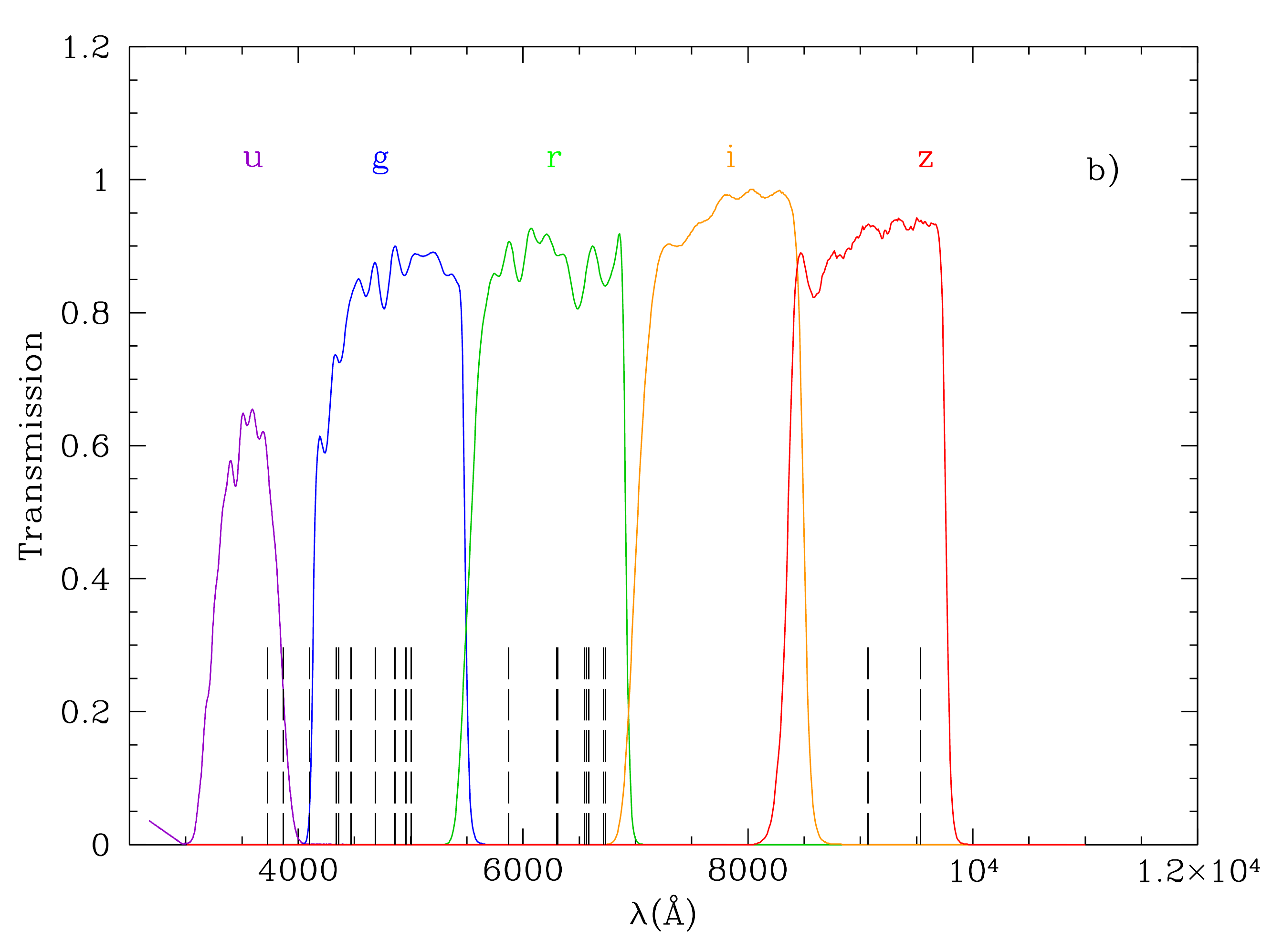}}
\caption{Transmission curves of a) U, B, V, R and I broad-band Johnson filters and b) u, g, r, i and z SDSS system filters. The figure also shows 
the location of the emission lines contributing to these filters.}
\label{filters}
\end{figure}

The filters included in this work are the optical broadband associated
with Johnson's system (UBVRI) and the Sloan SDSS ugriz system.  The
transmission curves for both are shown in Fig.~\ref{filters}.  Table
2, for which we show an example, and given in electronic format only,
contains the transmission values as a function of wavelength for
Johnson and SDSS filters.

\begin{table*}
\caption{Transmission of the broad band filters at the rest wavelength of the selected emission lines}
\begin{tabular}{lccccccccccc}
\hline
Emission line & T at U & T at B & T at B & T at V & T at R & T at I & T at u & T at g & T at r & T at i & T at z \\
$  (\AA)$ & & for U-B & for B-V & & & & & & & & \\
\hline
 3727[OII]    &      0.9966 &  0.0329 &  0.0438 &  0.0000 &  0.0000 &  0.0000 &  0.5781 &  0.0000 &  0.0000 &  0.0000&   0.0000\\
 3869[NeIII]  &      0.7837 &  0.3321 &  0.3911 &  0.0000 &  0.0000 &  0.0000 &  0.2263 &  0.0000 &  0.0000 &  0.0000&   0.0000\\
 4101 \hdelta &      0.0557 &  0.9556 &  0.9962 &  0.0000 &  0.0000 &  0.0000 &  0.0009 &  0.0493 &  0.0000 &  0.0000&   0.0000\\
 4340 \hgamma &      0.0000 &  0.9983 &  0.9612 &  0.0000 &  0.0000 &  0.0000 &  0.0005 &  0.7353 &  0.0000 &  0.0000&   0.0000\\
 4363[OIII]   &      0.0000 &  0.9947 &  0.9507 &  0.0000 &  0.0000 &  0.0000 &  0.0000 &  0.7253 &  0.0000 &  0.0000&   0.0000\\
 4471 HeI     &      0.0000 &  0.9405 &  0.8768 &  0.0000 &  0.0000 &  0.0000 &  0.0000 &  0.8237 &  0.0000 &  0.0000&   0.0000\\
 4686 HeII    &      0.0000 &  0.7372 &  0.6621 &  0.0000 &  0.0000 &  0.0000 &  0.0000 &  0.8748 &  0.0000 &  0.0000&   0.0000\\
 4861 \hbeta  &      0.0000 &  0.5556 &  0.4859 &  0.0943 &  0.0000 &  0.0000 &  0.0000 &  0.9004 &  0.0000 &  0.0000&   0.0000\\
 4959[OIII]   &      0.0000 &  0.4474 &  0.3873 &  0.3265 &  0.0000 &  0.0000 &  0.0000 &  0.8577 &  0.0000 &  0.0000&   0.0000\\
 5007[OIII]   &      0.0000 &  0.3934 &  0.3380 &  0.4829 &  0.0000 &  0.0000 &  0.0000 &  0.8820 &  0.0000 &  0.0000&   0.0000\\
 5871 HeI     &      0.0000 &  0.0000 &  0.0000 &  0.4638 &  0.5828 &  0.0000 &  0.0000 &  0.0000 &  0.9063 &  0.0000&   0.0000\\
 6300[OI]     &      0.0000 &  0.0000 &  0.0000 &  0.1200 &  0.8394 &  0.0000 &  0.0000 &  0.0000 &  0.8861 &  0.0000&   0.0000\\
 6312[SIII]   &      0.0000 &  0.0000 &  0.0000 &  0.1134 &  0.8447 &  0.0000 &  0.0000 &  0.0000 &  0.8855 &  0.0000&   0.0000\\
 6548[NII]    &      0.0000 &  0.0000 &  0.0000 &  0.0273 &  0.9263 &  0.0000 &  0.0000 &  0.0000 &  0.8456 &  0.0000&   0.0000\\
 6563 \halpha &      0.0000 &  0.0000 &  0.0000 &  0.0250 &  0.9304 &  0.0000 &  0.0000 &  0.0000 &  0.8625 &  0.0000&   0.0000\\
 6584[NII]    &      0.0000 &  0.0000 &  0.0000 &  0.0225 &  0.9359 &  0.0000 &  0.0000 &  0.0000 &  0.8842 &  0.0000&   0.0000\\
 6716[SII]    &      0.0000 &  0.0000 &  0.0000 &  0.0154 &  0.9622 &  0.0000 &  0.0000 &  0.0000 &  0.8417 &  0.0001&   0.0000\\
 6731[SII]    &      0.0000 &  0.0000 &  0.0000 &  0.0148 &  0.9654 &  0.0000 &  0.0000 &  0.0000 &  0.8404 &  0.0002&   0.0000\\
 9069[SIII]   &      0.0000 &  0.0000 &  0.0000 &  0.0000 &  0.0314 &  0.9060 &  0.0000 &  0.0000 &  0.0000 &  0.0003&   0.9326\\
 9532[SIII]   &      0.0000 &  0.0000 &  0.0000 &  0.0000 &  0.0000 &  0.6210 &  0.0000 &  0.0000 &  0.0000 &  0.0003&   0.9381\\
\hline
\label{transmision}
\end{tabular}
\end{table*}

The emission lines considered in this work are: 3727[OII], 3869[Ne III], 
4101 \hdelta, 4340 \hgamma, 4363[OIII], 4471 HeI, 4686 HeII, 4861 
\hbeta, 4959[OIII], 5007[OIII], 5871 HeI, 6300[OI], 6312[SIII], 
6548[NII], 6563\halpha, 6584[NII], 6716[SII], 6731[SII], 9069[SIII] and 
9532[SIII]. The transmission of the selected broad band filters at these 
wavelengths is given in Table~\ref{transmision}. Fig.~\ref{filters} 
shows the filter transmission curves. Panel a) shows the UBVRI Johnson 
filters while panel b) shows the ugriz SDSS filters. We have marked in 
both panels the position of the 20 selected emission lines at $z$=0. We 
remind the reader that the use of these models is only valid for local, 
low-redshift, systems. For more distant objects, the shift of the lines 
has to be taken into account since the transmission of the filter at the 
wavelength of a given emission line (and therefore the contribution to 
the integrated colour) will also vary with redshift. We will discuss 
these effects in a future work.

The magnitudes have been calculated as:
\begin{equation}
 m = -2.5 log \int_{\lambda_{1}}^{\lambda_{2}} L_{\lambda}
 d\lambda  + \sum_{i=1}^{20} T_{i} \times L_{i} + C
\end{equation} 

\noindent
where $\lambda_{1}$ and $\lambda_{2}$ are the pass-band limits in each 
filter, L$_{\lambda}$ is the stellar SED luminosity, L$_{i}$ is the 
integrated luminosity in the narrow line for the line $i$ and T$_{i}$ 
the line filter transmission.  In this calculation we assume that the 
line width is much narrower than the broad band filter pass-band. For 
Johnson's filters, C is the constant for flux calibration in the Vega 
system. According to the \citet{gir02} prescriptions, Vega has been 
taken as the average of \citet{lcb97} models for Z $=$ 0.004 and Z $=$ 
0.008 at T = 9500K (Vega's data: Z $=$ 0.006, m$_{bol} = $ 0.3319, BC 
$=$ -0.25 and V=0.58). C values are 2.19 (U), 3.29 (B when calculating 
U-B), 3.27 (B when calculating B-V), 2.54 (V), 2.76 (R) and 2.11 (I) 
when luminosities are in L$_{\odot}$ units. In the SDSS system the constant is always -48.60. 
Details about calculation following \citet*{gir04} and 
\citet{smith02} can be seen in Paper I.  Colours are the differences  between the two selected magnitudes.

\section{Model Results for SSPs}

\subsection{Results}

We have computed the colours from the pure continuum SSP SEDs (stellar + 
nebular), and the colours contaminated by the emission lines as described 
in \S2.3. Model results for Z $=$ 0.008 and some ages as an example 
are summarised in Table~\ref{result1} and Table~\ref{result2}, 
for Johnson and SDSS colours, respectively. The complete tables are 
computed for two values of the ionising gas electron density, n$_{e}$, 
since this parameter leads to important differences in the emission line 
spectrum. This is well known and widely discussed in Paper II. 
Table~\ref{result1} and Table~\ref{result2} are computed for n$_{e} = 
10$\,cm$^{-3}$ and for n$_{e} = 100$\,cm$^{-3}$.  These tables provide the 
computed colours (in Johnson and SDSS systems respectively) together with 
other cluster parameters that can be obtained from photometrical 
observations. Table ~\ref{result1} columns are: the metallicity Z, the 
logarithm of the age $\tau$ (in yr), the cluster mass $\rm M_{cl}$ (in  \Msun), 
the logarithm of the \hbeta\ and \halpha\ luminosities, L \hbeta\ and
L \halpha\ (both in erg.$s^{-1}$), the absolute magnitude V and colours
(U-B)$_{c}$, (B-V)$_{c}$, (V-R)$_{c}$ and (R-I)$_{c}$ contaminated
with the emission lines, the uncontaminated absolute magnitude V, and
colours U-B, B-V, V-R and R-I, (which obviously do not change with the
mass of the stellar cluster).  Table ~\ref{result2} columns are: the
metallicity Z, the logarithm of the age $\tau$ (in yr), the cluster
mass $\rm M_{cl}$ (in  \Msun), the equivalent widths
of \hbeta\ and \halpha, EW(\hbeta) and EW(\halpha) in \AA, the
absolute magnitude g and colours
(u-g)$_{c}$, (g-r)$_{c}$, (r-i)$_{c}$ and (i-z)$_{c}$ contaminated
with the emission lines, the uncontaminated absolute magnitude g and
colours u-g, g-r, r-i and i-z. The whole table 
with results for all ages and metallicities will be given in electronic 
format.

\subsection{Colour Evolution}

\begin{figure*}
\centering
\resizebox{\hsize}{!}{\includegraphics[height=0.485\textheight,angle=0]{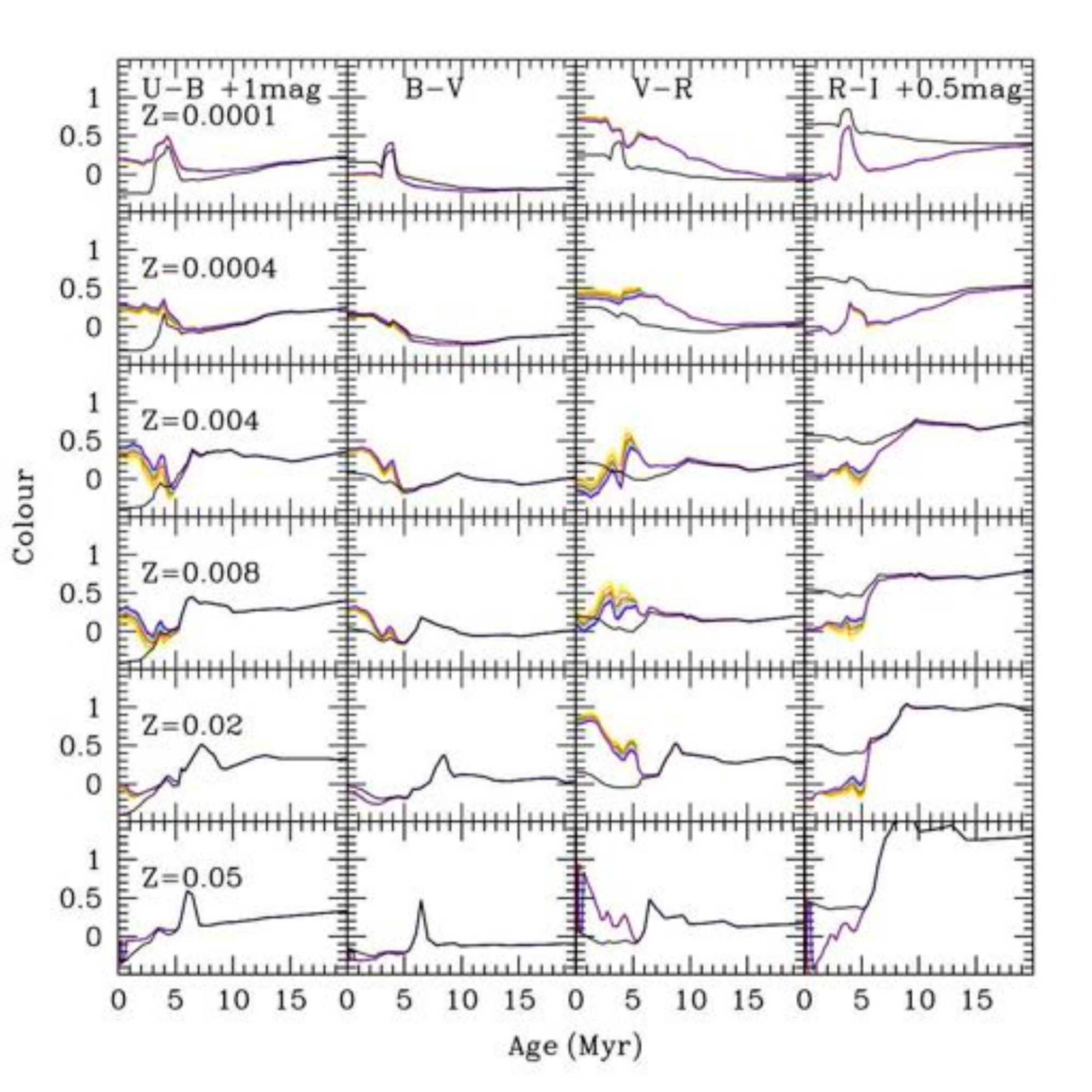}}
\caption{Photometric evolution of colours ~U-B, B-V, V-R and R-I {\sl
vs} cluster age ( in\,Myr) for $\rm Z = 0.0001$, 0.0004,
0.004, 0.008, 0.02 and 0.05 from top to bottom panels as labelled.  Each
colour corresponds to a different cluster mass  as in 
previous figures.}
\label{color_age}
\end{figure*}

Fig~\ref{color_age} shows the results of the Johnson colours' evolution 
with and without the emission line contribution for instantaneous young 
bursts (SSPs) between 1 and 20\,Myr in age.  We show four colours as a 
function of the cluster age (in\,Myr): U-B (left columns); B-V 
(left-middle columns); V-R (right-middle columns) and R-I (right 
columns). A different metallicity (Z = 0.0001, 0.0004, 0.004, 0.008, 
0.02, and 0.05) is displayed from top to bottom, as labelled in the U-B 
diagram of each row.

Colours from the clusters without the emission line contamination are 
plotted with a solid black line.  Colours including the contribution of 
the emission line spectrum are plotted with different colours, according 
to cluster mass - 0.12, 0.20, 0.40, 0.60, 1.00, 1.50 and 2.00 $\times$ 
10$^5$\, \Msun, from the lighter colour (yellow) to the darkest one 
(purple), using the same coding as in the previous figures.  As explained 
before, a different cluster mass implies not only a variation in the 
number of ionising photons but also a change in the size of the 
associated H{\sc ii} region due to the assumption (in our models) that 
the bubble radius depends on the total mechanical energy deposited by 
the cluster's winds and supernovae.

This figure shows that colours have very different values when the 
emission line contribution is included in the calculation. These 
variations in the expected colours for a young stellar cluster are 
especially important for low-to-intermediate metallicities (Z = 0.0004 
to 0.008); this seems reasonable, as the emission lines in the visible 
have the strongest intensities in this abundance range.

The contaminated colours U-B and B-V for all metallicities, and all
colours for high-metallicity (Z $>$ 0.008) evolve with age, reaching the
uncontaminated colours as soon as the emission lines disappear near
5.5\,Myr (this value depends on the metallicity, as discussed in
Paper II). However, V-R and R-I contaminated colours at low metallicity
still show variations with respect to the uncontaminated colours even up
to ages of  $\sim$20\,Myr. 

\begin{figure*}
\centering
\resizebox{\hsize}{!}{\includegraphics[angle=0]{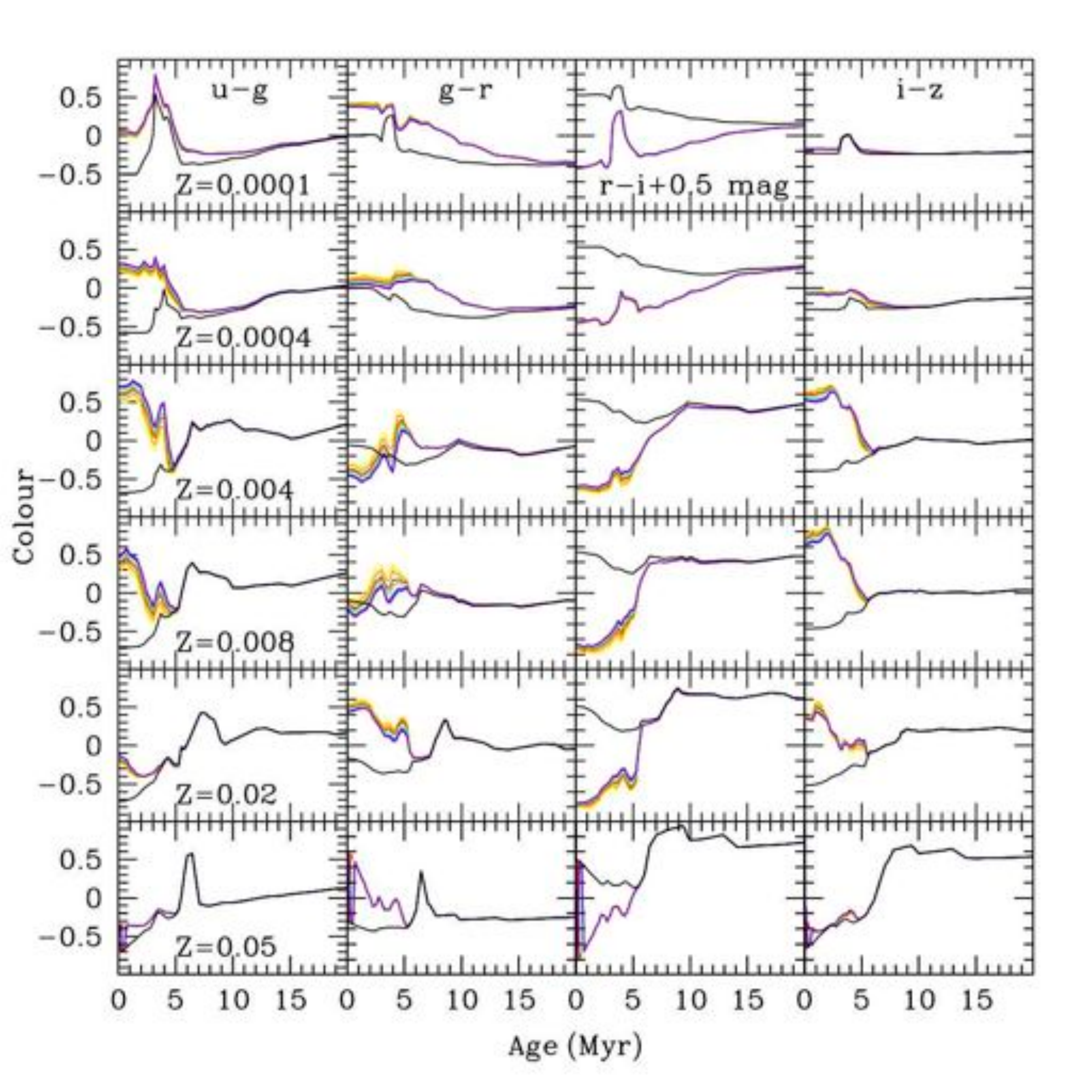}}
\caption{Photometric evolution of colours ~u-g, g-r, r-i and i-z {\sl
vs} cluster age ( in \,Myr)  for $\rm Z = 0.0001$, 0.0004,
0.004, 0.008, 0.02 and 0.05 from top to bottom panels as labelled. Each
colour corresponds to a different cluster mass  as in previous figures.}
\label{color_sdss_age}
\end{figure*}
This is because these colours are contaminated by 
hydrogen emission lines, and the R-band is the most highly contaminated. 
This indicates the presence of ionising photons - i.e., a young population -
which produces \halpha\ emission that falls almost in the centre of the
R-band, near to the maximum of the filter transmission curve, producing
a redder V-R colour simultaneous to a bluer R-I. 
\begin{table*} 
\caption{Johnson colours evolution results for the modelled stellar
clusters of $\rm Z = 0.008$ at some selected ages for n$_{e}$ = 10
cm$^{-3}$. Complete table 4 for all ages and metallicities and for n$_{e}$ = 10 and 100 cm$^{-3}$ can be found in the online version}
\begin{tabular}{crrccrrrrrrrrrr} 
\multicolumn{15}{c}{n$_{e}$ = 10 cm$^{-3}$}\\
\hline 
Z & log$\tau$ & $\rm M_{cl}$ & $\rm L H\beta$ & $\rm L H\alpha$ & V$_{c}$ & 
(U-B)$_{c}$ & (B-V)$_{c}$ & (V-R)$_{c}$ & (R-I)$_{c}$ & V & U-B & B-V & 
V-R & R-I \\
   &   (yr)    &  (\Msun ) &  (erg s$^{-1}$) &  (erg s$^{-1}$) & &  &  & & & & & & \\
\hline
0.008 &  6.30  &  1.2 10$^{4}$  &38.23&38.70&-10.407& -1.123  & 0.130 & 0.431 &-0.437 & -9.648&    -1.356& -0.015 & 0.159 & 0.019\\
0.008 &  6.30  &    2 10$^{4}$  &38.45&38.92&-10.998& -1.060  & 0.151 & 0.365 &-0.413 &-10.203&    -1.356& -0.015 & 0.159 & 0.019\\
0.008 &  6.30  &    4 10$^{4}$  &38.75&39.22&-11.793& -0.984  & 0.174 & 0.290 &-0.391 &-10.955&    -1.356& -0.015 & 0.159 & 0.019\\
0.008 &  6.30  &    6 10$^{4}$  &38.93&39.40&-12.256& -0.944  & 0.186 & 0.253 &-0.384 &-11.395&    -1.356& -0.015 & 0.159 & 0.019\\
0.008 &  6.30  &    1 10$^{5}$  &39.15&39.62&-12.835& -0.899  & 0.199 & 0.212 &-0.378 &-11.950&    -1.356& -0.015 & 0.159 & 0.019\\
0.008 &  6.30  &  1.5 10$^{5}$  &39.33&39.80&-13.293& -0.869  & 0.208 & 0.183 &-0.377 &-12.390&    -1.356& -0.015 & 0.159 & 0.019\\
0.008 &  6.30  &    2 10$^{5}$  &39.45&39.92&-13.618& -0.848  & 0.215 & 0.163 &-0.378 &-12.703&    -1.356& -0.015 & 0.159 & 0.019\\
0.008 &  6.40  &  1.2 10$^{4}$  &38.15&38.62&-10.166& -1.243  &-0.028 & 0.608 &-0.493 & -9.715&    -1.288& -0.075 & 0.104 &-0.012\\
0.008 &  6.40  &    2 10$^{4}$  &38.37&38.84&-10.753& -1.191  &-0.009 & 0.540 &-0.459 &-10.270&    -1.288& -0.075 & 0.104 &-0.012\\
0.008 &  6.40  &    4 10$^{4}$  &38.67&39.15&-11.545& -1.124  & 0.015 & 0.462 &-0.427 &-11.022&    -1.288& -0.075 & 0.104 &-0.012\\
0.008 &  6.40  &    6 10$^{4}$  &38.85&39.32&-12.005& -1.088  & 0.028 & 0.423 &-0.413 &-11.462&    -1.288& -0.075 & 0.104 &-0.012\\
0.008 &  6.40  &    1 10$^{5}$  &39.07&39.55&-12.582& -1.047  & 0.041 & 0.381 &-0.400 &-12.017&    -1.288& -0.075 & 0.104 &-0.012\\
0.008 &  6.40  &  1.5 10$^{5}$  &39.25&39.72&-13.038& -1.018  & 0.051 & 0.351 &-0.392 &-12.457&    -1.288& -0.075 & 0.104 &-0.012\\
0.008 &  6.40  &    2 10$^{5}$  &39.37&39.85&-13.362& -0.998  & 0.057 & 0.332 &-0.389 &-12.770&    -1.288& -0.075 & 0.104 &-0.012\\
\hline
\label{result1}
\end{tabular}
\end{table*}
\begin{table*}
\caption{SDSS colour evolution results for the modelled SSP of $\rm Z
= 0.008$ at some selected ages for all ages and metallicities and for n$_{e}$ = 10 cm$^{-3}$. Complete table
5, for n$_{e}$ = 10 and  100 cm$^{-3}$ can be found in the online version}
\begin{tabular}{crrrrrrrrrrrrrr}
\multicolumn{15}{c}{n$_{e}$ = 10 cm$^{-3}$}\\
\hline
 Z & log$\tau$ &  $\rm M_{cl}$ & EW $H\beta$ & EW $H\alpha$  & g$_{c}$ & (u-g)$_{c}$ & (g-r)$_{c}$ & (r-i)$_{c}$  & 
 (i-z)$_{c}$ &g  & u-g &     g-r &     r-i &     i-z \\
   &   (yr)    &  (\Msun ) &  (\AA) &  (\AA) & &  & & & & & &  &\\
\hline
 0.008 & 6.30& 1.2 10$^{4}$ &   409.97 & 2466.30&  -10.249 & -0.020  &  0.136& -1.278&  0.863 &  -9.280  &  -0.678& -0.154& -0.019& -0.445\\  
 0.008 & 6.30&   2 10$^{4}$ &   409.97 & 2466.30&  -10.840 &  0.053  &  0.067& -1.245&  0.852 &  -9.835  &  -0.678& -0.154& -0.019& -0.445\\  
 0.008 & 6.30&   4 10$^{4}$ &   409.97 & 2466.30&  -11.634 &  0.139  & -0.011& -1.208&  0.833 &  -10.587 &  -0.678& -0.154& -0.019& -0.445\\  
 0.008 & 6.30&   6 10$^{4}$ &   409.97 & 2466.30&  -12.098 &  0.186  & -0.051& -1.192&  0.822 &  -11.027 &  -0.678& -0.154& -0.019& -0.445\\  
 0.008 & 6.30&   1 10$^{5}$ &   409.97 & 2466.30&  -12.677 &  0.236  & -0.094& -1.174&  0.803 &  -11.582 &  -0.678& -0.154& -0.019& -0.445\\  
 0.008 & 6.30& 1.5 10$^{5}$ &   409.97 & 2466.30&  -13.134 &  0.270  & -0.124& -1.160&  0.784 &  -12.022 &  -0.678& -0.154& -0.019& -0.445\\  
 0.008 & 6.30&   2 10$^{5}$ &   409.97 & 2466.30&  -13.459 &  0.294  & -0.144& -1.153&  0.771 &  -12.335 &  -0.678& -0.154& -0.019& -0.445\\  
 0.008 & 6.40& 1.2 10$^{4}$ &   314.46 & 2027.40&  -10.006 & -0.287  &  0.314& -1.238&  0.697 &  -9.382  &  -0.617& -0.218& -0.082& -0.420\\  
 0.008 & 6.40&   2 10$^{4}$ &   314.46 & 2027.40&  -10.594 & -0.222  &  0.241& -1.198&  0.693 &  -9.937  &  -0.617& -0.218& -0.082& -0.420\\  
 0.008 & 6.40&   4 10$^{4}$ &   314.46 & 2027.40&  -11.387 & -0.142  &  0.158& -1.155&  0.682 &  -10.689 &  -0.617& -0.218& -0.082& -0.420\\  
 0.008 & 6.40&   6 10$^{4}$ &   314.46 & 2027.40&  -11.846 & -0.101  &  0.118& -1.134&  0.672 &  -11.129 &  -0.617& -0.218& -0.082& -0.420\\  
 0.008 & 6.40&   1 10$^{5}$ &   314.46 & 2027.40&  -12.423 & -0.054  &  0.073& -1.112&  0.660 &  -11.684 &  -0.617& -0.218& -0.082& -0.420\\  
 0.008 & 6.40& 1.5 10$^{5}$ &   314.46 & 2027.40&  -12.879 & -0.020  &  0.041& -1.095&  0.648 &  -12.124 &  -0.617& -0.218& -0.082& -0.420\\  
 0.008 & 6.40&   2 10$^{5}$ &   314.46 & 2027.40&  -13.204 &  0.004  &  0.021& -1.088&  0.641 &  -12.437 &  -0.617& -0.218& -0.082& -0.420\\  
\hline
\end{tabular}
\label{result2}
\end{table*}

\begin{figure*}
\resizebox{\hsize}{!}{\includegraphics[width=\textwidth,angle=-90]{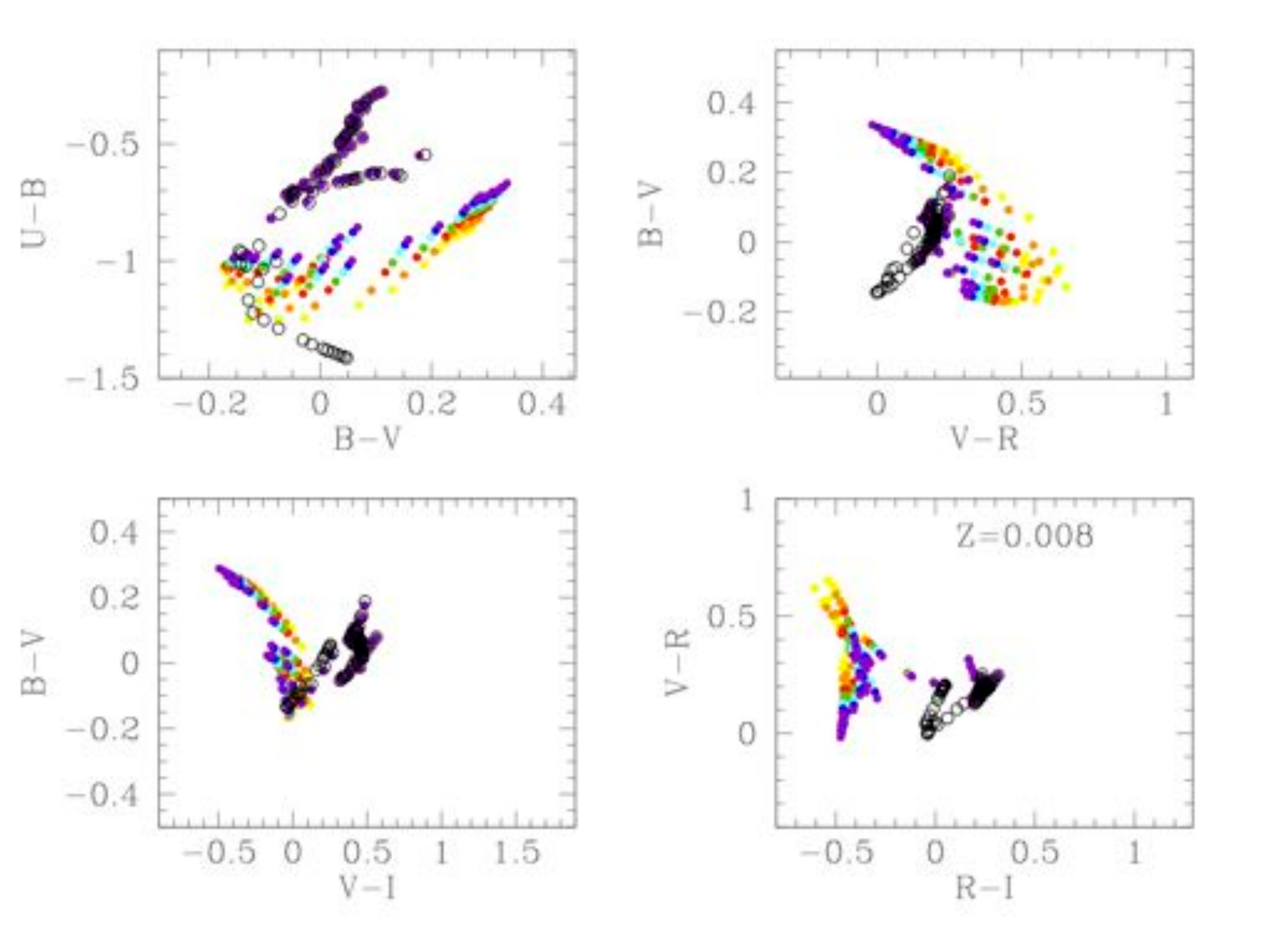}}
\caption{Colour-Colour diagrams for young clusters.  Panel d) Results for Z=0.008 : top-left) U-B {\sl vs}  B-V; top-right) B-V {\sl vs} V-R; 
bottom-left)B -V {\sl vs} V-I and right-bottom) V-R {\sl vs} R-I.  Open and full dots are the results without and with emission lines. 
Each colour corresponds to a different cluster mass . The same figures for the other metallicities: $\rm Z =$ 0.0001, 0.0004, 0.004,  0.02 and 0.05,
panels a), b), c), e) and f)  are given in electronic format.}
\label{color_color}
\end{figure*}

\begin{figure*}
\resizebox{\hsize}{!}{\includegraphics[width=\textwidth,angle=-90]{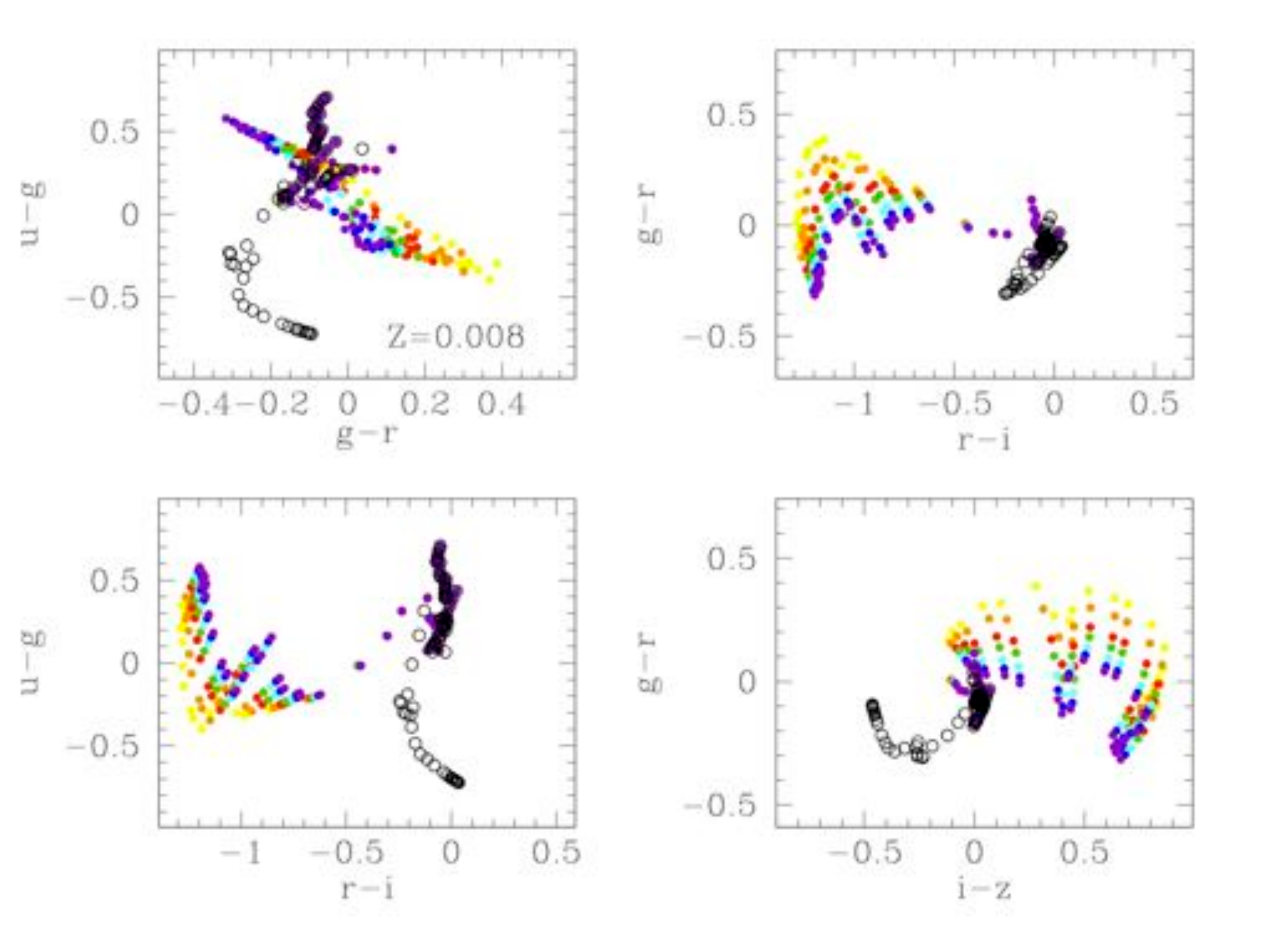}}
\caption{Colour-Colour diagrams for young clusters. Panel d) Results for Z=0.008 : top-left) u-g {\sl vs}  g-r; top-right) g-r{\sl vs} r-i; bottom-left) u-g{\sl vs} r-i and bottom-right) r-i {\sl vs} i-z. R
Open and full dots are the results without and with emission lines. Each colour corresponds to a different cluster mass . 
The same figures for the other metallicities: $\rm Z =$ 0.0001, 0.0004, 0.004,  0.02 and 0.05, panels a), b), c), e) and f)  are given in electronic format.}
\label{color_color_sdss}
\end{figure*}

We note again that \halpha\ emission is expected even for ages beyond 
5.5\,Myr (when other emission lines cannot be produced) and is 
detectable up until ages of 10 to 20\,Myr, depending upon metallicity. 
Up to now, arguments for a reddened colour had been based upon the claim 
for an old underlying population or a dust excess. We show here that a 
reddened V-R colour, with a contemporaneous blue R-I, in a SSP, can be 
explained in a new way: with the presence of a young population,
which possesses sufficient numbers of ionising photons to produce 
\halpha\ emission.

In a similar manner, Fig.~\ref{color_sdss_age} shows the effect of these 
emission lines in the SDSS filter system. Differences in colours, with 
and without the contribution of the emission lines, are even more 
substantial than in the Johnson system, reaching in some colours, 
difference of up to one magnitude.

\subsection{Colour-Colour Diagrams}

Fig~\ref{color_color} shows four Johnson colour-colour diagrams for SSPs 
up to 20\,Myr in age: 
U-B {\sl vs} B-V (top left);  B-V {\sl vs} V-R (top right); 
B-V {\sl vs} V-I (bottom left) and V-R {\sl vs} R-I (bottom right).  
We show results for Z $=0.008$ , panel d) and the corresponding ones to other metallicities,
Fig.11a, 11b, 11c, 11e and 11f will be in electronic format. Different colours correspond to different 
cluster masses, as in Fig~\ref{color_age}. Black open ones correspond to the youngest ages including
the nebular continuum contribution but not the emission lines.

In panels of  the U-B {\sl vs} B-V diagram we observe that 
when young stellar populations (contaminated and uncontaminated) colours 
are plotted, the points fall out of the location expected when standard 
SSP synthetic colours without the emission line (or continuum) 
contribution are used. The nebular continuum moves the points out of the 
standard sequence for old (age $\ge$ 20\,Myr, see Fig.\,13) stellar populations. 
Moreover, when the emission lines contribution is included in the colour 
calculation, the sequence shows almost the original slope, but moved to 
an almost parallel line for $Z< 0.02$, with higher B-V for 
similar U-B.

Observing the B-V {\sl vs} V-R diagram, the situation changes, since the 
contaminated colours fall in the same region of the plot, albeit in an 
orthogonal sense, in most cases. This behaviour appears independent of 
the stellar cluster mass, especially for $Z=0.004$ and $Z=0.008$. This 
characteristic might allow one to predict the strength of a new 
starburst overlaid upon an underlying and older stellar population.  
Further, one might determine the age of this burst from the distance of 
a given point to the canonical line of old ages. This effect is stronger 
when mixed population models (\S4) are interpreted.

In the B-V {\sl vs} V-I diagrams, the youngest stellar populations show 
colours with the same trend as the uncontaminated ones for some 
metallicity cases, while appearing orthogonal for others. On the one hand, the 
\halpha\ emission reddens the V-I colour. On other, if the metallicity is 
intermediate (Z=0.004-0.008), the oxygen emission lines also increase the 
luminosity in the V-band, since the nebular cooling is mostly done 
through [OIII] lines. The resulting populated area is very different from 
the one with uncontaminated SSP colours.

Finally, the V-R {\sl vs} R-I diagram is the one most affected by the 
contribution of \halpha\ emission to the R-band luminosity (which 
dominates over the effect of sulphur lines contributing to the I-band). 
This diagram may therefore be useful to check if these contamination 
effects (due to a young ionising population) is present regardless of 
the cluster physical properties, since all the metallicities and cluster 
mass change the colour-colour diagram with respect to the canonical 
uncontaminated models. Fig.~\ref{color_color_sdss} is similar to 
Fig~\ref{color_color} for SDSS colour-colour diagrams.

In summary, our remarkable conclusions from Fig~\ref{color_color} and 
Fig.~\ref{color_color_sdss} are: 1) The youngest stellar population 
models - therefore, those contaminated most by emission lines - do not 
follow the generic trend shown by SSP colours calculated in the absence 
of this emission line contribution, and 2) these young regions' colours 
fall in a well-defined region for each colour-colour diagnostic diagram. 
In some cases (e.g. B-V {\sl vs} V-R), this region of contaminated 
colours is located orthogonally to the region where the canonical 
uncontaminated colours reside. It is clear that emission lines change 
strongly the location of the models in any given colour-colour diagram.  
We emphasise that most observers use canonical 
(uncontaminated and probably without the nebular continuum contribution) models 
to compare with their star-forming region photometry and use these to 
derive the properties of the unveiled young stellar population. In some 
cases, a reddened colour is often interpreted as implying the existence 
of an old population.  As we have demonstrated in the previous 
paragraphs and shown in the accompanying figures, we can interpret 
reddening in some colours, instead, actually as a proof of the existence 
of a young population. In other cases the lack of good fitting between 
the classical photometrical models (uncontaminated colours) and the 
observations is interpreted as extinction (reddened colours are 
interpreted as dust contamination).  This can lead to an overestimation 
of dust and extinction when this value is derived from photometric data 
in H{\sc ii} regions, even after correcting for an underlying old 
component. In summary, the interpretation has to be done using 
contaminated models and of course the impact will depend on the colours 
chosen.

\begin{figure*}
\subfigure{\includegraphics[width=0.62\textwidth,angle=-90]{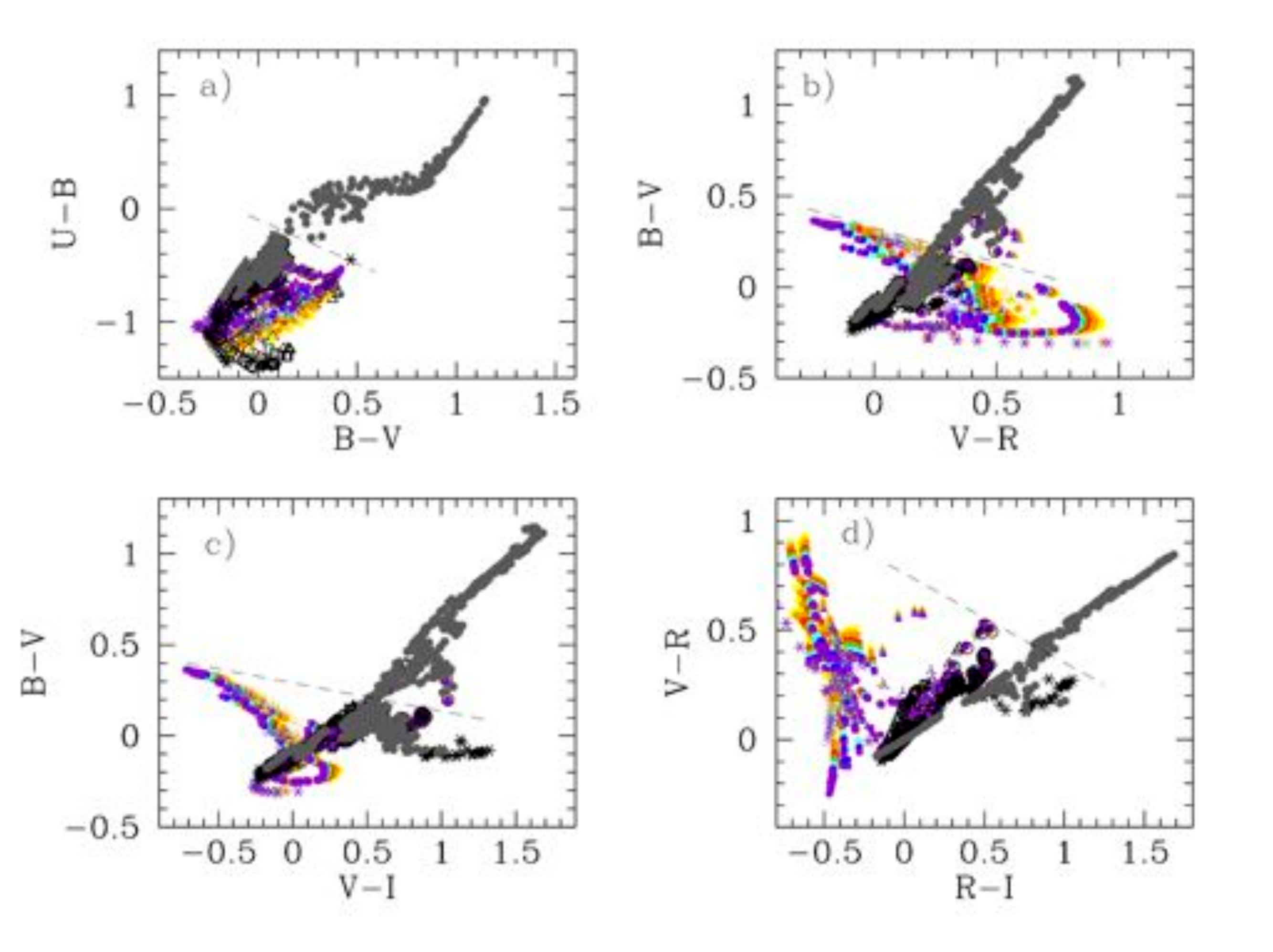}}
\subfigure{\includegraphics[width=0.62\textwidth,angle=-90]{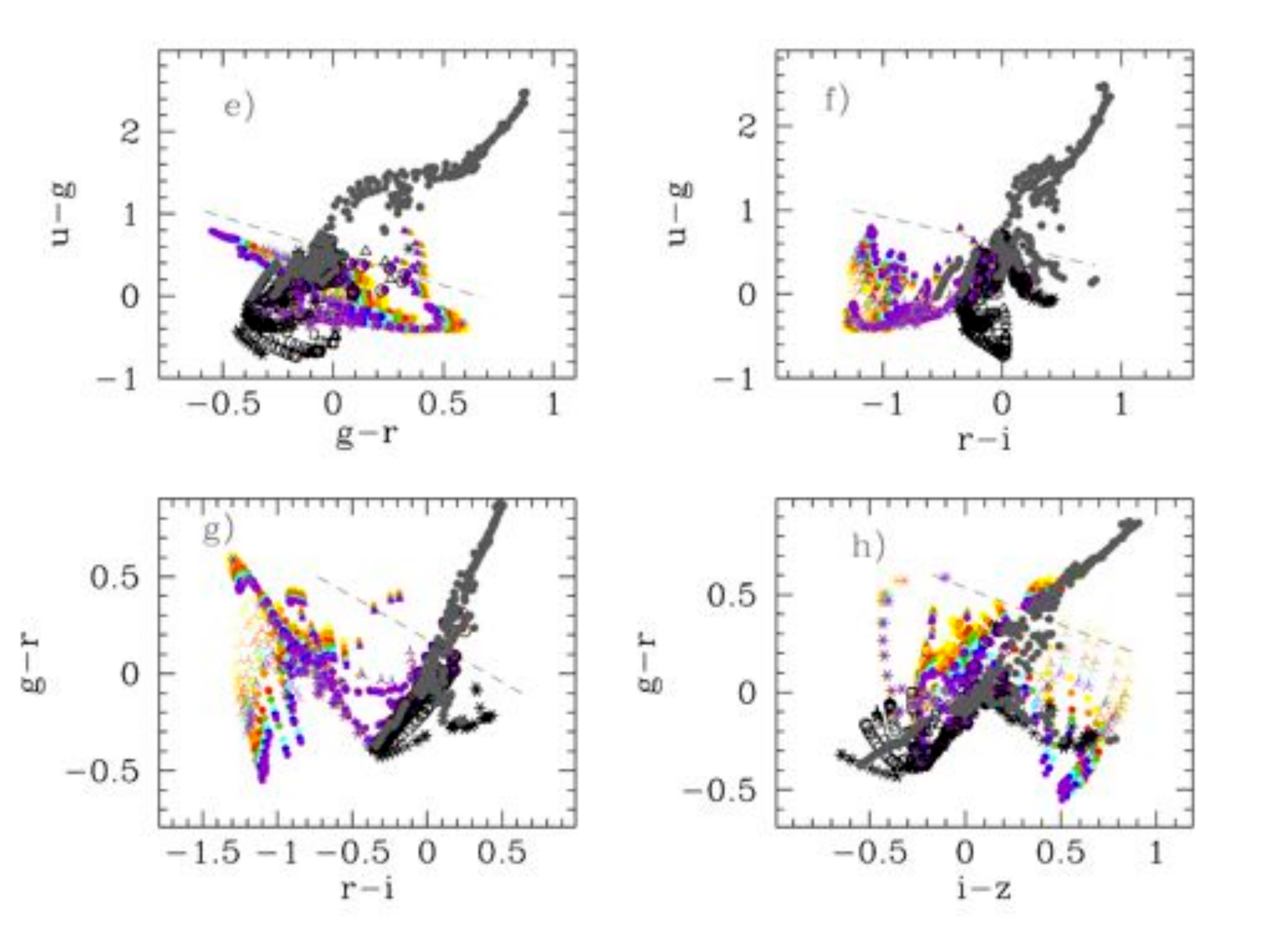}}
\caption{Colour-colour diagrams evolution for all ages and all
metallicities. Panels show a) U-B vs B-V; b) B-V vs V-R;
c) B-V vs V-I , d) V-R vs R-I, e) u-g vs g-r; f) u-g vs r-i; g) g-r vs
r-i and h) g-r vs i-z. Dashed black lines separate the old stellar
population region ($\tau > 20$ \,Myr) from the zone of young ones. Colours represent different 
cluster masses, as in previous figures. Grey full dots correspond to the stellar population colours without 
any nebular contribution. The black symbols represent colours including the nebular continuum contribution
but not the emission lines one. All cluster metallicities and ages have been included in the figure.}
\label{color_color_all}
\end{figure*}

Finally, and as an extension of Fig.~\ref{color_color} and 
Fig.~\ref{color_color_sdss}, we have used all the models together in 
Fig.~\ref{color_color_all}. This figure shows the eight different 
colour-colour diagrams (4 for Johnson and 4 for SDSS filters). Models from 
different metallicities have been plotted together. Large 
black symbols correspond to the young stellar cluster colours without 
contamination of emission line spectra.  Coloured symbols correspond to 
values that include the emission line contribution for the youngest ages 
(between 1.0 and 5.0\,Myr). Different cluster masses (which, we remind, 
in our models implies different H{\sc ii} regions' inner radii and 
therefore different emission line spectrum), have been represented with 
the same colours as seen previously in Fig~\ref{color_age} and 
Fig~\ref{color_sdss_age}. Finally, together with the young clusters, we 
have included the whole evolutionary sequence, that is all metallicities 
in the range $\rm Z =$ 0.0001 - 0.05 and all ages from 20 \,Myr up to 
15\,Gyr, obtained with {\sc PopStar} in Paper I, as grey full dots. 
This figure shows a clear age sequence for SSP ages older than 10\,Myr 
in all colour-colour diagrams. This sequence places pure young populations 
(ages below 10-20\,Myr) and pure old SSPs in completely different areas in 
the plot since the youngest fall out of the old sequence. We have 
plotted a dashed line in each colour-colour diagram to separate the zone 
where young stellar clusters would reside from the one where the old 
stellar populations fall.  When the young populations with the 
contaminated colours are included, the diagrams change again since some 
colours are redder, more similar to the ones of older populations, but 
others are bluer, and therefore points are located in different diagram 
regions. The problem is even worse when mixed populations are studied as 
discussed in \S4.

\begin{figure*}
\centering
\subfigure{\includegraphics[width=0.62\textwidth,angle=-90]{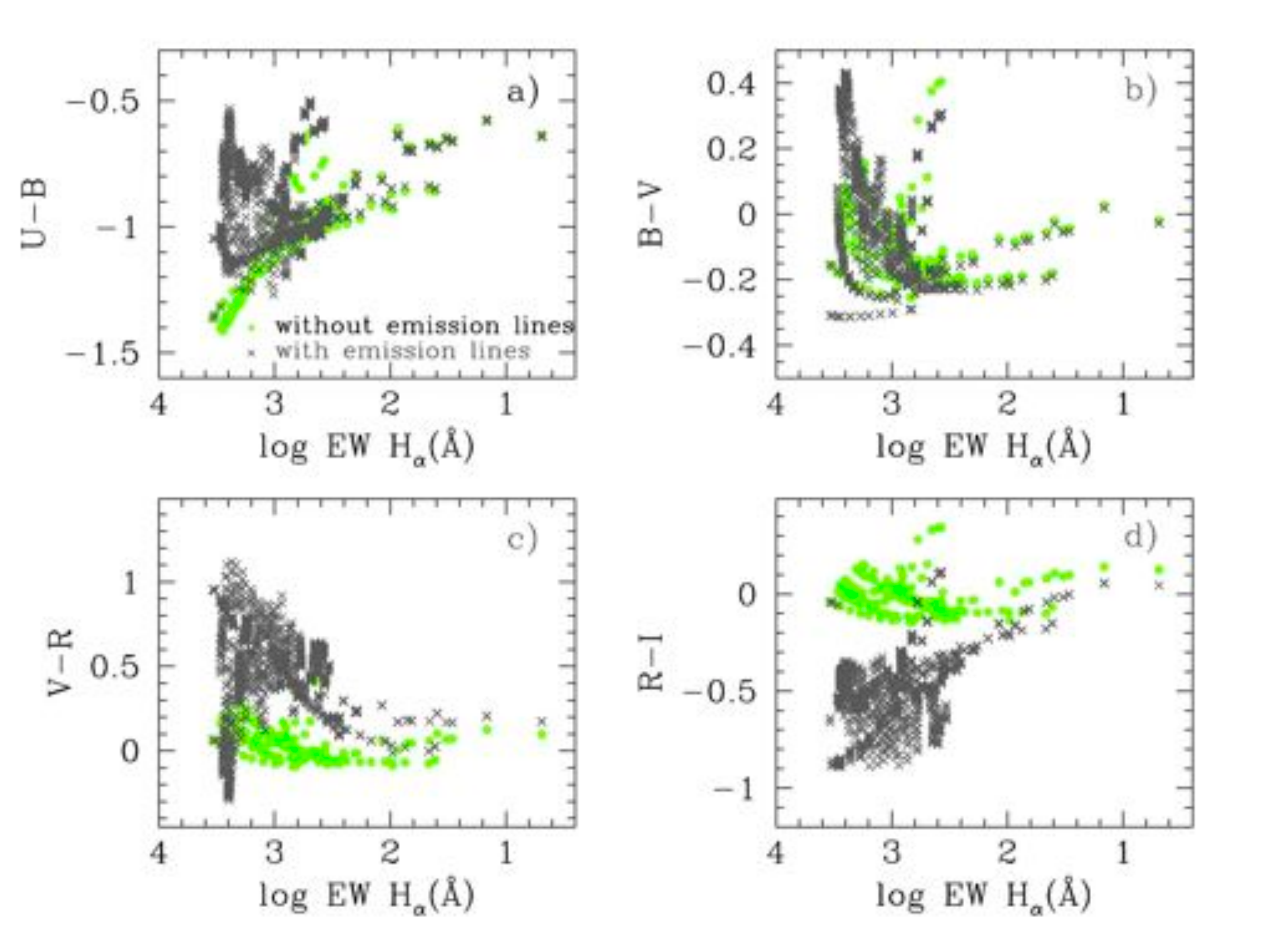}}
\subfigure{\includegraphics[width=0.62\textwidth,angle=-90]{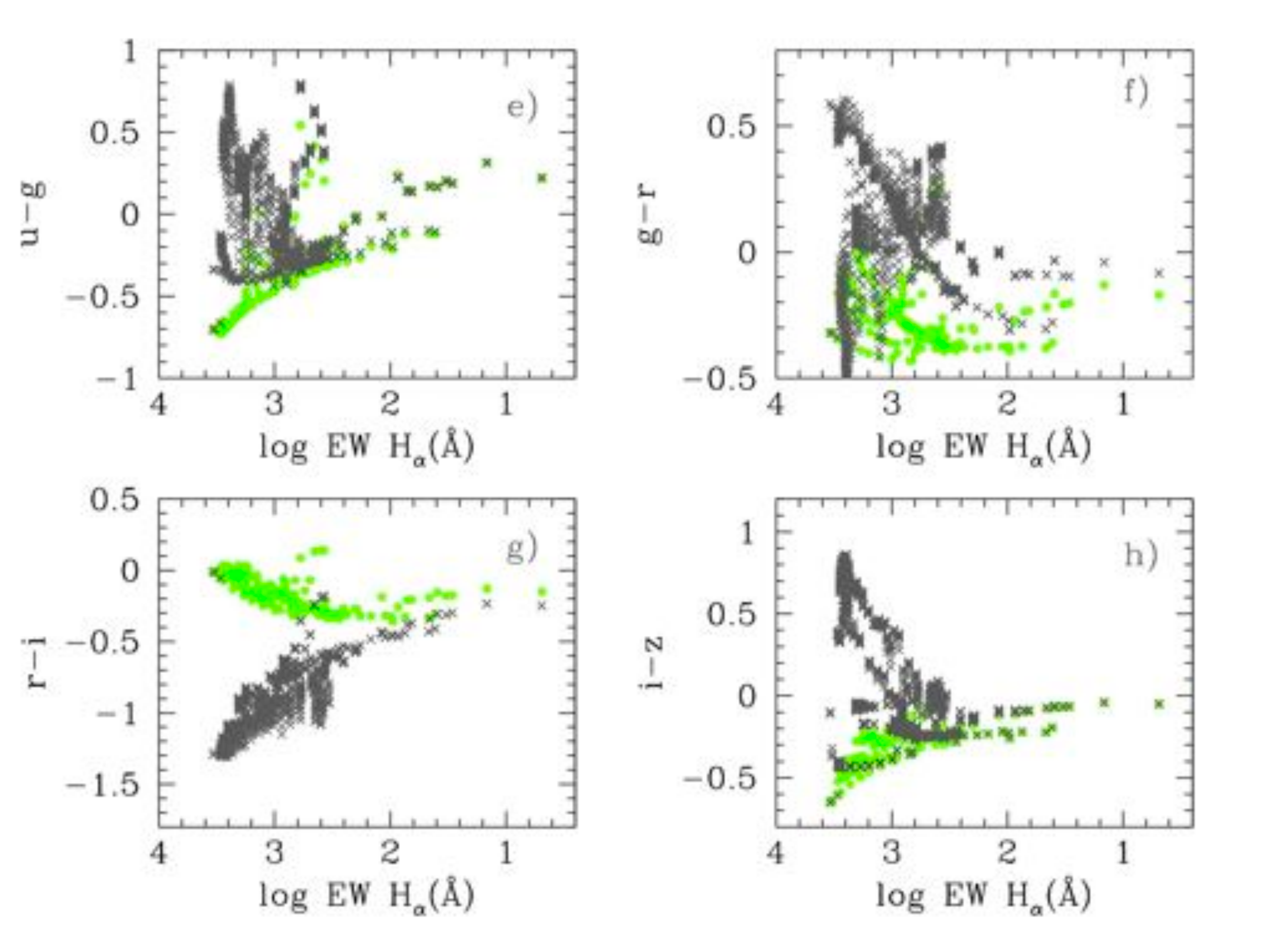}}
\caption{Colours vs log of H$\alpha$ equivalent width diagrams: a) U-B {\sl vs} log EW(\halpha)
; b)  B-V {\sl vs} log EW(\halpha); c) V-R {\sl vs} log EW(\halpha); d) R-I {\sl vs} log EW(\halpha); e) u-g {\sl vs} log EW(\halpha); 
f)  g-r {\sl vs} log EW(\halpha); g) r-i {\sl vs} log EW(\halpha) and h)  i-z {\sl vs} log EW(\halpha). 
In all panels full (green) dots and (black) crosses are the results without and with emission lines,
respectively. All cluster masses, metallicities, and ages have been included in the figure.}
\label{ewha_color1}
\end{figure*}

\subsection{Colours vs Other Photometric Parameters}

We have used in this work not only the information coming from the 
colours, but also that associated with other photometric parameters. 
Fig~\ref{ewha_color1} shows the relationship between colours (both 
Johnson and SDSS) and equivalent widths of \halpha. We have plotted all 
masses, metallicities and ages together in the same plot.

The original colours without the emission line contribution are the full
green dots. These models define a clear sequence in all plots. Colours 
including the contribution of emission lines are the black crosses. 
The equivalent width in this case does not change with the emission 
lines, while the colours do. This causes the points to move up or down 
with respect to their original position. Panel a) shows that in a given 
time, U-B becomes redder and the points move up slightly. This occurs 
when EW(\halpha) is still quite high (young ages).

The same occurs in panel c) with V-R, although in this case the 
reddening extends also for low values of EW(\halpha). Changes in panel 
b), the B-V colour, are less evident because emission lines contaminate 
both B- and V-band in a comparable proportion. Panel d) shows how the 
R-I points move down (towards bluer colours), since R is a highly 
contaminated band. We cannot find a clear metallicity dependence in 
these plots.

The figure of EW H$_{\alpha}$ {\sl vs} U-B is particularly interesting 
because most of the data cannot be reproduced with pure single stellar 
populations as \citet[][ and references therein]{mmetal08} showed.  The 
usual explanation was the existence of an underlying older 
stellar population which, in addition to the recent starburst, could 
reproduce simultaneously a red U-B color together with the observed 
values of the Balmer lines equivalent widths. We claim that,  even 
without adding an old population, an U-B reddening due to 
the strong contribution of the emission lines in this color is predicted. The effect 
is a larger dispersion in the plot, erasing the original trend in the 
plot U-B color {\sl vs} EW(\halpha) which showed an age sequence from 
the high equivalent width and blue colour to the low equivalent width and 
red colour.

Panels e) to h) represent the same kind of plots but for the SDSS 
colours. A clear separation of the colour sequences without and with 
emission lines is seen, mainly for the r-i colour, where \halpha\ has a 
stronger contribution. These plots may be useful to separate young (ages 
$\tau < 20$\,Myr and clear \halpha\ emission) from old stellar 
populations.

\section{Mixed Populations}

All the models presented in the previous sections assume that there is a 
single cluster producing the colours of the observed region. In more 
massive objects where different stellar populations coexist, the 
contribution of the youngest cluster (responsible for the ionisation and 
the emission lines) mass to the total stellar mass can change the final 
broadband magnitudes and therefore the colours. The more massive the 
young population, the more important the contribution of emission lines 
to the integrated colours. On the contrary, objects where the young 
population contribution is small can be well-fit by canonical synthetic 
colours since the contribution of the emission lines to the total 
luminosity and therefore to the colours will be minimal.

The calculation of the mixed population (combination of young and old 
clusters) colours requires a good reference grid of models that allows 
one to derive the physical properties of composite stellar systems when 
only photometric information is available. This section is devoted to 
these composite models in which we have a young ionising stellar burst 
superimposed upon an underlying old population. The difference with 
earlier work in the field (see Paper I for a review of Evolutionary 
Synthesis Models) is that the emission lines coming from the ionisation 
and the nebular continuum, produced both by the young cluster, are 
included when computing the integrated colours. We remark the fact that 
these models have been computed in a consistent manner by taking into 
account the cluster evolution (and therefore the mechanical energy) to 
derive the region's photometric size and therefore the radius that, 
together with the number of ionising photons and SED shape, determines 
the ionisation parameter responsible for the emission line luminosities. 
These emission line luminosities are the ones included in the colour 
determination after weighting the luminosities (from both old and young 
populations) according to their relative masses.
\begin{table*}
\centering
\caption{Example of Johnson and SDSS colour evolution table for two mixed stellar populations. The complete table is in electronic format.}
\begin{tabular}{ccccrcccrccccc}
\hline
Z$_{o}$ & log$\tau_{o}$ & Z$_{y}$ & log$\tau_{y}$  & F &  $\rm M_{cl}$ & R$_{out}$ &  logLH$_{\alpha}$ & EWH$_{\alpha}$ & V$_{c}$ & U-B$_{c}$ & B-V$_{c}$ & V-R$_{c}$ & R-I$_{c}$ \\

   &   yr    &  &  yr & & M$_{\odot}$  & pc&  $\rm ers.s^{-1}$ & \AA  & &  & & &   \\
\hline
 0.008  & 8.00 &0.0001 & 6.00 &    0 &  1.2 $10 ^{4}$ & 126.45& 38.81& 1757.00 &-10.315 & -0.850 & -0.011 &  0.736 & -0.549  \\     
 0.008  & 8.00 &0.0001 & 6.00 &    1 &  1.2 $10 ^{4}$ & 126.45& 38.81& 1643.81 &-10.380 & -0.827 & -0.004 &  0.711 & -0.509  \\      
 0.008  & 8.00 &0.0001 & 6.00 &   10 &  1.2 $10 ^{4}$ & 126.45& 38.81& 1030.71 &-10.838 & -0.680 &  0.034 &  0.559 & -0.277  \\      
 0.008  & 8.00 &0.0001 & 6.00 &  100 &  1.2 $10 ^{4}$ & 126.45& 38.81&  172.46 &-12.456 & -0.388 &  0.092 &  0.285 &  0.105  \\      
 0.008  & 8.00 &0.0001 & 6.00 & 1000 &  1.2 $10 ^{4}$ & 126.45& 38.81&  -39.88 &-14.810 & -0.290 &  0.108 &  0.202 &  0.214  \\      
 0.008  & 8.00 &0.0001 & 6.00 & 5000 &  1.2 $10 ^{4}$ & 126.45& 38.81&  -61.54 &-16.544 & -0.280 &  0.110 &  0.193 &  0.226  \\      
 0.008  & 8.00 &0.0001 & 6.00 &    0 &  2.0 $10 ^{4}$ & 163.68& 39.03& 1757.00 &-10.878 & -0.845 & -0.007 &  0.726 & -0.548  \\
 0.008  & 8.00 &0.0001 & 6.00 &    1 &  2.0 $10 ^{4}$ & 163.68& 39.03& 1643.60 &-10.943 & -0.822 & -0.001 &  0.701 & -0.508  \\
 0.008  & 8.00 &0.0001 & 6.00 &   10 &  2.0 $10 ^{4}$ & 163.68& 39.03& 1029.86 &-11.397 & -0.677 &  0.036 &  0.552 & -0.276  \\  
 0.008  & 8.00 &0.0001 & 6.00 &  100 &  2.0 $10 ^{4}$ & 163.68& 39.03&  172.05 &-13.012 & -0.387 &  0.093 &  0.284 &  0.105  \\
 0.008  & 8.00 &0.0001 & 6.00 & 1000 &  2.0 $10 ^{4}$ & 163.68& 39.03&  -39.93 &-15.365 & -0.290 &  0.108 &  0.202 &  0.214  \\ 
 0.008  & 8.00 &0.0001 & 6.00 & 5000 &  2.0 $10 ^{4}$ & 163.68& 39.03&  -61.55 &-17.099 & -0.280 &  0.110 &  0.193 &  0.226  \\      
\hline
\end{tabular}
\label{mix}
\end{table*}
\begin{table*}
\setcounter{table}{5}
\caption{Cont. Johnson and SDSS colour evolution for two mixed stellar populations}
\begin{tabular}{ccccccccccccccc}
\hline
V & U-B & B-V & V-R & R-I  & g$_{c}$ & u-g$_{c}$ & g-r$_{c}$ & r-i$_{c}$ & i-z$_{c}$ & g & u-g & g-r & r-i & i-z  \\
   &     &     &     &       &      &            &           &          &           &  &      &     &     &      \\
\hline
-10.076 & -1.245 &  0.158 &  0.253 &  0.152& -10.124 & -0.010  & 0.424 & -0.890 & -0.167 & -9.619  &-0.499 &  0.009 &  0.030 & -0.228 \\      
-10.157 & -1.196 &  0.154 &  0.249 &  0.157& -10.177 &  0.015  & 0.405 & -0.851 & -0.154 & -9.703  &-0.443 &  0.004 &  0.024 & -0.211 \\      
-10.696 & -0.913 &  0.137 &  0.226 &  0.185& -10.567 &  0.182  & 0.285 & -0.618 & -0.091 &-10.259  &-0.114 & -0.019 & -0.005 & -0.123 \\      
-12.425 & -0.437 &  0.115 &  0.198 &  0.220& -12.077 &  0.550  & 0.040 & -0.199 & -0.011 &-12.008  & 0.485 & -0.049 & -0.042 & -0.017 \\      
-14.807 & -0.296 &  0.111 &  0.192 &  0.228& -14.402 &  0.686  &-0.044 & -0.070 &  0.008 &-14.394  & 0.679 & -0.055 & -0.051 &  0.007 \\      
-16.543 & -0.281 &  0.110 &  0.191 &  0.229& -16.133 &  0.702  &-0.054 & -0.056 &  0.010 &-16.131  & 0.701 & -0.056 & -0.052 &  0.009 \\      
-10.631 & -1.245 &  0.158 &  0.253 &  0.152& -10.687 & -0.001  & 0.414 & -0.888 & -0.168 &-10.174  &-0.499 &  0.009 &  0.030 & -0.228 \\    
-10.712 & -1.196 &  0.154 &  0.249 &  0.157& -10.740 &  0.024  & 0.396 & -0.849 & -0.155 &-10.258  &-0.444 &  0.004 &  0.024 & -0.211 \\    
-11.251 & -0.913 &  0.137 &  0.226 &  0.185& -11.128 &  0.188  & 0.279 & -0.616 & -0.091 &-10.814  &-0.114 & -0.019 & -0.005 & -0.124 \\    
-12.980 & -0.437 &  0.115 &  0.198 &  0.220& -12.633 &  0.552  & 0.038 & -0.199 & -0.011 &-12.563  & 0.485 & -0.049 & -0.042 & -0.017 \\    
-15.362 & -0.296 &  0.111 &  0.192 &  0.228& -14.957 &  0.687  &-0.045 & -0.070 &  0.008 &-14.949  & 0.679 & -0.055 & -0.051 &  0.007 \\    
-17.098 & -0.281 &  0.110 &  0.191 &  0.229& -16.687 &  0.702  &-0.054 & -0.056 &  0.010 &-16.686  & 0.701 & -0.056 & -0.052 &  0.009 \\    
 \end{tabular}
\label{mix2}
\end{table*}
\normalsize

This paper is focused on showing the importance of the emission line 
contributions in mixed population colours for nearby H{\sc ii} and 
starburst regions at redshift zero. The effect will be similar for more 
distant galaxies, where the emission line contribution to the different 
bands will change with redshift for the standard systems (e.g. Johnson, 
SDSS, etc.). 

This can impact upon estimations of some M/L relationships 
and/or dust extinction in systems at different redshifts dominated by 
young ionising star forming bursts. It will also be more important at 
increasing redshift than at redshift zero, since the star formation 
becomes more violent and the mass of the young ionising population 
becomes larger. 

When a detailed study of the photometric evolution at 
high-redshift is performed, it will be necessary to take into account 
that a large number of star-forming regions could be contributing in the 
different pass-bands, changing the canonical and widely accepted 
colour-colour theoretical diagrams for synthetic populations. This will be 
the object of future work in this series.

We have calculated the colours of a system composed of two populations: 
one older than 100 \,Myr (log$\tau(yr) \ge 8.00$) and one 
younger than this same limit (log$\tau < 8.00$). The stellar mass of the 
young population in this model grid takes the same values as in the 
SSP models: 0.12, 0.20, 0.40, 0.60, 1.0, 1.5 and 2.0 $\times 10^{5}$ 
M$_{\odot}$. For each model or composite system, we assume an old stellar 
population with a mass on the zero time main sequence defined by a 
factor $\rm F=M_{old}/M_{young}$. We have taken 6 possible values for 
this grid, F=0, 1, 10, 100, 1000 and 5000. Colours are computed by using 
the total luminosity emitted by both the old population and the young 
stellar population, including the emission lines contribution and the 
nebular continuum, as we show in the previous sections.

We have obtained a table for each old stellar population defined by its 
age and metallicity, where all possible combinations with the young 
stellar population (scanning the grid in mass, age and metallicity of 
the young cluster) are included. We show in Table~\ref{mix} an example 
of this type of result. The complete set of tables for all ages and 
metallicities of the old stellar population are available in electronic 
format (and they are also available at the {\sc POPSTAR} web page, 
http://www.fractal-es.com/PopStar). Different columns in 
Table~\ref{mix} are: column 1 and 2, the metallicity and age of the old 
population; columns 3 and 4, the metallicity and age of the young 
cluster; column 5, the factor F or mass ratio between old and young 
populations; column 6, the young population's mass; 
column 7, the outer radius (or observed photometrical radius) of 
the region in pc; column 8, the logarithm of the luminosity of \halpha\  
in erg.s$^{-1}$; column 9, the equivalent with of \halpha, EW (\halpha), 
in \AA; columns 10 to 14 are the magnitude V and colours U-B, B-V, V-R 
and R-I respectively, corresponding to Johnson system, and 
in which the  emission lines contribution has been included, columns 15 to 19 are the 
same quantities than in columns 10 to 14 but without the emission line 
contribution (canonical system). Similarly, columns 20 to 24 are the 
magnitudes g and colours u-g, g-r, r-i and i-z (of the SDSS system) 
including the emission lines contribution while columns 25 to 29 are the 
same quantities without the emission line effect.

\subsection{Colour Evolution}

Fig~\ref{colorsmixj_age} shows two examples of these mixed populations.  
We show the time evolution of the same colours as in the previous SSP 
figures for two cases: Case (1) in the four top panels, we consider a 
stellar system composed of an old stellar population of 10 \,Gyr and $\rm Z = 0.0001$
 plus a young stellar population of 10$^{5}$\, \Msun\ and $\rm Z = 0.0004$. 
 The figure shows the colour evolution with the young cluster age, 
ranging from 0 to 20\,Myr. Case (2) in the four bottom panels considers 
the mixing of an old stellar population of 10\,Gyr and $\rm Z=0.004$ with a 
young cluster mass of 10$^{5}$ \Msun\ and metallicity of $\rm Z = 0.008$. In 
all cases the dashed lines are the colours calculated without the 
emission line contribution, while the solid lines correspond to the 
total colour including this contribution. Different colours indicate 
different F values, ranging from 0 (black, and meaning a pure young 
population) to 5000 (red, meaning a quite important mass contribution of 
the underlying old component, in this case 5 $\times 10^{8}$\,\Msun\ 
since the young cluster mass is 10$^{5}$ \Msun). These F values and the 
legend have been labelled in the V-R {\sl vs} R-I diagrams. The higher 
the value of F, the smaller the emission lines impact on the integrated 
colour, as expected. In the plots, larger differences between dashed and 
solid lines can be appreciated for small F values while both lines are 
almost coincident for large F values. The cases with $\rm F = 0$ correspond to 
a pure young population and therefore to the case discussed in \S3 of 
this paper. It can also seen that as the young population evolves, the 
contribution of the lines tends to disappear, and the dashed and solid 
lines tend towards coincidence.

When F is in the range 1 (magenta line) to 10 (blue line), the old 
population contributes in a small proportion and the young population 
dominates the system. In these cases, colours U-B, V-R and R-I for the 
first evolutionary phases of the young population (age $< 5$\,Myr) 
present a behaviour close to the one with F around 100 and no-emission 
lines contribution (green dashed lines). It means that using a mixture 
of two populations without taking into account the emission line 
contribution, can lead to a misinterpretation, by assigning a stronger 
contribution of the underlying population than the actual one. Thus, for 
this particular example, the error determination can be as high as one 
order of magnitude in the F value, implying a similar error for the 
determination of the mass of the underlying population (assuming that we 
can constrain the young population mass through other photometric 
values, like the \halpha\ luminosity).

Only the colour B-V and somehow U-B seem to provide similar estimations 
for the old population's age, regardless of the models used (with or 
without emission lines). A similar behaviour is observed in the Case 
(2), displayed in the four bottom panels, which indicates that this 
effect is quite independent of the metallicity for these examples. Of 
course, these conclusions can change for other combinations of 
metallicities and for younger ages of the old population. The potential 
confusion will become larger when a stronger starburst is overlapping 
with an intermediate-age population. We have plotted this extreme case 
with a very old underlying population to demonstrate that even in this 
system we can find important differences in the inferred physical 
properties when using canonical models instead of the ones with the 
emission line contribution.

Fig.~\ref{colorsmixs_age} shows the same example of mixed populations, 
but for SDSS colours.  The plots for case (1) show the same effect of 
possible misinterpretation in the colours u-g, g-r, and r-i when using 
uncontaminated colours instead the ones including the emission line 
contribution.  Again, differences larger than one order of magnitude in 
the burst strength can be inferred. From the lower panels (case 2), all 
the colours, and especially r-i and i-z, would lead to misinterpretation.
We cannot provide plots for 
the whole range of age, mass, and metallicity of the contributing 
populations, but given a specific set of photometric observations, a 
good representation of the reality may be obtained by fitting the 
different parameters to the theoretical models, following for example a 
mean square error (MSE) as a measure of estimator quality. We provide 
the complete set of tables and models in electronic format to allow 
these computations.

\subsection{Colour-Colour Diagrams}

\begin{figure*}
\centering
\subfigure{\includegraphics[width=0.62\textwidth,angle=-90]{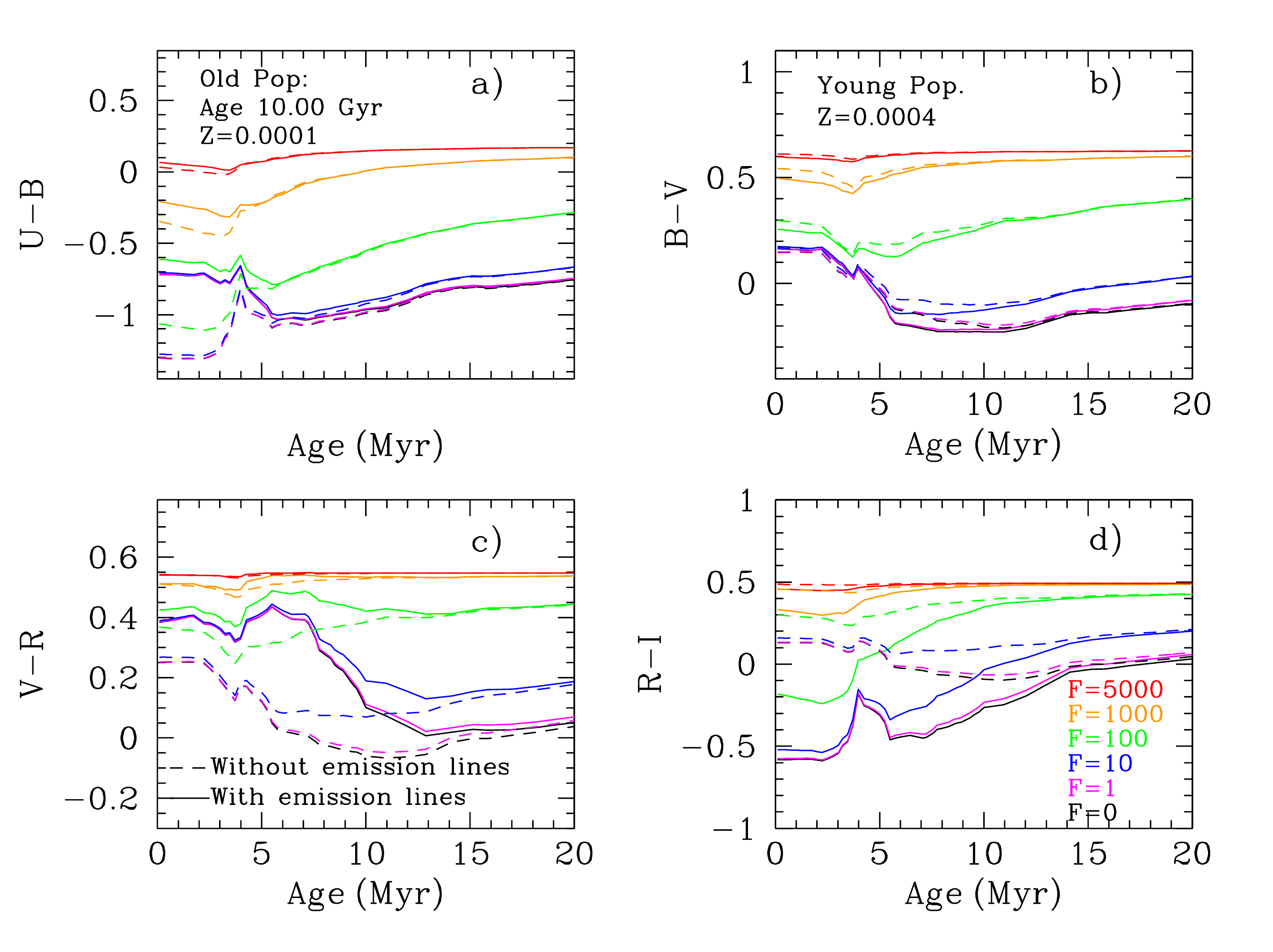}}
\subfigure{\includegraphics[width=0.62\textwidth,angle=-90]{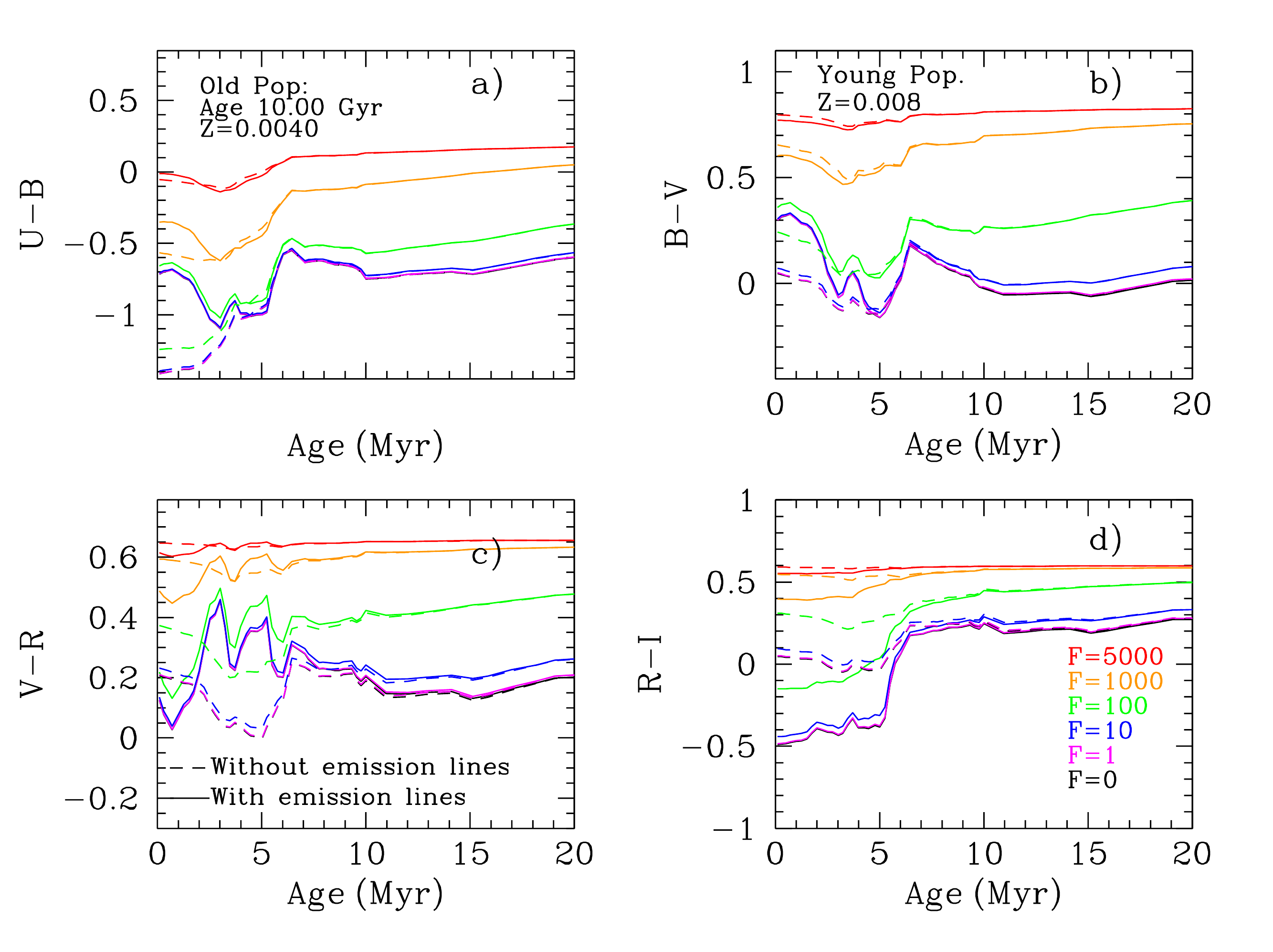}}
\caption{Examples of Johnson colour evolution with a young cluster age in a mixed population. This is composed of an old 10\,Gyr underlying cluster plus a young burst ($\tau < 20$\,Myr and clear \halpha\ emission). Each colour represents a different mass contribution of the old population to the young burst (F, being mass-old / mass-young) as labelled. 
Synthetic colours with and without the emission lines contribution are plotted as solid and dashed lines respectively.
Two metallicity cases have been chosen as examples: top panels use $\rm Z=0.0001$ and $\rm Z=0.0004$ for old and young populations, while bottom panels get $\rm Z=0.004$ and $\rm Z=0.008$,
 as labelled in the plots.}
\label{colorsmixj_age}
\end{figure*}

\begin{figure*}
\centering
\subfigure{\includegraphics[width=0.62\textwidth,angle=-90]{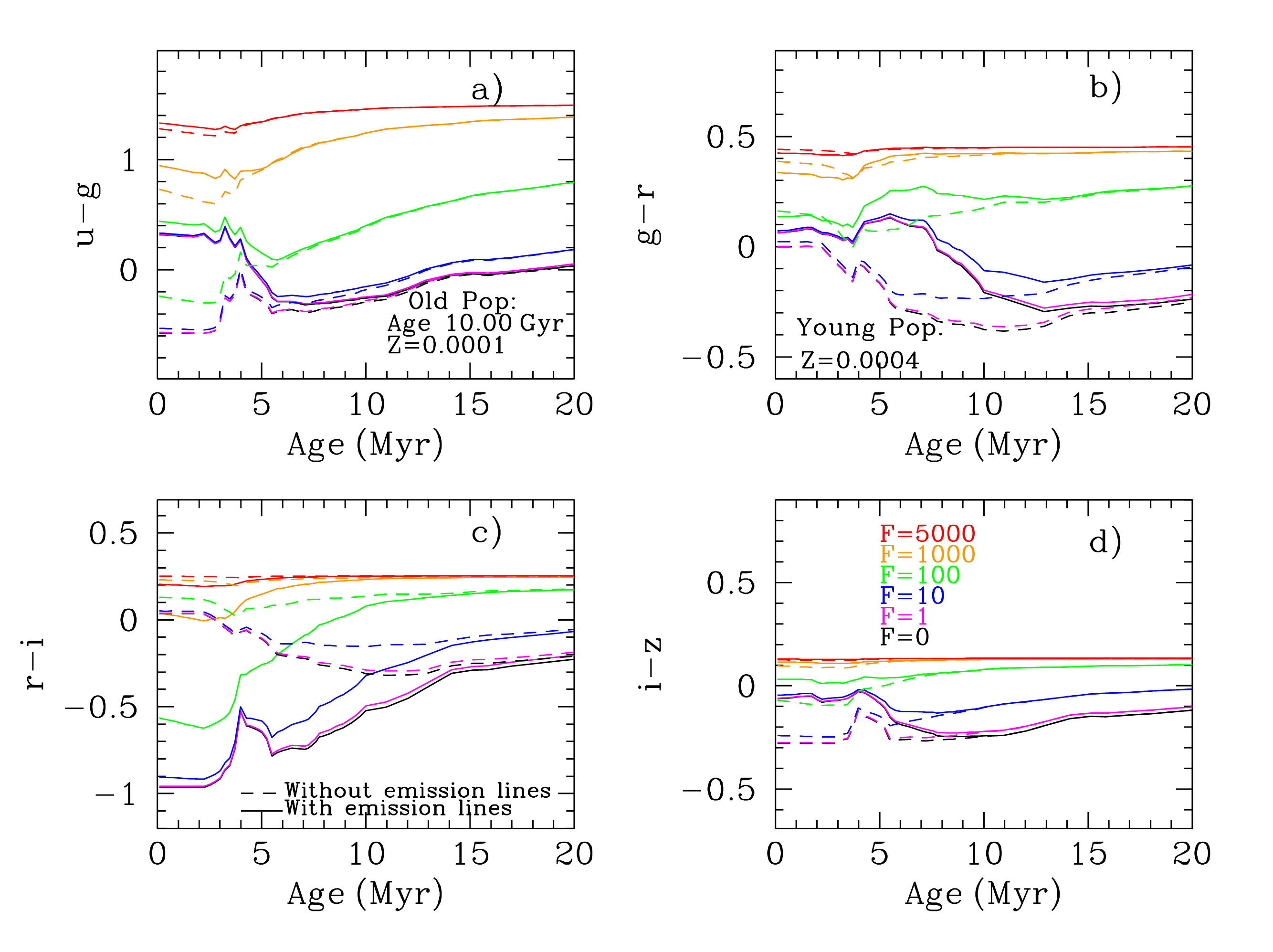}}
\subfigure{\includegraphics[width=0.62\textwidth,angle=-90]{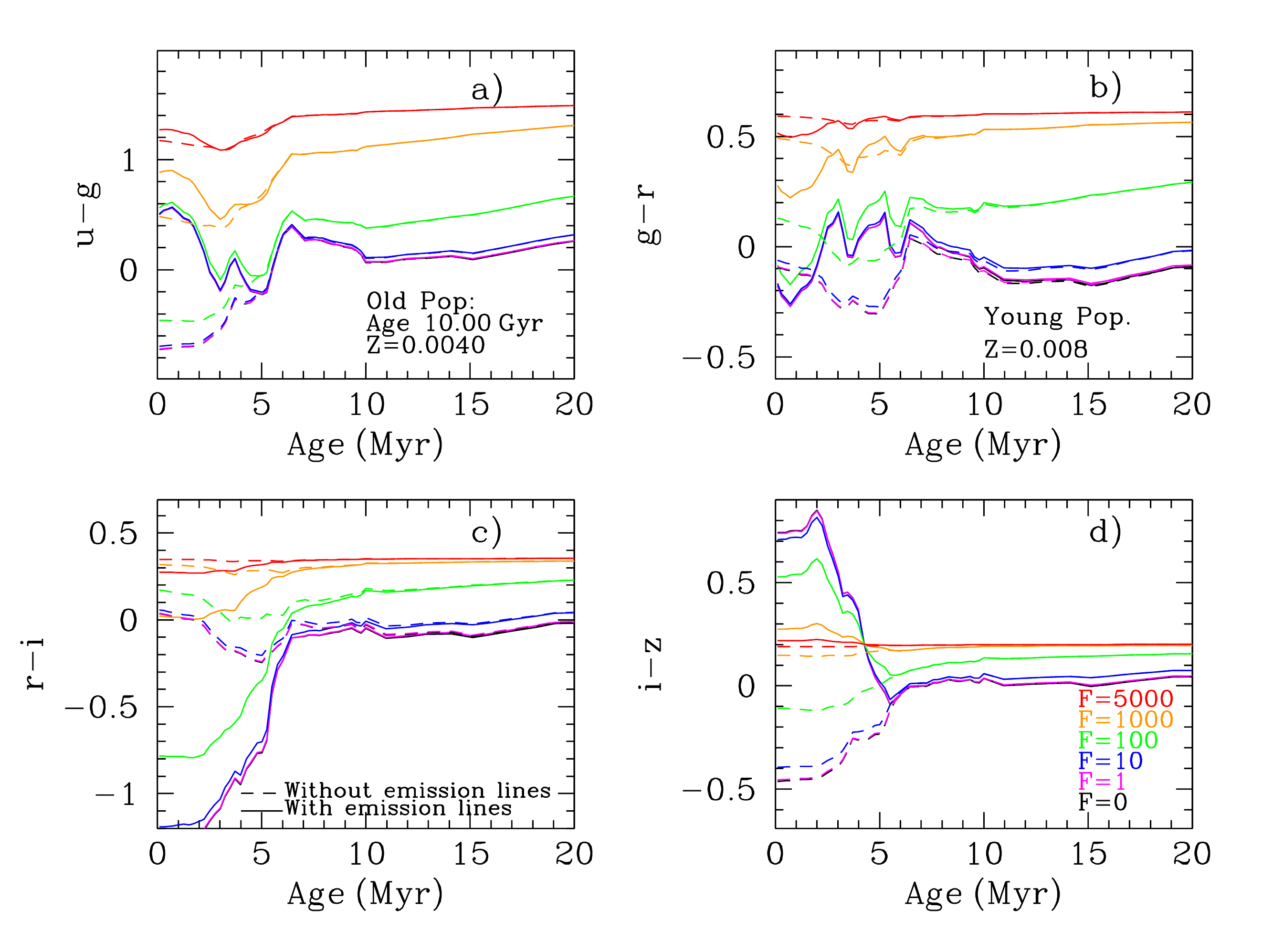}}   
\caption{Examples of SDSS colour evolution with a young cluster age for a mixed population. This is composed by an old 10\,Gyr underlying cluster plus a young burst (Age $\tau < 20$\,Myr and clear \halpha\ emission).  Each colour represents a different mass contributions of the old population to the young burst (F, being mass-old / mass-young) as labelled.
 Synthetic colours with and without the emission lines contribution are plotted as solid and dashed lines respectively.
Two metallicity cases have been chosen as examples: top panels use $\rm Z=0.0001$ and $\rm Z=0.0004$ for old and young population, while bottom panels get $\rm Z=0.004$ and $\rm Z=0.008$, as labelled in the plots.}
\label{colorsmixs_age}
\end{figure*}

\begin{figure*}
\centering
\subfigure{\includegraphics[width=0.61\textwidth,angle=-90]{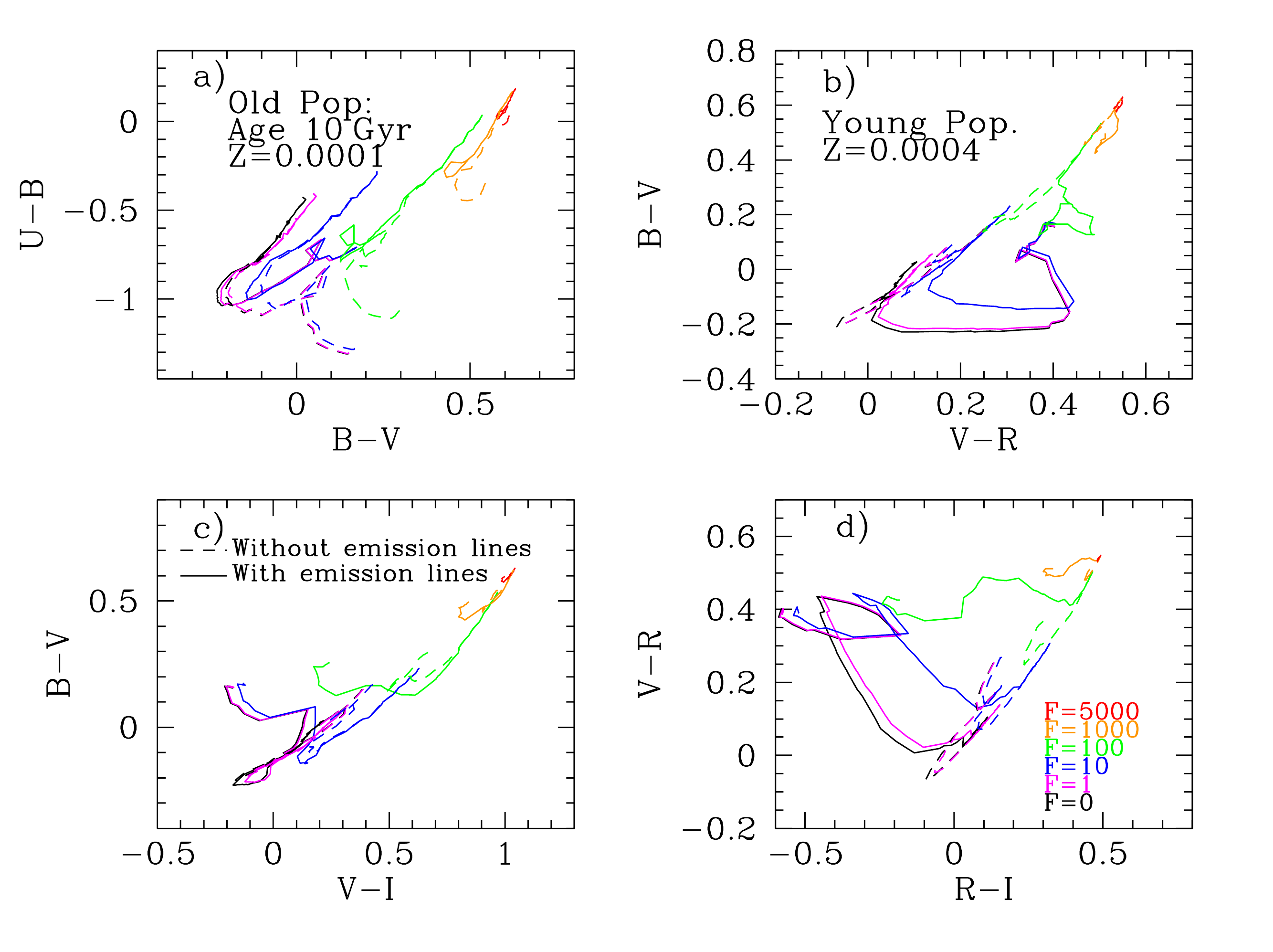}}
\subfigure{\includegraphics[width=0.61\textwidth,angle=-90]{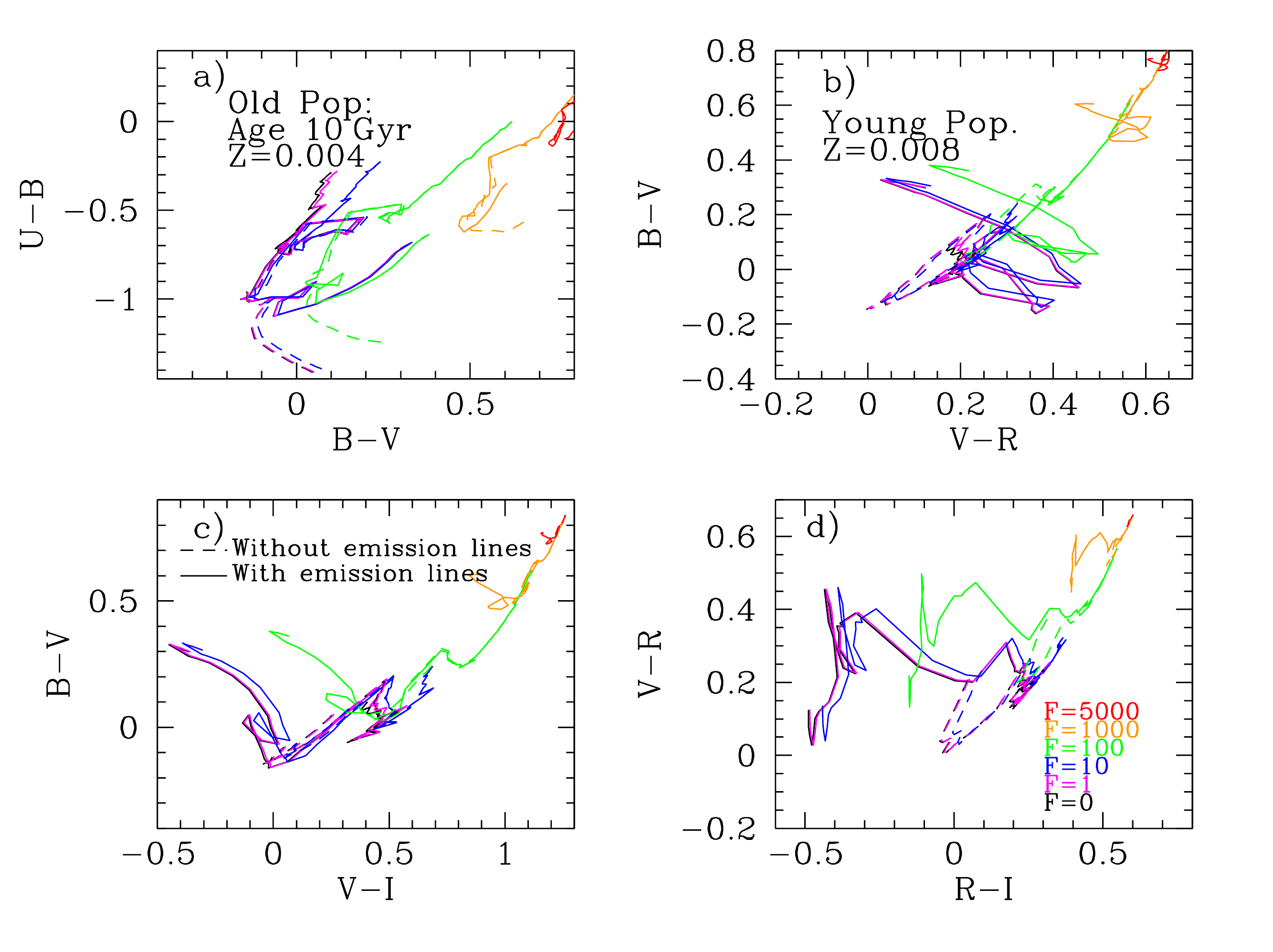}}
\caption{Examples of Johnson colour-colour evolution for a mixed population. This is composed by an old 10\,Gyr underlying cluster plus a young burst ($\tau < 20$\,Myr and clear \halpha\ emission). Different coloured lines represent a different mass contribution of the old population to the young burst (F, being mass-old / mass-young). Synthetic colours with and without the emission lines contribution are plotted as solid and dashed lines respectively.  Two metallicity cases have been chosen as examples: top panels use $\rm Z=0.0001$ and $\rm Z=0.0004$ for old and young population respectively while bottom panels get $\rm Z=0.0004$ and $\rm Z=0.0008$, as labelled in the plots.}
\label{colorsmixj_colormixj}
\end{figure*}

\begin{figure*}
\centering
\subfigure{\includegraphics[width=0.61\textwidth,angle=-90]{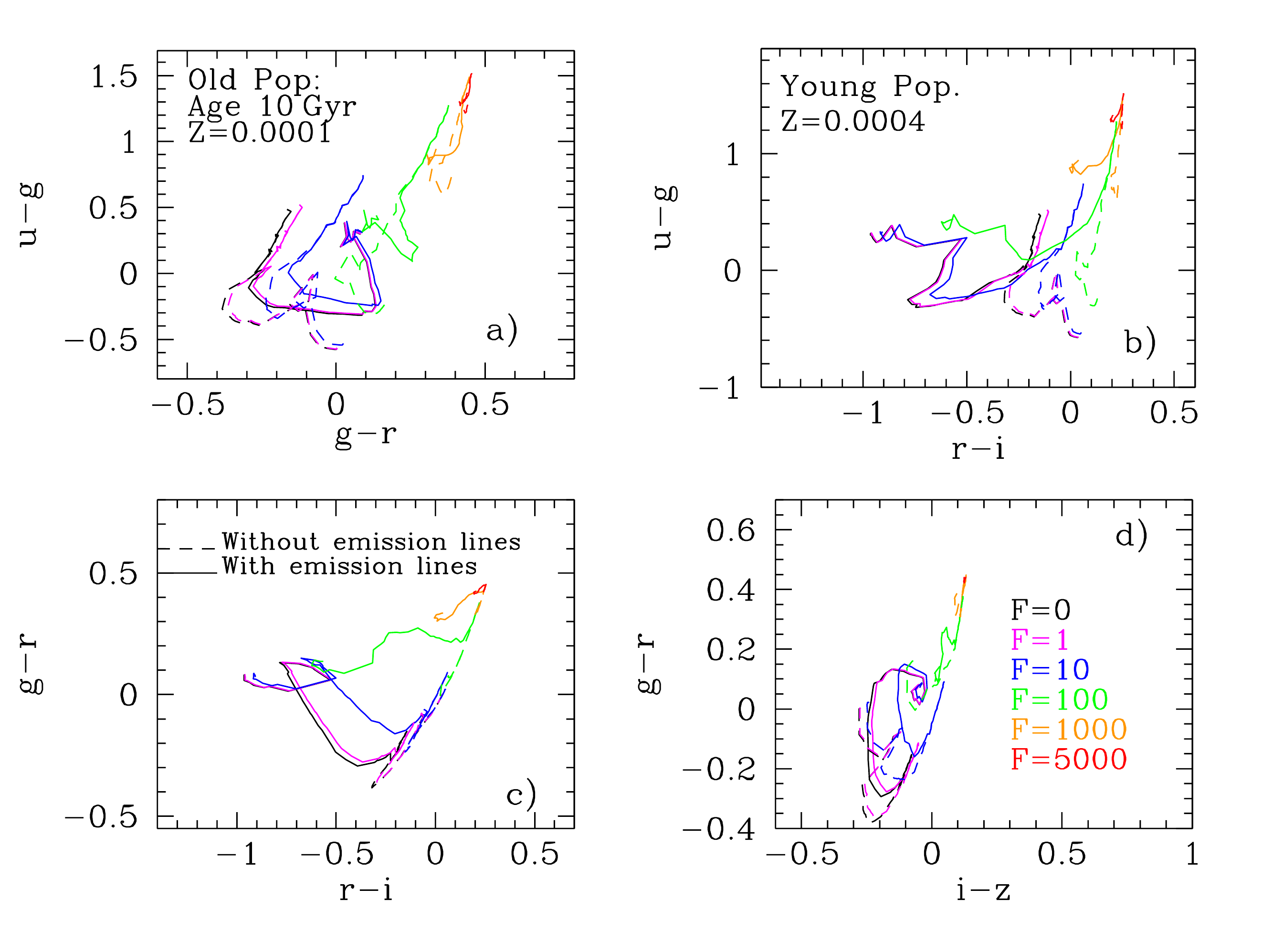}}
\subfigure{\includegraphics[width=0.61\textwidth,angle=-90]{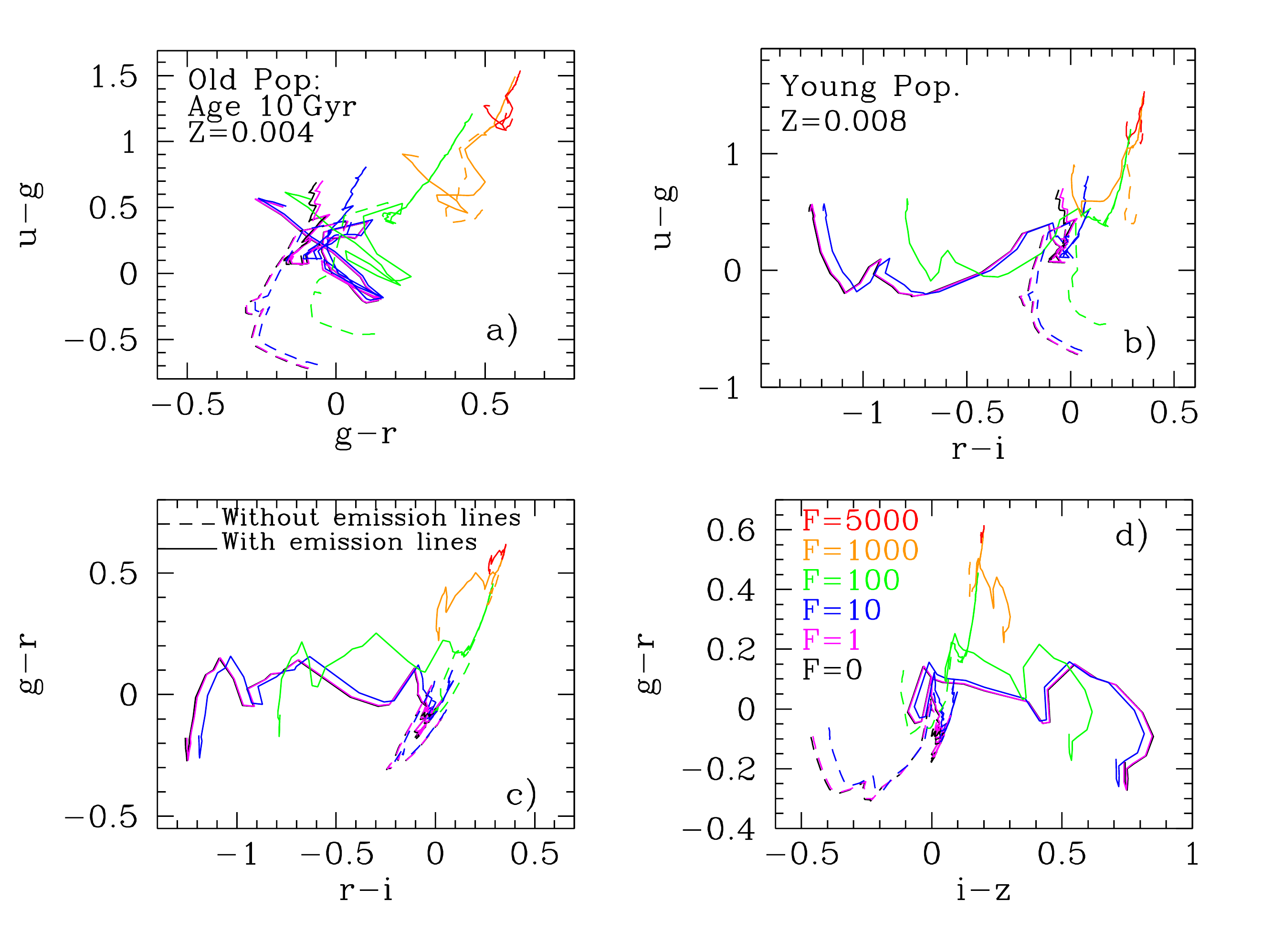}}
\caption{Examples of SDSS colour-colour evolution for a mixed population. This is composed by an old 10\,Gyr underlying cluster plus a young burst ($\tau < 20$\,Myr and clear \halpha\ emission). Different coloured lines represent a different mass contribution of the old population to the young burst (F, being mass-old / mass-young). Synthetic colours with and without the emission lines contribution are plotted as solid and dashed lines respectively.   Two metallicity cases have been chosen as examples: top panels use $\rm Z=0.0001$ and $\rm Z=0.0004$ for old and young population respectively while bottom panels get Z=0.0004 and Z=0.0008, as labelled in the plots.}
\label{colorsmixs_colormixs}
\end{figure*}

The same mix of stellar populations as before are represented in 
Fig.\ref{colorsmixj_colormixj} and Fig.\ref{colorsmixs_colormixs} as 
colour-colour diagrams. As in Fig.~\ref{colorsmixj_age} and 
Fig.~\ref{colorsmixs_age}, different colour lines represent different 
ratios in mass, F, between old and young stellar populations. Thus red 
lines give the colour-colour results when the old stellar population 
dominates in mass, with a value of 10$^{8}$\, \Msun\ over the young one 
of 10$^{5}$\,\Msun\, while the black lines represent the evolution of a 
simple young stellar population without underlying population. 
All the other colours (orange, green, blue and magenta) represent intermediate 
cases. 
  
The dashed lines are the results without the emission line 
contribution while the solid lines show our results including the 
emission line contamination.  
In these figures, the model results given 
by the solid lines (emission lines included) can be bluer or redder, 
compared to the models of the dashed-lines (no emission lines). This 
will depend also on F.  For a given value of F, models on the solid 
lines (with lines) will be close or far from the canonical solid line 
(they can be placed also on an orthogonal line) depending on the age of 
the young population as discussed in \S3. Our results predict a large 
dispersion among the observational values when plotting them in 
colour-colour plots, explaining in a natural way this fact. Previous work 
often proposed a reddening excess due to dust to explain the discrepancy 
between the synthetic colours and the observed ones.

\subsection{Colours {\sl vs} \halpha\ Equivalent Widths}

\begin{figure*}
\centering
\subfigure{\includegraphics[width=0.61\textwidth,angle=-90]{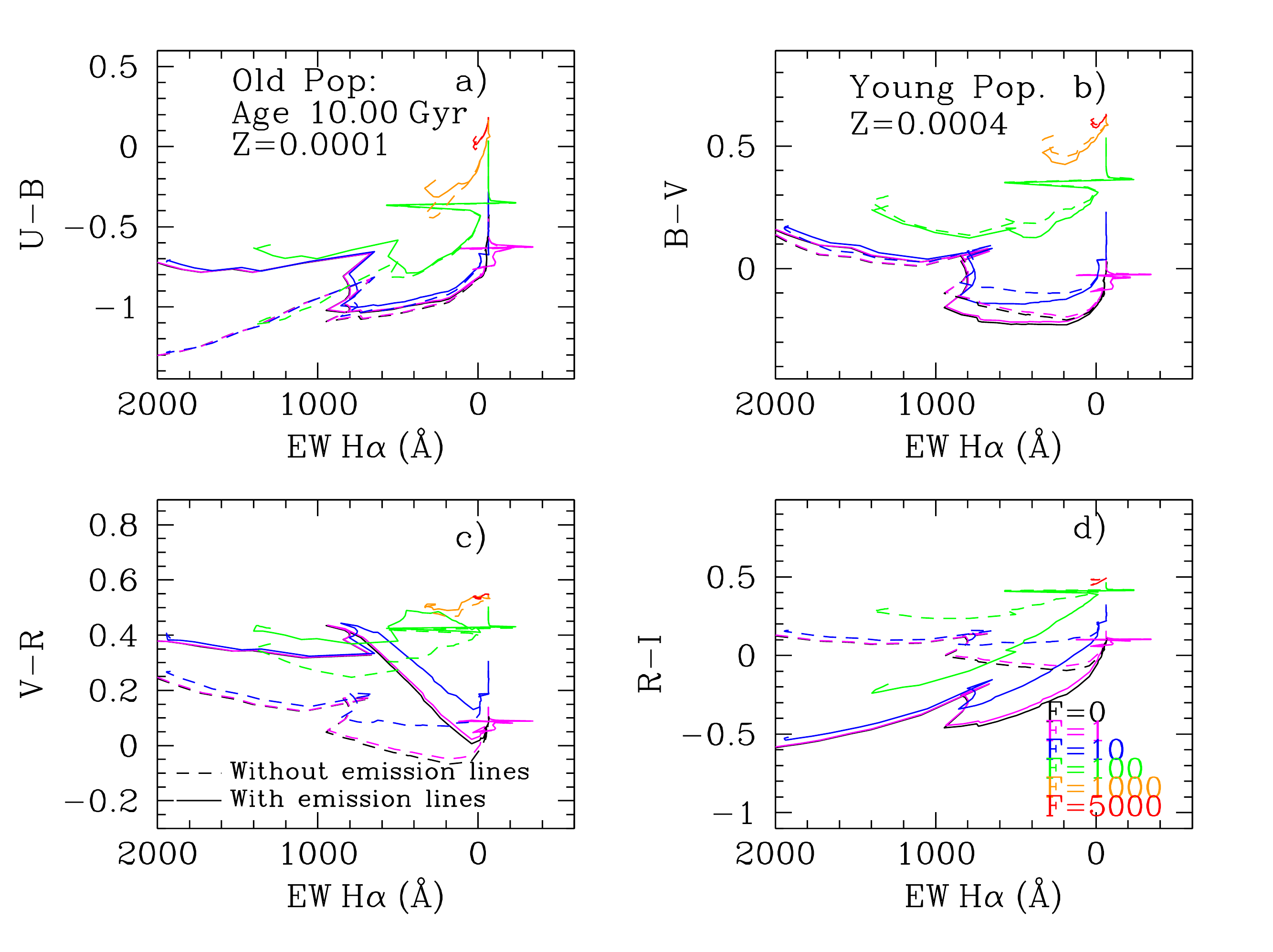}}
\subfigure{\includegraphics[width=0.61\textwidth,angle=-90]{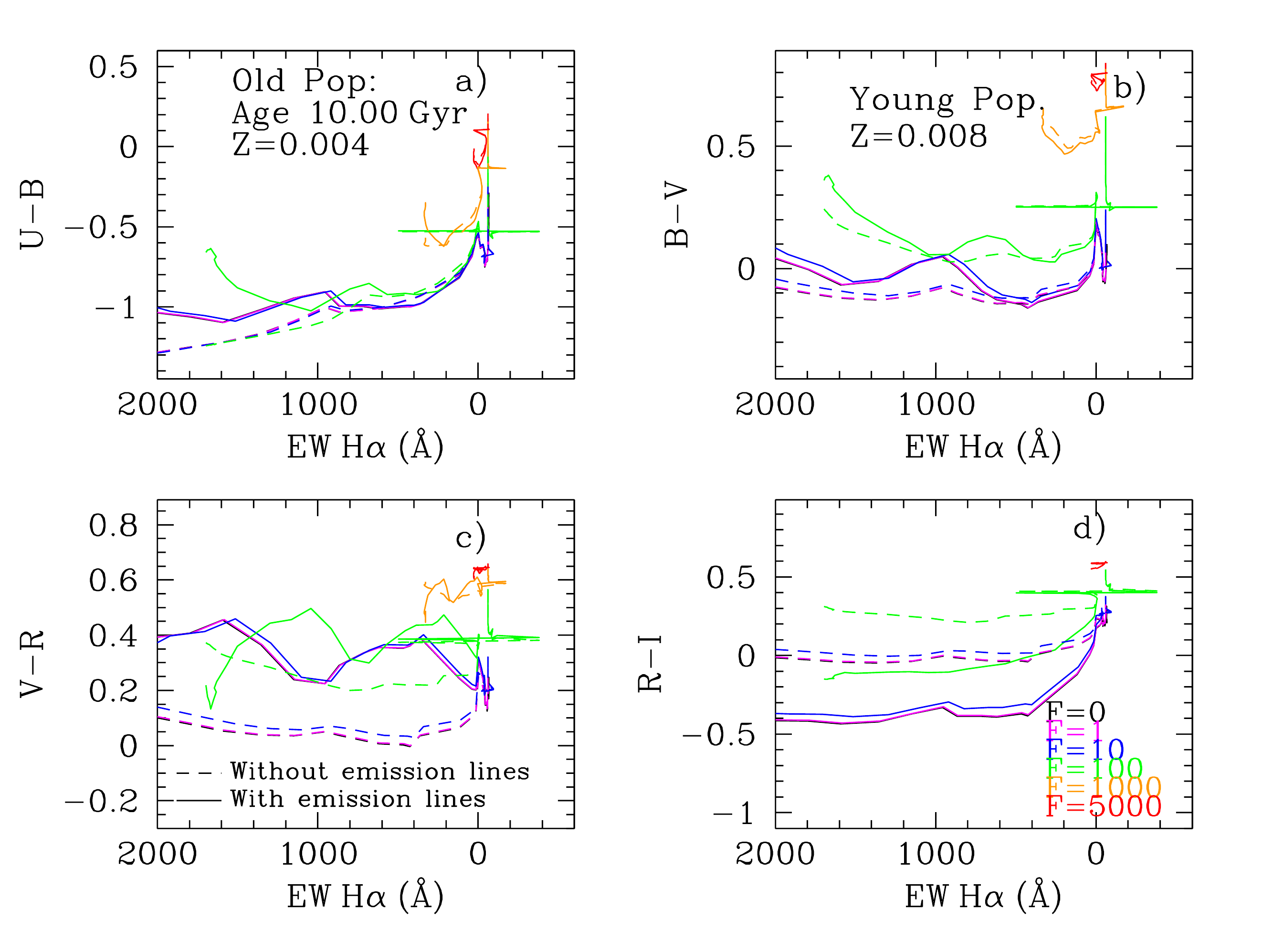}}
\caption{Examples of different Johnson colours - as labelled in the panels - plotted against the Equivalent Width of H$_{\alpha}$ in emission for a mixed population. This is composed by an old 10\,Gyr underlying cluster plus a young burst ($\tau < 20$\,Myr and clear \halpha\ emission). Each coloured line represents a different mass contribution of the old population to the young burst (F, being mass-old / mass-young). Synthetic colours with and without the emission lines contribution are plotted as solid and dashed lines respectively.  Two metallicity cases have been chosen as examples: top panels use $\rm Z=0.0001$ and $\rm Z=0.0004$ for old and young population respectively while bottom panels get $\rm Z=0.004$ and $\rm Z=0.008$, as labelled in the plots.}
\label{colorsmixj_ewha}
\end{figure*}

\begin{figure*}
\centering
\subfigure{\includegraphics[width=0.61\textwidth,angle=-90]{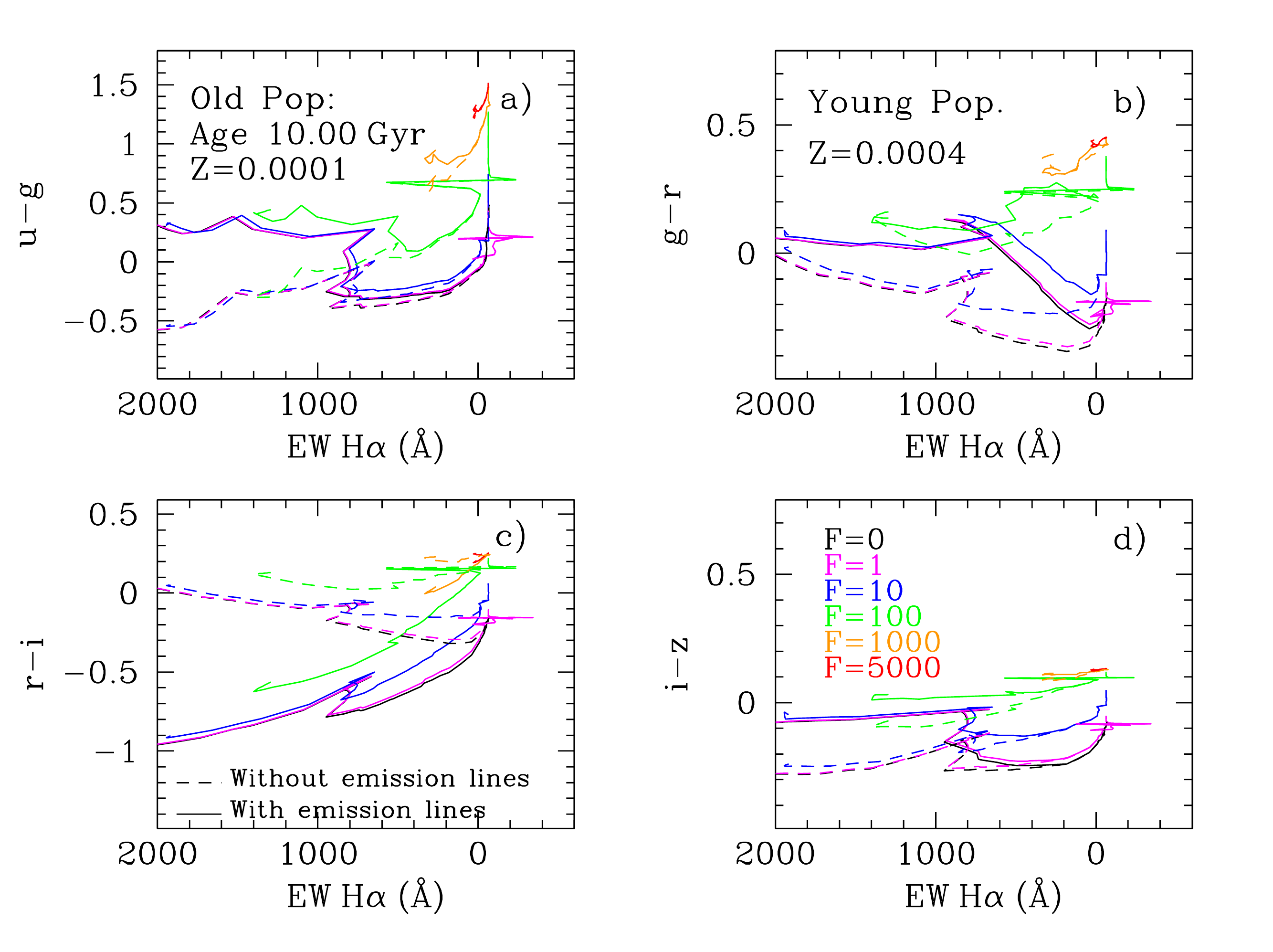}}
\subfigure{\includegraphics[width=0.61\textwidth,angle=-90]{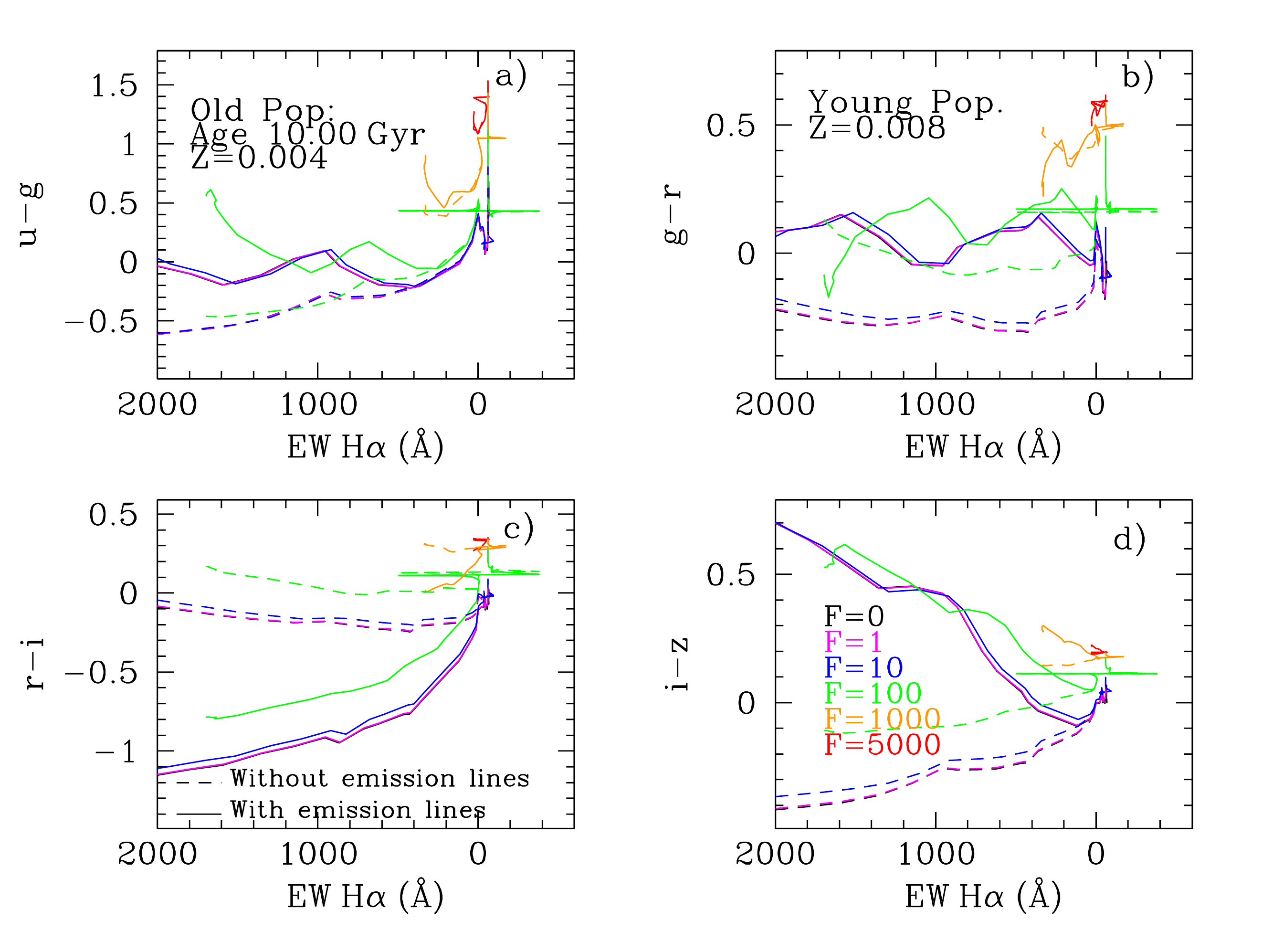}}
\caption{Examples of different SDSS colours - as labelled in the panels - plotted against the Equivalent Width of H$_{\alpha}$ in emission for a mixed population. This is composed by an old 10\,Gyr underlying cluster plus a young burst ($\tau < 20$\,Myr and clear \halpha\ emission). Each coloured line represents a different mass contribution of the old population to the young burst (F, being mass-old / mass-young). Synthetic colours with and without the emission lines contribution are plotted as solid and dashed lines respectively.  Two metallicity cases have been chosen as examples: top panels use $\rm Z=0.0001 $ and $\rm Z=0.0004$ for old and young population respectively while bottom panels get $\rm Z=0.004$ and $\rm Z=0.008$, as labelled in the plots.}
\label{colorsmixs_ewha}
\end{figure*}

Finally, Fig.~\ref{colorsmixj_ewha} and Fig.~\ref{colorsmixs_ewha} show 
the synthetic colours for the same combinations of stellar populations 
described in the previous sub-sections {\sl vs} the equivalent width of 
\halpha\ with the same meaning of colours and line types for Johnson and 
SDSS colours respectively. As we have seen before, the emission lines do 
colours bluer or redder, and therefore the resulting solid lines fall in 
a different region than the dashed lines (canonical models without the 
emission line contribution) for a given value of the \halpha\ equivalent 
width. In the example corresponding to Fig.~\ref{colorsmixj_ewha} Case 1 
(upper panels) B-V (panel b) shows that differences between solid and 
dashed lines are small, so that estimations based on this type of plot 
would be similar using the colours with or without emission lines.  
However this is not the case for the other colours (panels a, c and d for 
U-B, V-R and R-I respectively) for which we suggest that the colours 
including emission lines are redder (for U-B and V-R) or bluer (for R-I) 
than the ones synthesised without this contribution. A similar behaviour 
is found in Case 2 (lower four panels).

As in previous figures, different colour lines correspond to different 
values of F, from 0 (black, pure young population) to 5000 (red, with a 
strong mass contribution of the underlying population). For F = 0 models 
(single population), the evolution shows blue colours and high values of 
EW(\halpha) for the younger clusters, and a progressively decrease of 
EW(\halpha) and colour reddening as far as the cluster evolves. Models on 
dashed lines do not present simultaneously red colours and high values of 
EW (\halpha) as observed in H{\sc ii} galaxies for example. This is not 
the case for the models with the emission lines, in which we can find 
red colours and high values of EW(\halpha) in the range of several 
hundreds or even few thousands, at early ages, without the need for 
interpreting this as a dust excess.

\subsection{Comparison with Observational Data}

\begin{figure*}
\centering
\resizebox{\hsize}{!}{\includegraphics[width=0.40\textwidth,angle=0]{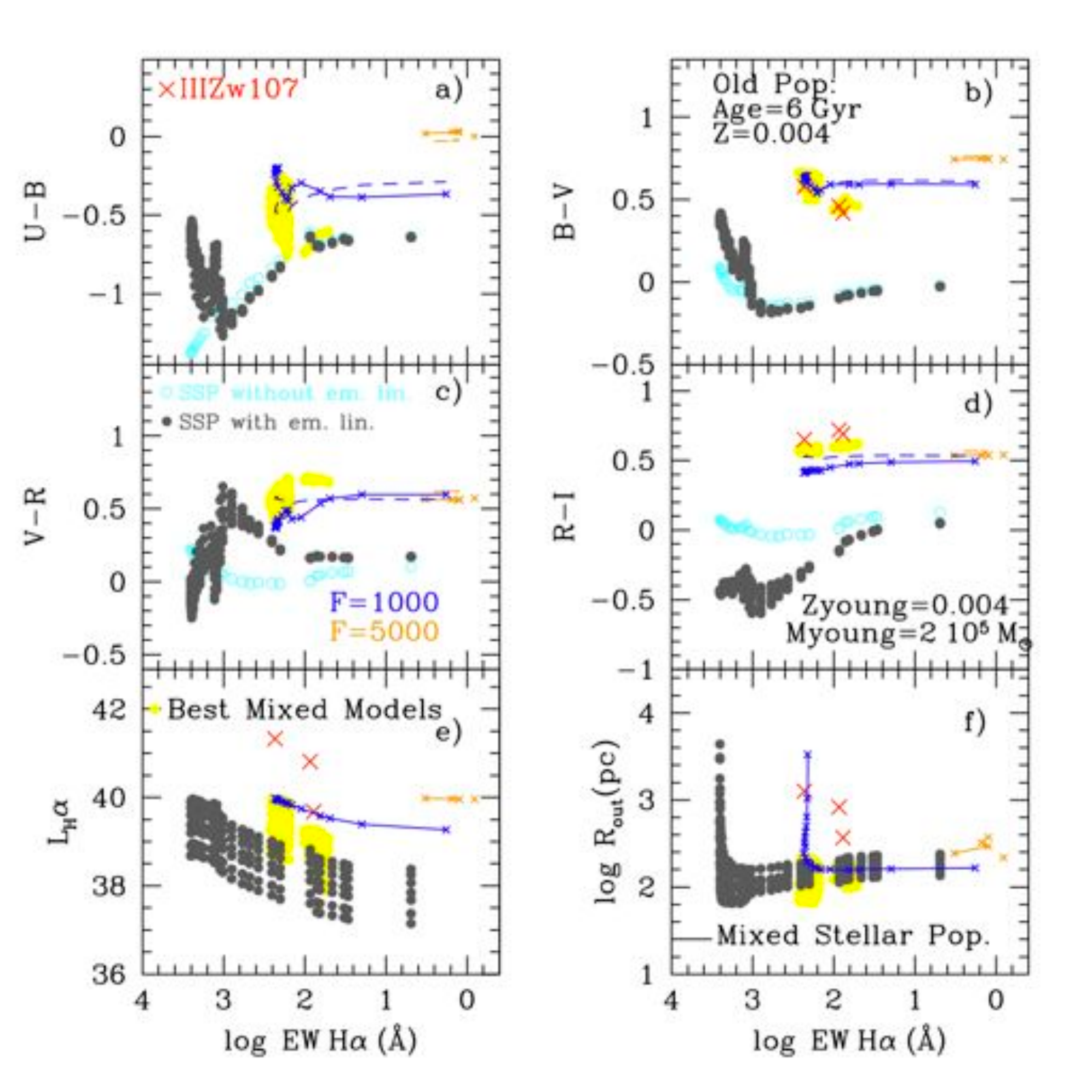}}
\caption{Colours U-B, B-V, V-R and R-I, logarithm of \halpha luminosity (erg.s$^{-1}$) and logarithm of 
the H{\sc ii} region radius, R$_{out}$( in pc), {\sl vs} EW(\halpha). Observed data for star-forming regions in the galaxy IIIZw107 (\citet{ismael}) are plotted as red crosses. SSPs models with 
$\rm Z=0.004$ are shown with open (cyan) and full (black) dots.  The yellow region is the locus of models with $\rm Zyoung=0.004$, the estimated metallicity from visible spectroscopy of the observed star-forming regions, that are able to reproduce simultaneously the observed colours and EW(\halpha). From these 2-population mixed models we have selected those ones 
with an old population of the same metallicity, $\rm Z=0.004$, and 6\,Gyr old and two values of F, $\rm F = 1000$,  whose evolution with the young population age is plotted as a blue line, 
and $\rm F = 5000$, plotted as an orange line, to show as example the evolution of a mixed population.}
\label{color_color_obs1}
\end{figure*}

\begin{figure*}
\resizebox{\hsize}{!}{\includegraphics[width=0.40\textwidth,angle=0]{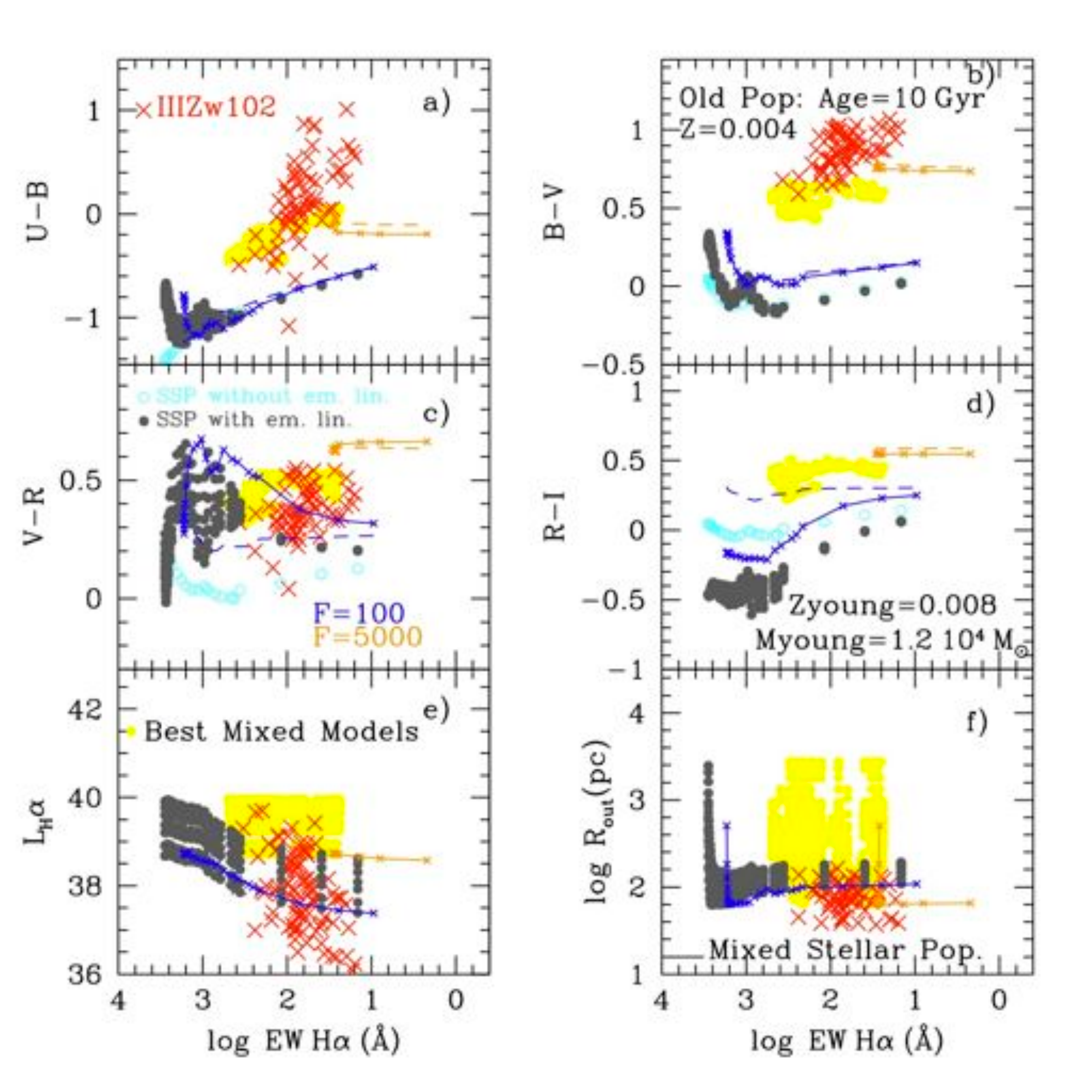}}
\caption{Colours U-B, B-V, V-R and R-I, logarithm of \halpha luminosity (in erg.s$^{-1}$) and logarithm of 
the H{\sc ii} region radius, R$_{out}$( in pc), {\sl vs} EW(\halpha). Observed data for star-forming regions in the galaxy IIIZw102 (\citet{ismael}) are plotted as red crosses. SSPs models with 
$\rm Z=0.008$ are shown with open (cyan) and full (black) dots.  The yellow region is the locus of models with $\rm Zyoung=0.008$, the estimated metallicity from visible spectroscopy of the observed star-forming regions that are able to reproduce simultaneously the observed colours and EW(\halpha). From the 2-population mixed models we have selected those ones 
with an old population of $\rm Z=0.004$ and 10\,Gyr old and two values of F, $\rm F = 100$,  whose evolution with the young population age is plotted as a blue line; and $\rm F = 5000$, 
plotted as an orange line, to show as example the evolution of a mixed population.}
\label{color_color_obs2}
\end{figure*}

In this section we test our models to see if they can reproduce the 
photometric observations from star forming regions where it is already 
proved that at least a young and an old stellar population exists there. 
We have analysed the H{\sc ii} region data of \citet{ismael}, where 
photometry exists for a well-observed sample of Blue Compact Dwarf (BCDs) galaxies with 
spectroscopic metallicity determinations.  This makes the sample 
suitable to compare with our models, in order to see if our predicted 
colours for mixed populations can reproduce these data. 
We have plotted different synthetic photometric 
parameters versus the equivalent width of \halpha, which is a proxy of 
the age, together with observations of H{\sc ii} regions for two 
galaxies: III~ Zw~ 107 and III ~Zw ~102 in Figs.~\ref{color_color_obs1} and 
\ref{color_color_obs2}, respectively. In both figures panels a) to d) 
plot the synthetic colours U-B, B-V, V-R and R-I respectively. Panel e) 
uses the \halpha luminosity and panel f) the photometric radius of the 
H{\sc ii} region, R$_{out}$.

Observed regions in Figs.~\ref{color_color_obs1} and \ref{color_color_obs2} correspond to data for 
the galaxies III Zw 107, and III Zw 102, respectively, plotted as red crosses. The metallicity for the first 
galaxy estimated from its oxygen abundance (itself derived using optical 
emission lines) is  12 $ + \log{[O/H]}=$  8.23, equivalent to $\rm Z \sim 0.004$ \citep{arls10}. This metallicity is also consistent with the appearance of WR features in spectroscopy observations.  
Similarly for IIIZw 102, the estimated abundance 
(from spectroscopic oxygen abundances from 3D spectroscopy ) is  12 $ + \log{[O/H]}=$ 8.49 \citep{cai12}, equivalent to $\rm Z \sim 0.008$. 
\footnote{Solar abundance in our models is taken as: 12 $ + \log{[O/H]}=$ 8.67, equivalent to $\rm Z= 0.015$, \citep{asp09}.}

In these two galaxies used as examples it is impossible to explain the observed colours without the existence of an older underlying stellar population,
as can be seen in Fig.~\ref{color_color_obs1} and Fig.~\ref{color_color_obs2}, where the synthetic colours  for the case of a 
SSP ($\rm F=0$) of a metallicity as the one estimated from the emission lines as given before, without the contribution of the emission lines 
(cyan open dots) and with the emission lines (black full dots), cannot reproduce the observed colours.  We can 
see that neither of the two set of models can reproduce the 
observations. In Fig~\ref{color_color_obs1} and and Fig.~\ref{color_color_obs2},  we cannot find any solution for colours U-B (panel a), B-V (panel b), V-R (panel c) and 
R-I (panel d), when data are available, and neither is it possible to find a population able to 
simultaneously fit the colours and the \halpha equivalent width. 

The presence of such an underlying old population is widely accepted in  BCGs, as the case of III Zw107 and III Zw102. Although their low metallicity and the very young stellar populations led 
some time ago to the proposal that these systems could be suffering their first burst of star formation \citep{ss70} more recent observations point to these galaxies seem to be not as young as it was originally thought. Photometric observations indicate that they host stellar populations, reaching in some cases a few Gyr \citep{tel97,leg00,tols03,cai03,vzee04,thuan05}.
There is now wide agreement about the existence of underlying populations with ages of several Gyr in BCGs and/or H{\sc ii} galaxies. 
Even the most metal-poor galaxy known, I Zw18, the best candidate to primordial galaxy, has started forming stars earlier than $\sim 1$ Gyr old \citep{ann13}. For this reason, the fit of a mixed population to the observed data, might improve the result.

Therefore we have selected among all the mixed stellar populations 
models computed in our grid and described in the previous section, those ones  whose 
colours and equivalent width EW (\halpha) can fit the observations within 
the error bars.  To do that we have performed a chi-square fit of our mixed models in which the metallicity of the young stellar population, Zyoung, is equal to the estimated one by the emission lines as given before.  The yellow region in each figure shows the results of this fit. These results represent all possible combinations of Zold+Age-old+Age-young able to reproduce the data within errors.   It is clear that there are solutions to the data. 

As an example we also plot in Fig~\ref{color_color_obs1} the evolution with the age of the young population for one of these combinations or mixed 
stellar populations, which has a young stellar mass of 2 $\times 
10^{5}$\,M$_{\odot}$, and $\rm Z=0.004$ with one 6\,Gyr old with $\rm Z=0.004$ and 
with two possible values for mass contributions: $\rm F=1000$ (dark blue 
lines) and $\rm F = 5000$ (orange lines). Solid lines represent models 
including emission lines while dashed lines are models without this 
contribution. We see that the three observed regions are better fit by 
the model with $\rm F = 1000$. In panels e) and f) we show the same yellow 
region. Panel e) shows that two of the observed regions have \halpha\ 
luminosities larger than those of the plotted models. This is likely due 
to the fact that our grid does not include cluster masses larger than  
$2 \times 10^{5}$ \Msun . At least (although obviously the interpretation would be different if other IMF is used)
 we can put a lower limit for the  mass of these H{\sc ii} regions.

Similar to Fig.~\ref{color_color_obs1} we have plotted 
Fig.~\ref{color_color_obs2} with our models and observed data of 
star-forming regions in IIIZw102. Again, most of 
observational points cannot be fit by the SSP. We represent the region 
of selected mixed models as yellow points as before. This region is not 
valid for all data and some observed parameters are not reproduced with 
our current set of models.  Perhaps larger F factors are necessary to 
fit these very red colours in regions where EW(\halpha) is still present,
or maybe the averaged abundance $\rm Z= 0.008$ is not representative for all
H{\sc ii} regions. Following the results of panel e), the young stellar population mass of 
these regions should be smaller that our lowest limit, 10$^{4}$\,\Msun. 
This would also explain the values of the observed radii also being 
smaller than the predicted ones. It is necessary to remain that the chi-square fit is performed using only the data of 
panels a) to d), not e) nor f). Nevertheless, the two-population mixed 
model is also a simplification since some of the regions are not 
spatially resolved and can contain more than two stellar populations 
with different ages. It has to be noticed that the worse fitting found in the colours U-B and B-V is fully consistent with the strong dust absorption found in this galaxy, as reported by 
\cite{cai12}

Fig.~\ref{color_color_obs1} and Fig.~\ref{color_color_obs2}  show that our models can fit real data of photometric observations of H{\sc ii} regions, obtaining the metallicity and the age for both the young and the old populations and reproducing several colours simultaneously when SSPs cannot do it. Moreover, the inclusion of emission lines in the models is needed to reproduce simultaneously all the colours in the two selected galaxies. It has to be notice that the two galaxies have very different values of the metallicity, which has been spectroscopically confirmed by different papers in the literature.

\begin{figure*}
\resizebox{\hsize}{!}{\includegraphics[width=0.40\textwidth,angle=0]{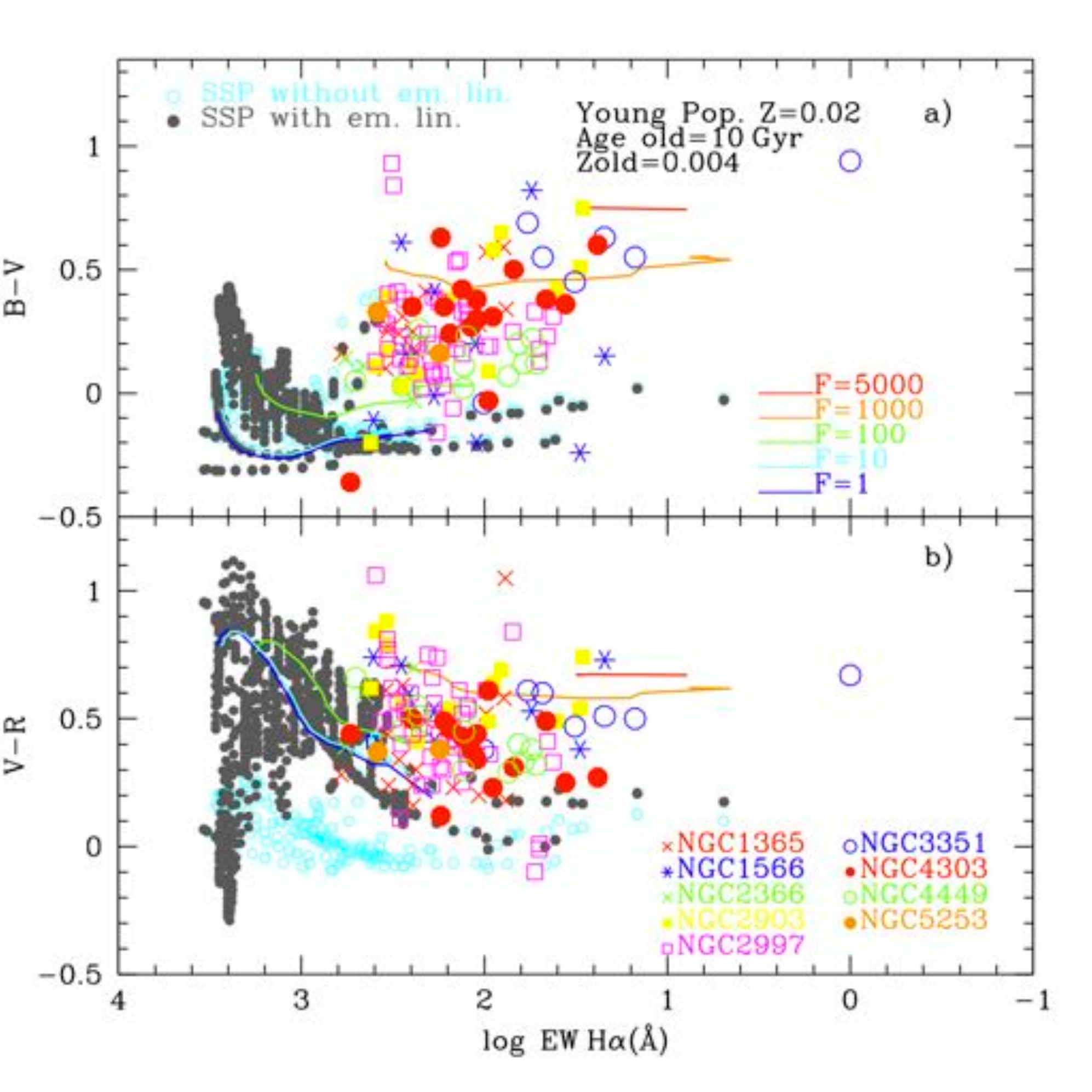}}
\caption{Colours B-V and V-R {\sl vs} EW(\halpha). Observed data are from \citet{mayya94} for
H{\sc ii} regions located in different galaxies as labelled in the lower plot. Two mixed populations,
an old one with $\rm Z = 0.004$ and 10,Gyr old and another younger one with $\rm Z=0.02$, and 
different ages over the evolutionary sequence, are represented as solid lines with different colours following the value of F
as labelled in the upper diagram. We can see that models could reproduce the observations.}
\label{color_color_obs}
\end{figure*}

Finally, we plot in Fig.~\ref{color_color_obs} data obtained by 
\citet{mayya94} for a sample of H{\sc ii} regions in external galaxies.  
Each galaxy's regions are plotted with a different code as labelled in 
panel b). Only two colours and the equivalent width EW(\halpha) are 
available for this sample.  In this figure, we plot as an example the 
model with $\rm Z = 0.02$ for the young stellar population and $\rm Z = 0.004$ for the 
old one, with an age of 10\,Gyr. Different factors F are shown with 
lines of different colours as labelled in the plot. We see that most of 
data could need values of F in the range 100 $-$ 1000, with only some 
exceptional points out of the region defined by these two lines. We 
might perform a similar selection technique to see which models are the 
best ones to fit these data, or use a mean square error (MSE) analysis 
to derive the most appropriate models for each galaxy or H{\sc ii} region, 
but this is beyond the scope of  this work. What we want to show with this figure is that mixed 
population models including emission lines can reproduce the 
observations of star-forming regions, while SSPs cannot, and models 
without emission lines do not offer the same result, and ultimately 
result in a misinterpretation of the physical stellar population 
parameters.

\section{Conclusions}

\begin{itemize}

\item We have calculated the same grid of magnitudes and colours as in 
Paper I with the PopStar code, but now incorporating the contribution 
from emission lines, using for that the intensities obtained in our 
Paper II, for a large grid of young stellar ionising clusters.  We 
provide tables with uncontaminated and contaminated colours.

\item Broadband filter magnitudes are contaminated by the emission lines 
coming from ionising nebulae surrounding young stellar clusters. This 
contamination modifies the magnitudes (ranging from 0.2 to 1.5 mag 
depending on the filter) in the visible and near-infrared bands. Colours 
U-B, B-V, V-R and R-I are modified by a value between 0.2 and 0.8 mag 
depending on the colour and on the metallicity of the stellar population. 
Similar results are found in SDSS filters and should be found in other 
photometric systems.

\item The synthetic contaminated colours for SSPs in the colour-colour 
diagrams fall outside the standard canonical sequence shown by the 
uncontaminated colours or those obtained for old stellar populations. In 
many cases they show orthogonal sequences to those ones.

\item We have also computed other photometric parameters, such as 
\halpha\ and \hbeta\ luminosities, equivalent widths of \halpha\ and 
\hbeta\ and radii of the corresponding H{\sc ii} regions.

\item The equivalent widths for \halpha and \hbeta\ do not depend only 
on the age. They also depend on the metallicity of the stellar 
population. The EW(\halpha) decreases with age until 5-7\,Myr if 
Z$>0.004$. For lower metallicities, EW\halpha\ maintains positive values 
until 15-18 Myr. The same occurs with \halpha\ and \hbeta\ luminosities, 
which have high values until $\sim 20$\, Myr for the lowest 
metallicities (Z$< 0.004$).

\item The outer calculated radii of low metal H{\sc ii} regions are 
quite large compared with those observed at intermediate metallicities. 
Since these regions are not present in the observational samples, and 
taking into account that they must show high \halpha\ luminosities too, 
we need to consider the impact of observational selection effects that 
might be able to lead the production of biased H{\sc ii} region samples.

\item An evolutionary track in the plane R$_{out}$-L(H$_{\alpha}$) shows 
that the radius of the region increases when the \halpha\ luminosity 
decreases. However a positive correlation L(H${_\alpha}) - R_{out}$ 
arises when all stellar cluster masses and metallicities are drawn 
together, so the observed correlation and its dispersion may be 
explained by a mix of stellar clusters of different ages, metallicities 
and masses.

\item We have also computed a grid of models where a recent ionising 
burst is mixed with an underlying and older host stellar population in 
different proportions.  We give magnitudes and colours for these mixed 
stellar populations. In some cases contaminated colours with a low 
contribution of an old stellar population are similar to the 
uncontaminated colours with a high proportion of old population. This is 
important when interpreting observations from regions where there exists 
an underlying stellar population, since the wrong ratio F will be 
estimated if the uncontaminated colours, instead the ones contaminated by 
emission lines, are used when comparing observational data and models to 
derive physical properties of the stellar populations.

\item Other photometric parameters have also been given for each mix of 
stellar populations as before: \halpha\ and \hbeta\ luminosities, 
equivalent widths for \halpha\ and \hbeta\ and radii of the 
corresponding H{\sc ii} regions. This allows one to compare the observed 
radii of H{\sc ii} regions, their colours, and their equivalent widths of 
\halpha\ with the models. We have checked with some test cases that our 
models can reproduce photometric observational data and that it is 
possible to find the best mix of stellar populations able to fit 
simultaneously several photometrical observations.

\item All these models have been computed at redshift zero but detailed 
models will be available soon for higher redshifts, taking into account 
their shift with $z$ and the contribution to the different standard 
filters in the visible and near-IR.

\end{itemize}

\section{Acknowledgments}

This work has been partially supported by FRACTAL SLNE, DGICYT grant 
AYA2007-67965-C03-02, AYA2010-21887-C04-02 and AYA2010-21887-C04-03, and 
partially funded by the Spanish MEC under the Consolider-Ingenio 2010 
Program grant CSD2006-00070: First Science with the GTC 
(http://www.iac.es/consolider-ingenio-gtc/).  Also, partial support from 
the Comunidad de Madrid under grant CAM S2009/ESP-1496 (AstroMadrid) is 
grateful. An anonymous referee is acknowledged by useful comments that improved
this work. We would like to thank Dr. B.K. Gibson warmly for his help reviewing this manuscript 
and correcting the English version.We would like to thank the anonymous
referee for suggestions that improved this paper.

\end{document}